\documentclass[twocolumn]{aastex62}

\usepackage{amsmath,amssymb}
\usepackage[mathscr]{euscript} 
\usepackage{natbib}
\usepackage{graphicx}
\usepackage{epstopdf}
\usepackage{placeins}
\usepackage{float}
\usepackage{comment}
\usepackage{xcolor}
\usepackage{soul}
\usepackage{silence}

\newcommand{\beq}{\begin{equation}}
\newcommand{\eeq}{\end{equation}}
\newcommand{\rmpi}{{\rm \pi}}
\newcommand{\rmi}{{\rm i}}
\newcommand{\rmd}{{\rm d}}

\newcommand{\pc}{{\rm pc}}
\newcommand{\yr}{{\rm yr}}
\newcommand{\Gyr}{{\rm Gyr}}
\newcommand{\msun}{\mbox{$M_{\odot} $}}

\newcommand{\bnabla}{\mbox{\boldmath $\nabla$}}

\newcommand{\bfx}{\mbox{\boldmath $x$}}
\newcommand{\bfv}{\mbox{\boldmath $v$}}
\newcommand{\bfr}{\mbox{\boldmath $r$}}
\newcommand{\bfp}{\mbox{\boldmath $p$}}
\newcommand{\bfw}{\mbox{\boldmath $w$}}
\newcommand{\bfI}{\mbox{\boldmath $I$}}
\newcommand{\bfl}{\mbox{\boldmath $l$}}
\newcommand{\scre}{{\cal E}}
\newcommand{\scrf}{{\cal F}}
\newcommand{\scrt}{{\cal T}}
\newcommand{\rp}{\mbox{$r_{\rm p}$}}
\newcommand{\rpsq}{\mbox{$r^2_{\rm p}$}}
\newcommand{\rstar}{\mbox{$r_*$}}

\newcommand{\Mp}{\mbox{$ M_{\rm p}$}}
\newcommand{\omp}{\mbox{$\Omega_{\rm p}$}}
\newcommand{\phip}{\mbox{$\Phi_{\rm p}$}}
\newcommand{\rmax}{\mbox{$r_{\rm max}$}}
\newcommand{\Imax}{\mbox{$I_{\rm max}$}}
\newcommand{\omb}{\mbox{$\Omega_b$}}

\newcommand{\Phitilda}{\mbox{$\widetilde{\Phi}$}}
\newcommand{\Ftilda}{\mbox{$\widetilde{F}$}}

\received{---}
\accepted{---}
\published{---}
\submitjournal{ApJ}

\shorttitle{STALLING OF GLOBULAR CLUSTERS}
\shortauthors{KAUR \& SRIDHAR}

\begin{document}

\title{STALLING OF GLOBULAR CLUSTER ORBITS IN DWARF GALAXIES}

\author{KARAMVEER KAUR}
\email{karamveer@rri.res.in}
\affiliation{Raman Research Institute, Sadashivanagar, Bangalore 560 080, India}

\author{S. SRIDHAR}
\email{ssridhar@rri.res.in}
\affiliation{Raman Research Institute, Sadashivanagar, Bangalore 560 080, India}

\begin{abstract}
We apply the Tremaine--Weinberg theory of dynamical friction to compute the orbital decay of a globular cluster (GC), on an initially circular orbit inside a cored spherical galaxy with isotropic stellar velocities. The retarding torque on the GC, $\,\scrt(\rp) < 0\,$, is a function of its orbital radius $\rp\,$. The torque is exerted by stars whose orbits are resonant with the GC's orbit, and given as a sum over the infinitely many possible resonances by the Lynden-Bell--Kalnajs (LBK) formula. We calculate the LBK torque $\scrt(\rp)$ and determine $\rp(t)$, for a GC of mass $\Mp = 2 \times 10^5~\!\msun$ and an Isochrone galaxy of core mass $M_{\rm c} = 4 \times 10^8~\!\msun$ and core radius $b=1000~\!\pc$. (i) When $\rp \gtrsim 300~\!\pc$ many strong resonances are active and, as expected, $\scrt\approx \scrt_{\text{C}}\,$, the classical Chandrasekhar torque. (ii) For $\rp < 300~\!\pc$, $\,\scrt$ comes mostly from stars nearly co-rotating with the GC, trailing or leading it slightly; Trailing resonances exert stronger torques. (iii) As $\rp$ decreases the number and strength of resonances drop, so $\,\vert\,\scrt\,\vert$ also decreases, with $\,\vert\,\scrt\,\vert < 10^{-2}\,\vert\,\scrt_{\text{C}}\,\vert\,$ at $\rp= \rstar\simeq (\Mp/M_{\rm c})^{1/5}\,b \simeq 220~\!\pc\,$, a characteristic `filtering' radius. (iv) Many resonances cease to exist inside $\rstar\,$; this includes all Leading and low-order Trailing ones. (v) The higher-order Trailing resonances inside $\rstar$ are very weak, with $\,\vert\,\scrt\,\vert < 10^{-4}\,\vert\,\scrt_{\text{C}}\,\vert\,$ at $\rp = 150~\!\pc$. (vi) Inspiral times for $\rp(t)$ to decay from $300~\!\pc$ to $\rstar$ far exceed $10~\!\Gyr$.
\end{abstract}


\section{Introduction}

A globular cluster (GC) orbiting a galaxy experiences dynamical friction, the drag exerted by the gravity of the wake it generates in the galaxy. \citet{c43} derived a formula for the drag on a perturber moving through an infinite and homogeneous sea of stars with isotropic velocity distribution. When applied as a local approximation to a GC of mass $\Mp$ moving with velocity $\bfv_{\rm p}$ inside a spherical galaxy, the Chandrasekhar dynamical friction formula for the drag force is   
\beq
\Mp\frac{\rmd \bfv_{\rm p}}{\rmd t} \;=\; -\,4\rmpi\,G^2M^2_{\rm p}\,\ln{\Lambda}\,\rho\!\left(\rp; v < v_{\rm p}\right)\,\frac{\bfv_{\rm p}}{v_{\rm p}^3}\,.
\label{ch-drag}
\eeq
Here $\rp$ is the GC's orbital radius; $\,\rho\!\left(\rp; v < v_{\rm p}\right)$ is the mass density at $\rp$ of stars and dark matter with speeds less than the GC's speed; and $\ln\Lambda$ is the Coulomb logarithm 
\citep{bt08}. This drag causes loss of the GC's orbital angular momentum, making it sink towards the galactic center. The effect is so strong in dwarf galaxies that a GC on an initially circular orbit is expected to sink to the galactic center within $\mbox{few}~\Gyr\,$ \citep{t76, hg98, olr00, v00, v01, gmr06}. But many dwarf galaxies host GCs that are old \citep[e.g.][]{dhg96, mlf98, lmf04}. A particularly good example is the Fornax dwarf spheroidal galaxy with five old metal-poor GCs \citep{bcz98, bcc99, mg03, sbf03, gcc07}, already noted in \citet{t76}. Why are these GCs observed so far away from their galactic centers? 

Work over the past two decades on this `dynamical friction problem' suggests that the orbits of GCs (or other compact masses) can indeed stall in the core regions of a dwarf galaxy. \citet{hg98} used the  Chandrasekhar formula for a King model halo to argue that dynamical friction weakens in the core region of a galaxy. Numerical simulations, exploring core-stalling as a function of the cored/cuspy nature of the galaxy's inner density profile, 
have shown that a nearly constant density core would result in core-stalling \citep{gmr06,rgm06,cdr12}. Analogous core-stalling of a supermassive black hole was studied by \citet{gm08}. Numerical simulations by \citet{i09,i11} are particularly revealing, with \citet{i11} --- hereafter In11 --- providing the deepest insights through the analysis of a single numerical experiment. A semi-analytic model, based on the Chandrasekhar formula, for cored galaxies has been offered by \citet{prg16}. The physical explanations advanced differ from each other, but all would agree that dynamical friction is highly suppressed, and can even be zero, in galaxies with a nearly constant density core. Both \citet{rgm06} and \citet{i11} have emphasized the role of `co-rotating' particles in the suppression of dynamical friction, but the term seems to refer to qualitatively different orbits. 
The goal of this paper is to seek a physical interpretation of the GC stalling phenomenon, in terms of the standard theory of dynamical friction in spherical stellar systems  due to \citet{tw84} --- hereafter TW84 --- explored further in \citet{w86,w89}.
 
%\smallskip
\noindent
{\bf Physical setting of TW84:} The stars and dark matter in the galaxy can be considered to be `collisionless' over Hubble timescales, so they respond similarly to gravitational fields \citep{bt08}. The galaxy is described by a mass distribution function (DF) in six dimensional position-velocity phase space. Each mass element (henceforth referred to as a `star') orbits in the combined gravitational fields of all the other stars and the GC. In the spherically symmetric potential of the unperturbed galaxy, a stellar orbit is a `rosette' confined to a plane, with radial and angular frequencies that are functions only of $E$ (the orbital energy per unit mass) and $L$ (the magnitude of the angular momentum per unit mass). The GC is initially on a circular orbit in the $x$-$y$ plane. Its gravitational attraction perturbs and rearranges the distribution of mass in the galaxy, with an associated change in the $z$-component of the angular momentum of the galaxy. The torque on the GC is equal and opposite to the rate at which angular momentum is absorbed by the galaxy. 

The rate of absorption of angular momentum by the galaxy can be positive or negative. When the perturbation is very weak (strictly infinitesimal) angular momentum is absorbed by those `rosette' orbits whose radial and angular frequencies are in resonance with the GC's orbital frequency.
Resonances are characterized by a triplet of integers, one each for the three frequencies. Each resonance is a five dimensional surface in phase space. As the three integers run over all possible values, the set of resonant surfaces covers phase space densely. Angular momentum exchanges
with the GC can be thought of as occurring on resonant surfaces. 
The sum of the torques exerted by all the resonances is equal to the 
net torque, referred to as the `LBK torque' by TW84, who generalized an earlier derivation by \citet{lbk72} for two dimensional discs. This is valid in the linear limit of perturbation theory.\footnote{TW84 also studied the non-linear dynamics of resonances and discussed orbit capture into resonant islands. This is beyond the scope of this paper.} The simplest model of a stable, spherical galaxy is an isotropic DF, $F(E)$ with $(\rmd F/\rmd E) < 0\,$, corresponding to the DF from which the initial conditions of In11 were drawn. TW84 note the important point that in this case the torque due to each resonance is always retarding. Therefore linear theory predicts that the GC will inspiral toward the galactic center. The questions of interest are: As the GC inspirals, what are the resonances available to it? How large are the resonant torques? What is the rate of orbital decay due to the net LBK torque? 

%\smallskip
\noindent
{\bf Going forward with TW84:} When the GC's orbit lies outside the core of the galaxy, there is a dense set of resonances available to it. TW84 note that, in the limit the resonances form a continuum, the LBK torque should reduce to the Chandrasekhar torque (with suitable choice of the Coulomb logarithm). The corresponding orbital decay can be seen in the
In11 simulation of a GC of mass $2 \times 10^5~\!\msun$, set on a circular orbit of initial radius of $750~\!\pc$, inside a spherical galaxy of mass $2\times 10^9~\!\msun$ and core radius $1000~\!\pc$. During the initial 4~Gyr the GC's orbital radius decreased from $750~\!\pc$ to $300~\!\pc$, in rough agreement with the action of the Chandrasekhar formula. Thereafter, its behavior departed drastically from the formula's prediction: the rate of decrease slows down dramatically and the radius stalls around a mean value of about $225~\!\pc$ until the end of the simulation at $10$~Gyr. We are interested in understanding this stalling phenomenon.

Let us imagine --- contrary to In11 --- that the GC has somehow managed to reach close to the galaxy's center. In a limiting sense, the GC has entered a constant density environment where the galaxy's potential $\propto r^2$ (an isotropic harmonic oscillator), so every stellar orbit is a centered 
ellipse with the same orbital frequency, independent of shape, size or orientation. Then either all stars are resonant (and the response is singular), or no star is resonant and the LBK torque would vanish. TW84 remark that in realistic systems the density is not quite constant so there will always be some resonances and the response will be finite. Hence a constant density core is a pathological case. But it does illustrate the point that, were the GC to reach the very central regions of the galaxy then there would likely be no resonant stars and hence no friction. This line of thought suggests the following physical picture of core-stalling in a realistic galaxy core with a varying central density profile. As the GC inspirals from a radius of $300~\!\pc$, there are progressively fewer strong resonances available for it to exchange angular momentum with the galaxy. Since only a small fraction of the core stars would then be involved in the resonant exchange, the LBK torque exerted on the GC could be so suppressed that the rate of orbital decay to the center may take much longer than a Hubble time. 

%\smallskip
\noindent
{\bf Plan of the paper:} \S~\!2 gives a brief account of the part of TW84 that is used in this paper. We set notation describing galaxies and linear perturbations, introduce action-angle variables especially adapted for studying core dynamics later, and give a short derivation of TW84's LBK torque formula, in the spirit of \citet{k71}. In \S~\!3 the unperturbed galaxy is represented by an Isochrone model, and described by action-angle variables in the rotating frame of the GC. Isochrone parameters are chosen by comparison with those used by In11. We compute the orbital decay of the GC according to the Chandrasekhar torque for the Isochrone --- this serves 
as a useful benchmark for later comparison with decay due to the LBK torque. The GC is modeled as a Plummer sphere, whose tidal potential needs to be expressed in terms of the Isochrone action-angle variables, using the formulas for the three dimensional orbit in space. Expressions for the resonant torques and the LBK torque are recorded.

\S~\!4 takes a close look at the structure of resonances in 
the Isochrone core. Core orbits are worked out and a natural small parameter is identified. The orbital and precessional frequencies are compared with the GC's orbital frequency, yielding a characteristic radius, $\rstar$.  Torques are written in dimensionless action-angle variables. Core resonances are classified as Co-rotating and non-Co-rotating, with the latter 
consisting of higher order, weaker resonances. Co-rotating resonances come in two types, Trailing and Leading, both of which are explored in \S~\!5 as functions of $\rp$, the orbital radius of the GC. The associated torque integrals, derived in the previous section, are now computed numerically. The progressive culling of low order resonances as $\rp$ decreases is followed in detail and the role of $\rstar$ as a `filtering' radius for resonances is clarified. Resonant torques are then summed over to obtain the net Trailing and Leading torques; the LBK torque is the sum of these two torques. 

In \S~\!6 we discuss Leading and Trailing torque profiles, compute the orbital decay of the GC according to the LBK torque, and compare this with the orbital decay due to the Chandrasekhar torque studied earlier. We conclude in \S~\!7.

\section{Tremaine--Weinberg theory}
\subsection{Collisionless dynamics of the galaxy} 

We begin with a brief account of the dynamical framework that is used in 
the construction of our model --- see \citet{bt08} for a comprehensive account. Let $\bfx$ be the position vector of a star and $\bfv$ its velocity vector, with respect to an inertial frame. The galaxy is described by a DF, $f(\bfx, \bfv, t)$, which is equal to the mass density, at time $t$, in the six dimensional $\{\bfx, \bfv\}$ phase space. $f$ is non-negative and normalized as 
\beq
\int\,\rmd\bfx\,\rmd\bfv\,f(\bfx, \bfv, t) \;=\; M,
\label{f-norm}
\eeq
where $M$ is the total mass of the galaxy. Time evolution of the DF is
governed by the collisionless Boltzmann equation (CBE):
\beq
\frac{\rmd f}{\rmd t} \;\equiv\; \frac{\partial f}{\partial t} \,+\,
\bfv\cdot\frac{\partial f}{\partial\bfx} \,-\, 
\frac{\partial\Phi^{\rm tot}}{\partial\bfx}\cdot\frac{\partial f}{\partial\bfv} \;=\; 0\,,
\nonumber
\eeq 
where $\Phi^{\rm tot}(\bfx, t)$ is the total gravitational potential, equal to the sum of the potentials due to the self-gravity of the galaxy and any external perturbing sources:
\beq
\Phi^{\rm tot}(\bfx, t) \;=\; \Phi(\bfx, t) \;+\; \Phi^{\rm ext}(\bfx, t)\,,  
\label{phisum}
\eeq
where 
\beq
\Phi(\bfx, t) \;=\;-G \int\,\rmd\bfx'\,\rmd\bfv'\,\frac{f(\bfx', \bfv', t)}
{\vert\bfx - \bfx'\vert}\,
\label{phiself}
\eeq
is the self-consistent Newtonian potential. The external potential, 
$\Phi^{\rm ext}(\bfx, t)$, depends on what dynamical process is being studied. It could be due to galactic bars, or spiral density waves, infalling objects such as satellite galaxies, GCs or massive black holes.

The CBE can be rewritten compactly as:
\beq
\frac{\partial f}{\partial t} \;+\; \left[\,f, H\right] \;=\; 0\,,
\label{cbe1}
\eeq
where $H$ is the Hamiltonian (equal to the orbital energy per unit mass),
\beq
H(\bfx, \bfv, t) \;=\; \frac{v^2}{2} \;+\; \Phi^{\rm tot}(\bfx, t)\,,
\label{ham1}
\eeq
and 
\beq
\left[\,f, H\right] \;=\; \frac{\partial f}{\partial\bfx}\cdot
\frac{\partial H}{\partial\bfv} \;-\; \frac{\partial f}{\partial\bfv}
\cdot\frac{\partial H}{\partial\bfx}
\label{pb1}
\eeq
is the Poisson Bracket between the phase space functions, $f$ and $H$.
This form of the CBE is particularly useful because the Poisson Bracket 
remains invariant when we later transform from the $\{\bfx, \bfv\}$ to 
other canonically conjugate variables.

An isolated, unperturbed galaxy is often described by a time-independent DF, $f_0(\bfx, \bfv)$. Let $\Phi_0(\bfx)$ be the galaxy's self-consistent potential, related to $f_0$ through equation~(\ref{phiself}). Then the unperturbed Hamiltonian is
\beq 
H_0(\bfx, \bfv) \;=\; \frac{v^2}{2} \;+\; \Phi_0(\bfx)\,.
\label{ham01}
\eeq
Since $f_0$ solves the CBE we must have $\left[f_0, H_0\right] = 0$, 
so $f_0$ is a function of the isolating integrals of motion of $H_0$ (the Jeans theorem). For any given function of the integrals, one needs to solve the self-consistent problem of equation~(\ref{phiself}), to determine $f_0$ as a function of $(\bfx, \bfv)$. 

Let $f_1(\bfx, \bfv, t)$ be a small perturbation to $f_0$, due to either fluctuations in the initial conditions or forced through weak external gravitational fields. In either case there is a corresponding perturbation to the self-gravitational potential, $\Phi_1(\bfx, t)$, which related to $f_1$ through equation~(\ref{phiself}). Including a weak external potential, $\Phi_1^{\rm ext}(\bfx, t)$, the perturbation to the Hamiltonian is $H_1(\bfx, t) = \Phi_1(\bfx, t) + \Phi_1^{\rm ext}(\bfx, t)$. The total DF, $f = f_0 + f_1$, must obey the CBE with total Hamiltonian, $H = H_0 + H_1\,$. In the limit of an infinitesimally small perturbation, $f_1$ obeys the linearized CBE (LCBE):
\beq
\frac{\partial f_1}{\partial t} \;+\; \left[\,f_1, H_0\right]
\;+\; \left[\,f_0,\, H_1\right] \;=\; 0\,, 
\label{lcbe1}
\eeq
where the second order term, $\left[\,f_1,\, H_1\right]$, has been neglected. It is necessary that the unperturbed galaxy be linearly stable in the absence of external perturbations. In other words, the LCBE with 
$H_1 = \Phi_1$ should not admit solutions, $f_1$, that are exponentially growing in time; then $f_0$ is said to be a linearly stable DF. 

While considering the response of such a stable DF to $\,\Phi_1^{\rm ext} 
\neq 0$, TW84 neglected `gravitational polarization' effects and set $\Phi_1 = 0$. Then $H_1 = \Phi^{\rm ext}_1$, and the LCBE governing the `passive response' of the galaxy to the perturber is 
\beq
\frac{\partial f_1}{\partial t} \;+\; \left[\,f_1, H_0\right]
\;+\; \left[\,f_0,\, \Phi^{\rm ext}_1\right] \;=\; 0\,, 
\label{lcbe-pra}
\eeq
There is no real justification for the neglect of $\Phi_1$, besides the fact 
that, by doing so, one has to deal only with a partial differential equation, instead of an integral equation. We follow TW84 and use 
equation~(\ref{lcbe-pra}) to compute the linear response of the 
galaxy.\footnote{\citet{k72} argued that gravitational polarization effects in a uniformly rotating sheet can suppress dynamical friction, and showed that the frictional force on a perturber on a circular orbit is zero. Indeed, polarization effects can be important. But the demonstration of the strict vanishing of frictional force is limited by the fact that the dynamical response of a uniformly rotating sheet is as pathological as the strictly constant density galactic core, discussed earlier in the Introduction.} 

\subsection{Unperturbed galaxy}

We use spherical polar coordinates to describe the unperturbed, spherical galaxy. Let $\bfr = (r, \theta, \phi)$ be the position vector with respect to the galactic center. The unperturbed potential, $\Phi_0(r)$, is a function of only $r$. The canonically conjugate momenta are $\bfp = (p_r, \,p_\theta, \,p_\phi)$, where $p_r = \dot{r}$ is the radial velocity, $\,p_\theta = r^2\dot{\theta}$, and $\,p_\phi = r^2\sin^2\theta\dot{\phi} = \hat{z}\cdot(\bfr\times\bfp) = L_z$ is the $z$-component of the angular momentum per unit mass. The magnitude of the angular momentum per unit mass is $L = \sqrt{p_\theta^2 + p_\phi^2/\sin^2\theta\,}\,$. The unperturbed Hamiltonian,
\beq
H_0(\bfr, \bfp) \;=\; \frac{1}{2}\left(p_r^2 + \frac{p_\theta^2}{r^2}
+ \frac{p_\phi^2}{r^2\sin^2\theta}\right) \;+\; \Phi_0(r)\,,
\label{ham02}
\eeq
which is equal to the energy per unit mass ($E$), governs the dynamics of a star. An orbit is confined to an invariant plane passing through the origin, 
which is inclined to the $x$-$y$ plane by angle $i$. $(E, L, L_z)$ are constant along the orbit, with $\cos i = L_z/L\,$. The radial and angular frequencies of the orbit in its plane, $\Omega_r(E, L)$ and $\Omega_\psi(E, L)$ respectively, are generally incommensurate. Hence a generic orbit describes a `rosette', as it goes through many periapse and apoapse passages, filling an annular disc. Since $(E, L, L_z)$ are also isolating integrals of motions, the Jeans theorem implies that any spherical unperturbed DF must be a function of these three integrals. A non-rotating galaxy with complete spherical symmetry in phase space must have a DF, $f_0 = F(E, L)$, that is independent of $L_z$; such a DF has zero streaming velocities with, in general, anisotropic velocity dispersion. The subclass, $f_0 = F(E)$, has isotropic velocity dispersion and is linearly stable to all perturbations when $(\rmd F/\rmd E) < 0$; these are the models relevant to the In11 simulation. 

We can transform  from $\{\bfr, \bfp\}$ to new canonical variables, $\{\bfw, \bfI\}$, where $\bfw = (w_1, w_2, w_3)$ are three coordinates called `angles', and $\bfI = (I_1, I_2, I_3)$ are their conjugate momenta called `actions' \citep[see][\S~3.5.2]{bt08}. There are many equivalent choices of action-angle variables. The standard `primitive' variables, also used by TW84, are:
\begin{subequations} 
\begin{align}
& I_1 = J_r = \mbox{radial action}\,,\quad  w_1 \;=\; \Omega_r (t - t_p) \\[1 ex]
& I_2 = L\,, \hspace{2.1cm}  w_2 \;=\; \chi \,+\,\Omega_\psi (t - t_p) \\[1ex]  
& I_3 = L_z\,, \hspace{2cm}  w_3 = h\,. 
\end{align}
\label{aa1}
\end{subequations}
Here $h$ is the longitude of the ascending node, and $\chi$ is the angle 
from the ascending node to a periapse that is visited at time $t_p$. 
An alternative set of actions and angles is:
\begin{subequations}
\begin{align}
& I_1 = 2J_r + L\,,\quad \; w_1 \;=\; \frac{\Omega_r}{2}(t - t_p) \\[1ex]
& I_2 = L\,, \hspace{0.7cm}   w_2 \;=\; \chi \,+\, \left(\!\Omega_\psi \,-\, \frac{\,\Omega_r\,}{2}\!\right)(t - t_p)
 \\[1ex]
& I_3 =  L_z\,, \hspace{1.4cm}  w_3 = h\,.
\end{align}
\label{aa2}
\end{subequations}
This proves particularly useful for the exploration of the dynamics of core stars, for which $\Omega_r \simeq 2\Omega_\psi\,$, making $w_2$ a slowly varying angle compared to $w_1\,$. This is the choice made from \S~\!3 onward. But in this section $\{\bfw, \bfI\}$ will stand for either set of variables, defined by equations~(\ref{aa1}) or (\ref{aa2}), both giving completely equivalent descriptions of dynamics over all of phase space. 

The Hamiltonian is independent of $L_z$, and can be written as $H_0(I_1, I_2)$. Hamilton's equations of motion, $\rmd\bfw/\rmd t = \partial H_0/\partial\bfI\,$ and  $\,\rmd\bfI/\rmd t = -\partial H_0/\partial\bfw\,$, imply that the actions $\bfI$ are constants of motion and the angles advance at constant rates, with $w_3 = h$ being fixed. The unperturbed DF, $f_0 = F_0(I_1, I_2)$ describes a spherically symmetric, non-rotating system with anisotropic velocity dispersion. The subclass with isotropic velocity dispersion will be written as $f_0 = F_0(H_0)$.

\subsection{Dynamics in the rotating frame of the perturber}

The galaxy is acted upon by an external, rotating potential perturbation of the form, $\Phi_1^{\rm ext}\left(r, \theta, \phi - \xi(t)\right)$, where $\rmd\xi(t)/\rmd t = \omp(t)$ is the time-dependent angular frequency of rotation about the $\hat{z}$-axis. This could arise from galactic bar or a GC on a circular orbit in the $x$-$y$ plane. TW84 studied the effect of bar-like perturbations, whose symmetry is such that it does not change the location of the center of mass of the galaxy. We are interested in the latter case, where the GC and the galaxy orbit each other in circles about a common and fixed center of mass. This motion of the center of the galaxy must be accounted for, by choosing $\Phi_1^{\rm ext}$ as the `tidal', rather than the `bare', potential of the perturber --- this is done in \S~3. But in either case, the potential perturbation is of the above form. In a frame that rotates with angular velocity $\hat{z}\omp(t)$ the perturbation would appear stationary, so this is the preferred frame for the formulation of stellar dynamics. 

Henceforth we use $\bfr = (r, \theta, \phi)$ for the position vector in the rotating frame, with origin at the center of the galaxy, and use $\bfp = (p_r, \,p_\theta, \,p_\phi = L_z)$ to denote its conjugate momentum. In the rotating frame the line of nodes of every unperturbed orbit (regardless of its size, shape or orientation) regresses at the common angular rate, $\omp(t)$. This can be taken into account by subtracting $\omp(t) I_3$ from $H_0$, to obtain the Jacobi Hamiltonian for unperturbed dynamics:
\beq
H_{\rm J0}(I_1, I_2, I_3, t) \;=\; H_0(I_1, I_2) \;-\; \omp(t) I_3\,.
\label{ham-jac}
\eeq
The unperturbed frequencies in the rotating frame are: 
\begin{subequations}
\begin{align}
&\frac{\rmd w_1}{\rmd t} \;\equiv\; \Omega_1(I_1, I_2) \;=\; \frac{\partial H_{\rm J0}}{\partial I_1} \;=\; \frac{\partial H_0}{\partial I_1}\,, \\[1ex]
&\frac{\rmd w_2}{\rmd t} \;\equiv\; \Omega_2(I_1, I_2) \;=\; \frac{\partial H_{\rm J0}}{\partial I_2}\;=\; \frac{\partial H_0}{\partial I_2}\,, \\[1ex]
&\frac{\rmd w_3}{\rmd t} \;\equiv\; \Omega_3 \;=\; \frac{\partial H_{\rm J0}}{\partial I_3} \;=\; -\omp(t)\,.
\end{align}
\label{freq2}
\end{subequations}
The time dependence of $\omp(t)$ is due to the back-reacting torque the galaxy exerts on the perturber, and is slow compared to orbital 
times.\footnote{$\omp(t)$ must be determined self-consistently for 
the coupled galaxy-perturber system.} Henceforth we drop the explicit time dependence in $\omp$, and treat $\,H_{\rm J0}(I_1, I_2, I_3)$ as an adiabatically varying Hamiltonian.

The perturbation is applied gradually in time, in order to eliminate 
transients in the response of the galaxy: the standard manner of ensuring this is to set $\Phi_1^{\rm ext} = \exp(\gamma t)\,\phip(\bfr)$, where 
$\gamma > 0$ can be taken to zero at the end of the calculation. Writing $f_1 = \exp(\gamma t)\,F_1(\bfr, \bfp)$, and substituting for $f_1$ and 
$\Phi_1^{\rm ext}$ in the passive response LCBE~(\ref{lcbe-pra}), we obtain
\beq
\gamma F_1 \;+\; \left[\,F_1, H_{\rm J0}\right]
\;+\; \left[\,F_0,\,\phip\right] \;=\; 0\,, 
\label{lcbe-tw}
\eeq
where we have replaced $H_0$ by the adiabatically varying Jacobi Hamiltonian of equation~(\ref{ham-jac}). The Poisson Brackets in terms of the  $\{\bfw, \bfI\}$ variables are: 
\begin{subequations}
\begin{align}
\left[\,F_1, H_{\rm J0}\right] &\;=\; \Omega_1\,\frac{\partial F_1}{\partial w_1} \;+\; \Omega_2\,\frac{\partial F_1}{\partial w_2} \;-\;
\omp\,\frac{\partial F_1}{\partial w_3}\,,
\label{term1}\\[1em] 
\left[\,F_0,\,\phip\right] &\;=\; - \frac{\partial F_0}{\partial I_1}\,\frac{\partial \phip}{\partial w_1} \;-\; \frac{\partial F_0}{\partial I_2}\,\frac{\partial \phip}{\partial w_2}\,.
\label{term2}
\end{align}
\end{subequations}
We expand $F_1$ and $\phip$ as Fourier series in the angles:
\begin{subequations}
\begin{align}
F_1 &\;=\; \sum_{l_1,l_2,l_3}\Ftilda_{l_1l_2l_3}(\bfI)\,\exp\left\{\rmi\left(\bfl\cdot\bfw\right)\right\}\,,
\label{f1-fou}\\[1ex]
\phip &\;=\; \sum_{l_1,l_2,l_3} \Phitilda_{l_1l_2l_3}(\bfI)\,\exp\left\{\rmi\left(\bfl\cdot\bfw\right)\right\}\,,
\label{phip-fou}
\end{align}
\end{subequations}
where the sum is over all integer triplets $(l_1,l_2,l_3)$, and $\bfl\cdot\bfw = l_1w_1 + l_2w_2 + l_3w_3\,$. Since $F_1$ and $\phip$ are real quantities, we must have $\Ftilda_{-l_1,-l_2,-l_3}= \Ftilda^{\,*}_{l_1l_2l_3}$ and $\Phitilda_{-l_1,-l_2,-l_3}= \Phitilda^{\,*}_{l_1l_2l_3}$. 
Substituting equations~(\ref{term1})---(\ref{phip-fou}) in equation~(\ref{lcbe-tw}) and solving for $\Ftilda_{l_1l_2l_3}\,$, we obtain
\beq
\begin{split}
\Ftilda_{l_1l_2l_3}(\bfI) \;=\; &\left(l_1\,\frac{\partial F_0}{\partial I_1} \;+\; l_2\,\frac{\partial F_0}{\partial I_2}\right) \times \\[1ex]
&\frac{\rmi\,\Phitilda_{l_1l_2l_3}(\bfI)}{\left[\,\gamma \;+\; \rmi\left(l_1\Omega_1 + l_2\Omega_2 - l_3\omp\right)\,\right]}
\end{split}
\label{f1-soln}
\eeq
as the linear response of the galaxy to the imposed perturbation,
when polarization effects are ignored.

\subsection{The Lynden-Bell and Kalnajs formula}

TW84 derive the LBK torque formula by computing the change in the $z$-component of the angular momentum of individual orbits, and then summing over the contributions of all orbits. Their method is an extension of \citet{lbk72} to spherical systems, and requires calculating the angular momentum changes to second order in the perturbation. However, the original derivation for flat discs by \citet{k71} only requires the first order change in the DF, and does not need computation of individual orbits to second order. Below we present a short and simple derivation of the LBK formula for spherical systems in the spirit of \citet{k71}. The $z$-component of the torque exerted by the galaxy on the perturber is equal and opposite to the torque exerted by the perturber on the galaxy. Hence the LBK torque on the perturber is:
\beq
\scrt = \int\!\rmd\bfr\,\rmd\bfp
\frac{\partial\Phi_1^{\rm ext}}{\partial\phi}\{f_0 + f_1\}
= \int\!\rmd\bfr\,\rmd\bfp
\frac{\partial\Phi_1^{\rm ext}}{\partial\phi}f_1\,.
\label{torq1}
\eeq
It is second order in the perturbation because $f_0$ is independent of 
$\phi$ and $\oint\rmd\phi\left(\partial \Phi_1^{\rm ext}/\partial\phi\right) 
= 0\,$. The decisive step is to note that $\,\left(\partial\Phi_1^{\rm ext}/\partial\phi\right) = -\left[\,p_\phi\,, \Phi_1^{\rm ext}\right] = 
-\left[\,L_z\,, \Phi_1^{\rm ext}\right]$, and use the invariance of the 
Poisson Bracket to rewrite equation~(\ref{torq1}) in terms of 
action-angle variables:
\begin{align}
&\scrt = - \!\int \!\! \rmd\bfw\rmd\bfI
\left[I_3,\,\Phi_1^{\rm ext}\right]\! f_1 =
\exp(2\gamma t)\!\int \!\! \rmd\bfw\rmd\bfI\,
\frac{\partial\phip}{\partial w_3}\,F_1\nonumber\\[1em]
&= 8\rmpi^3 \exp(2\gamma t)\!\!\!\sum_{l_1, l_2, l_3}\!\rmi l_3\!\! \int\!
\rmd\bfI\,\Phitilda_{l_1l_2l_3}(\bfI)\,\Ftilda^{\,*}_{l_1l_2l_3}(\bfI)\,.
\label{torq2}
\end{align}
Using equation~(\ref{f1-soln}) to express $\Ftilda^{\,*}_{l_1l_2l_3}$ in terms of $\Phitilda^{\,*}_{l_1l_2l_3}\,$,  
\beq
\begin{split}
\scrt = 8\rmpi^3 \exp(2\gamma t)&\!\!\sum_{l_1, l_2, l_3}\!l_3\int \rmd\bfI\,\left(l_1\,\frac{\partial F_0}{\partial I_1} + l_2\,\frac{\partial F_0}{\partial I_2}\right)\times \\[1ex] 
&\frac{\vert\Phitilda_{l_1l_2l_3}(\bfI)\vert^2}{\left[\,\gamma \;-\; \rmi\left(l_1\Omega_1 + l_2\Omega_2 - l_3\omp\right)\,\right]}\;.
\end{split}
\label{torq-kal}
\eeq
In the limit $\gamma \to 0_+\,$, we have \beq
\begin{split}
& \left[\,\gamma - \rmi\left(l_1\Omega_1 + l_2\Omega_2 - l_3\omp\right)\,\right]^{-1} \to \\[1ex]
& \rmi\left(l_1\Omega_1 + l_2\Omega_2 - l_3\omp\right)^{-1} + \rmpi\,\delta\!\left(l_1\Omega_1 + l_2\Omega_2 - l_3\omp\right)\,,
\nonumber
\end{split}
\eeq
where $\delta(\,)$ is the Dirac delta-function that picks out resonances.
The first term on the right side is pure imaginary and cannot contribute to the torque which is a real quantity; and indeed it does not, as can be seen by noting that the $(l_1, l_2, l_3)$ term is canceled by the $(-l_1, -l_2, -l_3)$ term because $\vert\Phitilda_{l_1l_2l_3}(\bfI)\vert^2 =
\vert\Phitilda_{-l_1,-l_2,-l_3}(\bfI)\vert^2$. On the other hand both terms add for the $\delta$-function contribution. Therefore we arrive at the LBK formula for the torque:\footnote{This is identical to equation~(66) of TW84, when $\Phitilda_{l_1l_2l_3}\to \Psi_{l_1l_2l_3}/2$,
to correspond to the differing conventions used in the definition of the Fourier-coefficients of the perturbing potential.}
\begin{subequations}
\begin{align}
\scrt &\;=\; \sum_{l_1,l_2 = -\infty}^{\infty}\;
\sum_{l_3 \,=\, 1}^{\infty}\;\; \scrt_{l_1l_2l_3}\;,\qquad\quad\mbox{where}
\label{torq-lbk}\\[1em]
\scrt_{l_1l_2l_3} &\;=\; 
16\rmpi^4\,l_3\!\!\int\rmd\bfI\left(l_1\,\frac{\partial F_0}{\partial I_1} 
+ l_2\,\frac{\partial F_0}{\partial I_2}\right) \times \notag\\[1ex]
& \hspace{1cm}  \delta\!\left(l_1\Omega_1 + l_2\Omega_2 - l_3\omp\right)\vert\Phitilda_{l_1l_2l_3}(\bfI)\vert^2\,.
\label{torq-res}
\end{align}
\end{subequations}
The total torque acting on the perturber is the sum of the torques 
exerted by all the resonant surfaces. Each resonant surface is a
five dimensional subspace of six dimensional phase space defined by the resonance condition,
\beq
l_1\Omega_1(I_1, I_2) \;+\; l_2\Omega_2(I_1, I_2) \;-\; l_3\omp \;=\; 0\,,
\label{res-cond}
\eeq
for any specified integer triplet, $(l_1, l_2, l_3 > 0)$. The resonance condition is independent of the third action, $I_3 = L_z\,$, because 
the unperturbed galaxy is spherical. 

The unperturbed galaxy relevant to the In11 simulation is represented by
a stable, spherical DF with isotropic velocity dispersions, of the 
form $F_0(E)$ with $\rmd F_0/\rmd E < 0\,$. TW84 make the important point 
that the LBK torque is always retarding. To see this, we note that
\beq 
\left(l_1\frac{\partial F_0}{\partial I_1} + l_2\frac{\partial F_0}{\partial I_2}\right) \;=\; (l_1\Omega_1 + l_2\Omega_2)\,\frac{\rmd F_0}{\rmd E} 
\;\to\; l_3\omp\frac{\rmd F_0}{\rmd E}
\nonumber
\eeq
because of the $\delta$-function.
Then the LBK torque is:
\begin{subequations}
\begin{align}
&\!\!\!\scrt \;=\; \sum_{l_1,l_2 = -\infty}^{\infty}\;
\sum_{l_3 \,=\, 1}^{\infty}\;\; \scrt_{l_1l_2l_3}\;,\qquad\quad\mbox{where}
\label{torq-fe}\\[1em] 
&\!\!\!\scrt_{l_1l_2l_3} = 
16\rmpi^4\,l_3^2\,\omp\int \rmd\bfI\;\frac{\rmd F_0}{\rmd E}\;\times \notag \\[1ex]
& \hspace{1.5cm} \delta\!\left(l_1\Omega_1 + l_2\Omega_2 - l_3\omp\right)\vert\Phitilda_{l_1l_2l_3}(\bfI)\vert^2\,.
\label{torq-res-fe}
\end{align}
\end{subequations}  
Therefore the torque due to each resonance, $\scrt_{l_1l_2l_3}\,$, has a sign that is opposite to $\omp$; the perturber always experiences a retarding torque. 

\section{Model of dynamical friction on a GC}

The unperturbed galaxy is chosen to have an Isochrone DF because it is a realistic representation of a stable spherical galaxy with isotropic velocity dispersion, with remarkably simple analytical representations of physical quantities \citep{h59a, h59b, h60, bt08}. The perturber is a GC on a circular orbit in the $x$-$y$ plane, modeled as a Plummer sphere of small core radius. The rate of decay of the GC's orbital radius is determined by the back-reacting torque exerted by the galaxy.  

\subsection{Isochrone galaxy model}

The gravitational potential of the Isochrone model is
\beq
\Phi_0(r) \;=\; -\,\frac{GM}{\,b + \sqrt{b^2 + r^2}\,}\,,
\label{phi-iso}
\eeq
where $M$ is the total mass of the galaxy, and $b$ is the core radius.
The mass density profile that gives rise to this potential, 
\beq
\begin{split}
&\rho_0(r) \;=\; \frac{1}{4\rmpi G}\bnabla^2\Phi_0(r) \\[1em]
& \,=
M\left[\frac{3(b^2 + r^2)\left(b + \sqrt{b^2 + r^2}\right)
-r^2\left(b + 3\sqrt{b^2 + r^2}\right)}
{4\rmpi \left(b + \sqrt{b^2 + r^2}\right)^3 (b^2 + r^2)^{3/2}}
\right],
\end{split}
\label{rho-iso}
\eeq
is a decreasing function of $r$: the central density is $\rho_0(0) = 3M/16\rmpi b^3$ and $\rho_0(r) \to bM/2\rmpi r^4$ as $r\to\infty$. The stellar mass enclosed within a radius $r$ is
\beq
\begin{split}
M_0(r) & \;=\; 4\rmpi\int_0^r\,\rmd r'\,r'^{\,2}\,\rho_0(r')\\
&\;=\; 
M\left[\frac{r^3}{\left(b + \sqrt{b^2 + r^2}\right)^2\sqrt{b^2 + r^2}\,}
\right]\,.
\end{split}
\label{mass-iso}
\eeq
A GC of mass $\Mp$ is on a circular orbit in the $x$-$y$ plane
of the galaxy. When the GC is at a radius $\rp$ from the galactic center, its angular frequency of rotation is
\beq
\omp(\rp) \;=\; \sqrt{\frac{G\left[\,M_0(\rp) \,+\, \Mp\,\right]}{\rp^3}\;}
\,. 
\label{omp-def}
\eeq
As in \S~2.3 we use $\bfr = (r, \theta, \phi)$ for the position vector in the rotating frame, with origin at the center of the galaxy, and use $\bfp = (p_r, \,p_\theta, \,p_\phi)$ to denote its conjugate momentum. The orbital energy per unit mass is, 
\beq
H_0 \;=\; \frac{1}{2}\left(p_r^2 + \frac{p_\theta^2}{r^2}
+ \frac{p_\phi^2}{r^2\sin^2\theta}\right) \;+\; \Phi_0(r)\,.
\label{h0-iso}
\eeq
The Jacobi Hamiltonian governing unperturbed dynamics in the rotating frame is
\beq
H_{\rm J0} \;=\; H_0 \;-\; \omp(\rp)\,p_\phi\,.
\label{hj0-iso}
\eeq

We now switch to the action-angle variables of equation~(\ref{aa2}). 
Dropping all indices, we have
\begin{subequations}
\begin{align}   
&\mbox{Actions} \;=\; \left(I, \,L, \,L_z\right),\quad \mbox{where} 
\quad I = 2J_r + L\,;\\[1em]
&\mbox{Angles} \;=\; \left(w, g, h\right),\quad \mbox{where}\\[1ex]
&\qquad w \;=\; \frac{\Omega_r}{2}(t - t_p)\,,\nonumber\\[1ex] 
&\qquad g \;=\; \chi + 
\left(\!\Omega_\psi \,-\, \frac{\,\Omega_r\,}{2}\!\right)(t - t_p)\,,\nonumber\\[1ex]
&\qquad h \;=\; \mbox{longitude of the ascending node.}\nonumber
\end{align}
\label{AA-final}
\end{subequations}
The radial and angular frequencies are 
\beq
\begin{split}
& \Omega_r(I, L) \;=\; \frac{8(GM)^2}{\left[\,I + \sqrt{I_b^2 + L^2\,}\,\right]^3}\,,\\[1ex]
& \Omega_\psi(I, L) \;=\; 
\frac{\Omega_r}{2}\left( 1 + \frac{L}{\sqrt{I_b^2 + L^2\,}}\right)\,,
\end{split}
\label{}
\eeq
where $I_b = \sqrt{4GMb\,} = \mbox{constant}$. The orbital energy per unit mass, $H_0 = E$, is
\beq
E(I, L) \;=\; -\,\frac{2(GM)^2}{\left[\,I + \sqrt{I_b^2 + L^2\,}\,\right]^2}\,.
\label{en-iso}
\eeq
Hence the Jacobi Hamiltonian governing dynamics in the rotating frame
is a simple function of the three actions:
\beq
H_{\rm J0}(I, L, L_z; \rp) \;=\; E(I, L) \;-\; \omp(\rp)\,L_z\,, 
\label{hj0}
\eeq
where we have included the dependence on $\rp(t)$, the radius of the GC's circular orbit. This is an adiabatically varying function 
of time (`orbital decay') which is calculated self-consistently in \S~\!6.
The three actions, $(I, L, L_z)$, are constants along an orbit. The unperturbed 
frequencies are
\begin{subequations}
\begin{align}
&\Omega_w(I, L) \;=\; 
\frac{\partial H_{\rm J0}}{\partial I} \;=\; \frac{4(GM)^2}{\left[\,I + \sqrt{I_b^2 + L^2\,}\,\right]^3}\,,
\label{omw}\\[1ex]
&\Omega_g(I, L) \;=\; 
\frac{\partial H_{\rm J0}}{\partial L} \;=\; \frac{L}{\sqrt{I_b^2 + L^2\,}}\,\Omega_w(I, L)\,,
\label{omg}\\[1em]
&\Omega_h(I, L) \;=\; 
\frac{\partial H_{\rm J0}}{\partial L_z} \;=\; -\,\omp(\rp)\,.
\label{omh}
\end{align}
\end{subequations}

The DF with isotropic velocity dispersion is:
\begin{align}
& F_0(E) =  \frac{M}{\sqrt{2}\,(2\rmpi)^3\,(GMb)^{3/2}}
\frac{\sqrt{\scre}}{\left[\,2\left(1 - \scre\right)\,\right]^4}\; \times 
\nonumber\\[1ex]
& \qquad \bigg[ 27 - 66\scre + 320\scre^2 - 240\scre^3 + 64\scre^4 
\nonumber\\
& \qquad\quad + 3\left(16\scre^2 + 28\scre - 9\right)
\frac{\arcsin{\sqrt{\scre}}}{\sqrt{\scre(1-\scre)}}\,
\bigg]\,,
\label{df-iso}
\end{align}
where $\scre = -Eb/GM$ is a dimensionless measure of the binding energy,
with $0 < \scre \leq 1/2\,$. $F_0(E)$ is a decreasing function of $E$, and hence has the desirable property of being linearly stable to perturbations. 

\subsubsection{Choice of parameters}

We now choose Isochrone parameters, central density $\rho_0(0)$ and 
core radius $b$, such that they are broadly consistent with In11, 
whose simulation used a \citet{b95} density profile for the galaxy: 
\beq
\rho_{\rm B}(r) \;=\; \frac{\rho_{\rm c}r_{\rm c}^3}
{\left(r_{\rm c} + r\right)\!\left(r_{\rm c}^2 + r^2\right)}\,,
\label{rho-bur}
\eeq
with core radius $r_{\rm c} = 1000~\!\pc$ and central density 
$\,\rho_{\rm c} = 0.1~\!\msun\pc^{-3}$. $\,\rho_{\rm c}$ was determined 
by requiring that the mass inside $300~\!\pc$ is about $10^7~\!\msun\,$, 
consistent with observations of dwarf galaxies \citep{sbk08}. 
Setting the Isochrone core radius $b = r_{\rm c} = 1000~\!\pc$, we
solve for $\rho_0(0)$ by setting $M_0(300~\!\pc) = 10^7~\!\msun\,$
in equation~(\ref{mass-iso}). This gives $\rho_0(0) = 0.096~\!\msun\pc^{-3}$, which is very close to $\rho_{\rm c}$. So our Isochrone model
has the same core radius and central density as the Burkert profile in In11.
We note that the total mass of the Isochrone is $M = 1.6 \times 10^9~\!\msun$, whereas the total mass in the Burkert profile is infinite because 
$\rho_{\rm B}(r) \propto r^{-3}$ for $r \gg  r_{\rm c}$. But this large $r$ behavior has no bearing on the core dynamics of interest to us. Indeed it proves more useful --- see \S~\!7 --- to define the `core' mass, 
\beq
M_{\rm c} \;=\; \frac{4\pi}{3}\rho_0(0)b^3 \;\simeq\;
\frac{4\pi}{3}\rho_{\rm c}r^3_{\rm c} \;\simeq\; 4\times 10^8\msun\,,
\label{core-mass}
\eeq  
which is the same for both galaxy profiles. 

\subsection{Expectation from the Chandrasekhar formula}

\begin{figure*}
\gridline{\fig{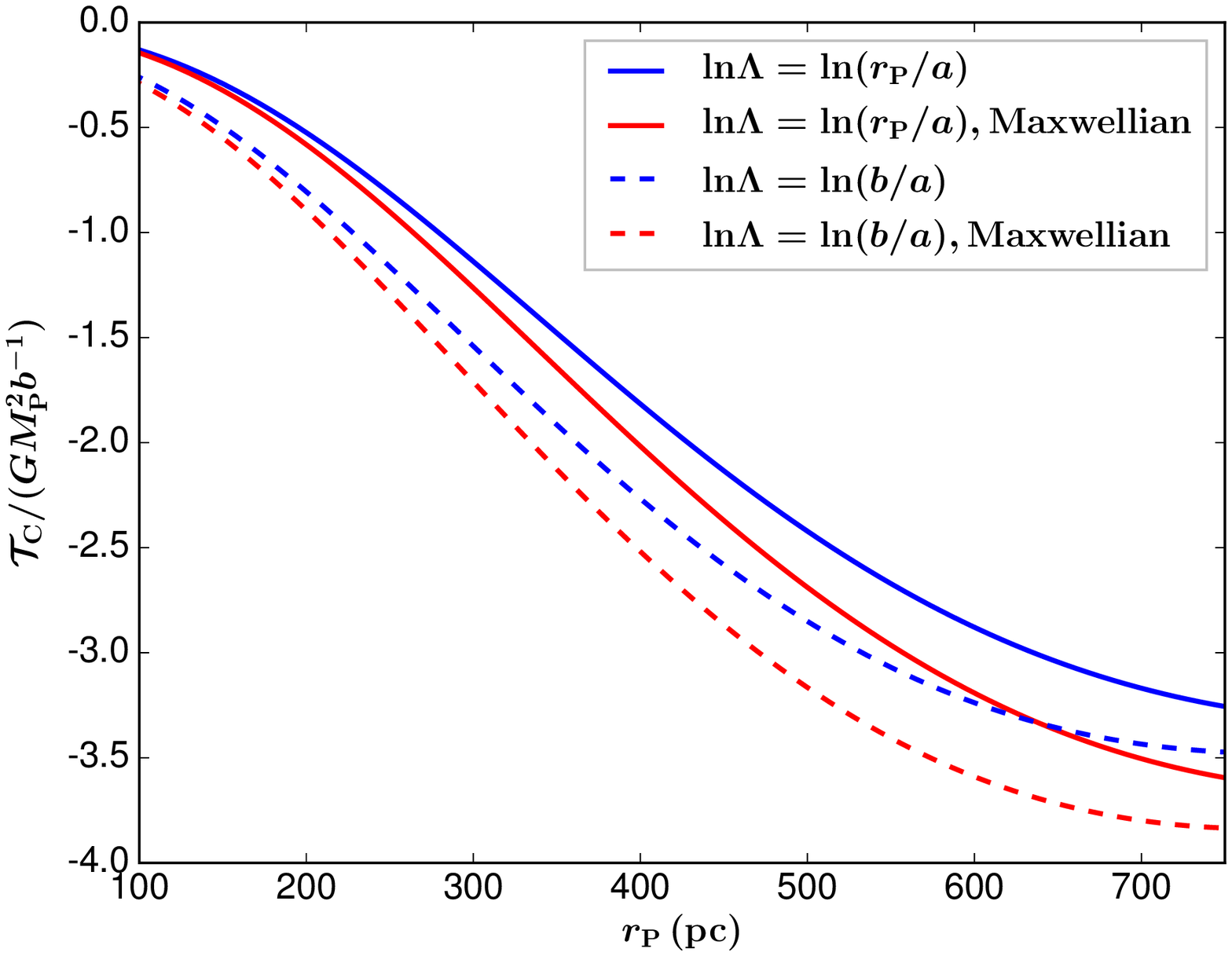}{0.52\textwidth}{(a) Torque profiles %torq_ch.eps
$\scrt_{\text{C}}(\rp)\,$.}
         \hspace{-0.4cm} \fig{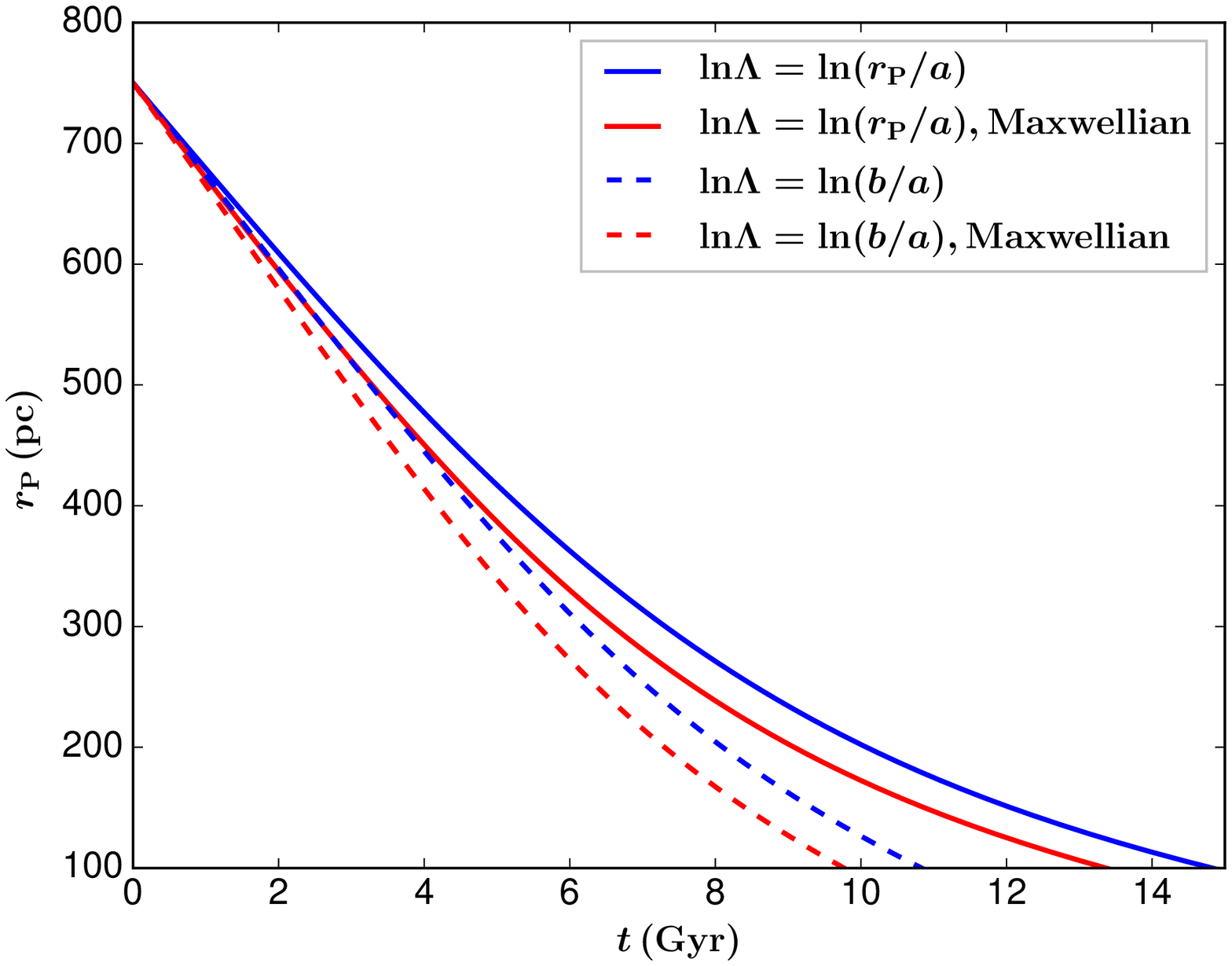}{0.52\textwidth}{(b) Orbital decay $\rp(t)$.} %rp_t-ch
          }\vspace{-0.0cm}
\caption{\emph{Torque profiles and orbital decay in the Isochrone model,
according to the Chandrasekhar formula}. Blue curves are for $\rho\!\left(\rp; v < \omp\rp\right)$ determined by integrating the Isochrone DF of equation~(\ref{df-iso}) over velocities. Red curves are for $\rho\!\left(\rp; v < \omp\rp\right)$
 determined by using the Isochrone density profile of equation~(\ref{rho-iso}) and a `local' Maxwellian distribution of velocities. Solid and dashed lines correspond to the two different choices of $\ln\Lambda$, as explained in the text.}
\vspace{0.2cm}
\label{decay-ch}
\end{figure*}

Before computing the orbital decay, $\rp(t)$, of a GC using the LBK torque, we need a benchmark in terms of what one may expect in an Isochrone galaxy, according to the Chandrasekhar formula. Similar to In11, we assume that the GC has mass $M_p = 2\times 10^5~\!\msun$ and core radius $a = 10~\!\pc$. Equation~(\ref{ch-drag}) implies that the rate of loss of the GC's orbital angular momentum is: 
\beq
\Mp\,\frac{\rmd}{\rmd t}\!\left(\omp\rpsq\right) \;=\; \scrt_{\text{C}}(\rp)\,,
\label{rp-eqn-ch}
\eeq
where  
\beq 
\scrt_{\text{C}}(\rp) = -\,4\rmpi G^2M^2_{\rm p}\ln{\Lambda}\,\rho\!\left(\rp; v < \omp\rp\right)\frac{1}{\Omega^{\,2}_{\rm p}\,\rp}
\label{torq-ch}
\eeq
is the Chandrasekhar torque. We need to determine the two quantities, 
$\ln{\Lambda}\,$ and $\rho\!\left(\rp; v < \omp\rp\right)$. This is a `local' approximation, so some sense needs to be made of the Coulomb logarithm. The standard choice is discussed in \citet{bt08}, and modifications have been suggested. We consider two extreme choices which should act as upper and lower bounds on what can be expected. These are $\ln{\Lambda} = \ln(b/a) = 4.6$ and $\ln{\Lambda} = \ln(\rp/a)$, which varies from $4.3$ to $3.4$ as $\rp$ varies from $750~\!\pc$ to $300~\!\pc$. The quantity $\rho\!\left(\rp; v < \omp\rp\right)$ is the mass density at $r$ of stars with speeds less than $\omp\rp$. The direct way to calculate this is by integrating the Isochrone DF of equation~(\ref{df-iso}) over velocities. But it is also traditional \citep[In11]{bt08} to simplify further by pushing the `local' nature of approximation thus: the DF is assumed to be the product of the galaxy's density profile (e.g. the Isochrone $\rho_0(r)$ of 
equation~\ref{rho-iso}) and a Maxwellian distribution of velocities with dispersion $\sigma(r)$ determined by the Jeans equations of hydrostatic equilibrium.

With two choices each, for $\rho\!\left(\rp; v < \omp\rp\right)$ and 
$\ln{\Lambda}\,$, we get four different functional forms for the Chandrasekhar torque --- see Figure~(\ref{decay-ch}a).  Substituting these in equation~(\ref{rp-eqn-ch}) and integrating with initial condition, 
$\rp = 750~\!\pc$ at $t = 0\,$,  we obtain the orbital decay of $\rp(t)$ for the four cases --- see Figure~(\ref{decay-ch}b). For $\,300~\!\pc < \rp < 750~\!\pc\,$, the torques differ from each other by factors less than $1.5\,$. The time for $\rp$ to decay from $750~\!\pc$ to $300~\!\pc$ varies from $5.6~\!\Gyr$ to $7.3~\!\Gyr$. After $10~\!\Gyr$ all models of the Chandrasekhar torque predict that the GC would be within $200~\!\pc$ of the center.  
 
 \pagebreak
\subsection{LBK torque}

The LBK torque of equations~(\ref{torq-fe}, \ref{torq-res-fe}) is:
\begin{subequations}
\begin{align}
&\scrt(\rp) \;=\; \sum_{n, \ell \,=\, -\infty}^{\infty}\;
\sum_{m \,=\, 1}^{\infty}\;\; \scrt_{n\ell m}(\rp)\;,\quad\mbox{where}
\label{torq-iso}\\[1em]
& \scrt_{n\ell m}(\rp) \;=\; 
16\rmpi^4\,m^2\,\omp\int_0^\infty\rmd I \int_0^I\rmd L 
\;\frac{\rmd F_0}{\rmd E}\, \times \nonumber\\[1ex]
&\delta\!\left(n\Omega_w + \ell\Omega_g - m\omp\right)
\int_{-L}^L\rmd L_z\,\vert\Phitilda_{n\ell m}(I,L,L_z)\vert^2\,.
\label{torq-res-iso}
\end{align}
\end{subequations}

We need to compute $\,\Phitilda_{n\ell m}(I,L,L_z)\,$ for a GC modeled  as a rigid Plummer sphere of mass $\Mp$ and core radius $a = 10~\!\pc$, as in In11. Its position vector with respect to the center of the galaxy, $\bfr_{\rm p}$, is quasi-stationary in the rotating frame. The perturbing potential experienced by a star at $\bfr$ is equal to the tidal potential due to the GC.\footnote{In the terminology of planetary dynamics $\phip$ is the `disturbing function', which is the sum of the `direct' and `indirect' terms \citep{md99}.}
\beq
\phip \;=\; \frac{-GM_p}{\sqrt{\;a^2 + \left\vert\bfr - \bfr_{\rm p}\right\vert^2}\;}  \;+\; \frac{\;GM_p\left(\bfr\cdot\bfr_{\rm p}\right)\;}
{\,\left(a^2 + \rpsq\right)^{3/2}\,}\;. 
\label{pert-tid}
\eeq

The function $\,\phip(w, g, h; I, L, L_z)\,$ can be obtained by writing 
$\bfr$ in terms of the action-angle variables of equation~(\ref{AA-final}). The orbital plane is determined by its constant inclination, $i = \arccos(L_z/L)$, and the longitude of ascending node, $h$. The radius ($r$) and angle in the orbital plane ($\psi$) can both be expressed in terms of $\{w,g,h; I, L, L_z\}$ by first defining three quantities \citep[see][\S~3.5.2]{bt08},
\begin{subequations}
\begin{align}
c(I, L) &\;=\; -\frac{GM}{2E} \;-\; b\,,
\label{c-def}\\[1ex]
e^2(I, L) &\;=\; 1 - \frac{L^2}{GMc}\left(1 + \frac{b}{c}\right)\,,
\label{e-def}\\[1ex]
w &\;=\; \frac{1}{2}\left(\,\eta \;-\; \frac{ec}{\,c + b\,}\,\sin\eta\right)\,.
\label{eta-def}
\end{align}
\end{subequations}
Here $c(I, L)\,$ is a length scale, $e(I, L)$ is an `eccentricity' and  
$\eta$ is an `eccentric anomaly'. Then 
\begin{subequations}
\begin{align}
&r^2 \;=\; \left[\,b \,+\, c(1 - e\cos\eta)\,\right]^2 - b^2\,,
\label{rsq-def}\\[1ex]
&\psi \;=\; g \;+\; \arctan\!\left(\sqrt{\frac{1+e}{1-e}\,}\,\tan(\eta/2)\right) \;+\;\nonumber\\[1ex]
&\quad\frac{L}{\sqrt{I_b^2 + L^2}}\left\{\arctan\!\left[
\sqrt{\frac{1 + e + 2b/c}{1 - e + 2b/c}}\,\tan(\eta/2)
\right] \,-\, w\right\}\label{psi-def}
\end{align}
\end{subequations}
give $(r, \psi)$ in terms of $(I, L, \eta, g)$, and equation~(\ref{eta-def}) can be used to express $\eta$ in terms of $(w, I, L)$ as required. The Fourier coefficients of $\phip$ of equation~(\ref{pert-tid}) can then be computed as: 
\beq
\begin{split}
\Phitilda_{n\ell m}(I,L,L_z) \;=\;
&\oint \frac{\rmd w}{2\rmpi}\,\frac{\rmd g}{2\rmpi}\,
\frac{\rmd h}{2 \rmpi}\;
\phip(w,g,h;I,L,L_z)\, \times \\
&\exp\left\{-\,\rmi\left(n w + \ell g + m h\right)\right\}\,.
\end{split}
\label{phi-tilda}
\eeq

\section{Resonances in the Isochrone core}

\subsection{Unperturbed orbits}

The $(n,\ell,m)$ resonance,
\beq
n\,\Omega_w(I, L) \;+\; \ell\,\Omega_g(I, L) \;-\; m\,\omp(\rp) \;=\; 0\,,
\label{res-iso}
\eeq
is a curve in the $(I, L)$ plane. Resonant stars with orbital sizes
comparable to the core radius have $\Omega_g$ comparable to $\Omega_w$. 
This implies that solutions of equations~(\ref{res-iso}) are possible for numerous triplets, $(n,\ell,m >0)$. So there are many resonances of comparable strengths in operation. TW84 argue that, in the limit the resonances form a continuum, the LBK torque should reduce to the Chandrasekhar torque. $\ln\Lambda = \ln(b_{\rm max}/b_{\rm min})$ where $b_{\rm max}$ and $b_{\rm min}$ are the maximum and minimum impact parameters of the encounter between the GC and stars. We set $b_{\rm max} = \rp$ the orbital radius of the GC \citep[see][chap. 8]{bt08} and $b_{\rm min} = a = 10~\!\pc\,$, to match the In11 value of the GC's core radius. Then $\ln\Lambda = \ln(\rp/a)$ varies from $4.3$ at $\rp = 750~\!\pc$ to $3.4$ at $\rp = 300~\!\pc$, which is approximately consistent with the constant best-fit value of $3.72$ used by In11. 

When $\rp = 300~\!\pc$, the Chandrasekhar formula is still approximately valid with $b_{\rm max} = 300~\pc$, as we confirm in \S~\!6. For $\rp < 300~\!\pc$ the rate of orbital decay slows down dramatically in In11. There is significant departure from the predictions of the Chandrasekhar torque until it breaks down completely when the cluster stalls at a mean orbital radius of about $225~\!\pc$, being affected by stars with orbital radii $\approx 400~\!\pc$ (i.e. stars with $b_{\rm max} \approx 200~\!\pc\,$). Hence, in order to describe dynamical friction for $\rp \leq 300~\!\pc$, we can focus attention on stars  whose orbital radii $\lesssim 600~\!\pc\,$. These stars oscillate well within the core radius of the galaxy, $b = 1000~\!\pc$, so $I \ll I_b\,$ (we always have $0\leq L \leq I$). Then equations~(\ref{c-def})---(\ref{eta-def}) reduce to 
\beq
c \,\simeq\, 2\left(\frac{I}{I_b}\right)b\,,\qquad
e^2 \,\simeq\, 1 \,-\, \frac{L^2}{I^2}\,,\qquad
w \,\simeq\, 2\eta\,.
\label{cw-core}  
\eeq
Also equations~(\ref{rsq-def}) and (\ref{psi-def}) now simplify:
\begin{subequations}
\begin{align}
& r^2 \,\simeq\, \frac{I}{\omb}\left[\,1 - e\cos(2w)\,\right]\,,\\[1ex]
& \psi \;\simeq\; g \;+\; \arctan\!\left(\sqrt{\frac{1+e}{1-e}\,}\,\tan w\right)\,,
\end{align}
\label{rpsi-core}
\end{subequations}
where 
\beq
\omb \;=\; \frac{1}{2}\sqrt{\frac{GM}{b^3}\,} \;=\; 
\sqrt{\frac{4\pi}{3}G\rho_0(0)}\,,
\label{omb-def}
\eeq
depends only on the central density. The mean-squared orbital radius is $r_{\rm rms}^{\,2} = I/\omb\,$. We consider the response of stars with $r_{\rm rms} \leq \rmax \simeq 632~\!\pc\,$. This provides a dynamically useful definition of `core stars' as those whose action variable $I \leq \Imax = \omb\,r^2_{\rm max}\,$. 
We define 
\beq
\varepsilon \;=\; \frac{\Imax}{I_b} \;=\; \frac{1}{4}\left(\frac{\rmax}{b}\right)^2 \;=\; \frac{1}{10}\,.
\label{eps-def}
\eeq
Since $I, L < \Imax$, both $I/I_b$ and $L/I_b$ are smaller than 
$\varepsilon$ in the core. So $\varepsilon$ is a natural small parameter of the problem. To first order in $\varepsilon$ the unperturbed frequencies of equations~(\ref{omw}) and (\ref{omh}) reduce to
\beq
\Omega_w(I) \;\simeq\; \omb\!\left(\!1 \,-\, 3\,\frac{I\,}{I_b\,}\!\right),\quad\;
\Omega_g(L) \;\simeq\; \omb\!\left(\!\frac{L\,}{I_b\,}\!\right),
\label{freq-core}
\eeq
The maximum error in this is $O(\varepsilon^2) = 1\%$.

A core star moves on an ellipse with orbital frequency $\Omega_w(I)$, while the apsides of the ellipse precess forward with frequency $\Omega_g(L)$. 
The frequencies are not constants but depend on $(I, L)$.\footnote{In an exactly constant density core we would have $\Omega_w = \omb = \mbox{constant}$ and $\Omega_g = 0$. But this exactly harmonic limit is pathological, as discussed in \S~\!1 and \S~\!7.} The orbital plane maintains a constant inclination $i = \arccos(L_z/L)$ with its line of nodes precessing with frequency $-\omp(\rp)\,$ (because the orbit is viewed in the rotating frame). 

As $I$ varies from $0$ to $\Imax\,$, $\,\Omega_w$ decreases from $\,\omb\,$ to $\,(1-3\varepsilon)\omb\,$, so $\,\omb$ is the central orbital frequency  with corresponding time period, $T_b = 2\rmpi/\omb \simeq 1.48\times 10^8~\yr\,$. As $L$ varies from $0$ to its maximum possible value of $\Imax\,$, $\,\Omega_g$ increases from $0\,$ to $\,\varepsilon\omb\,$.  Let $T_w = 2\rmpi/\Omega_w$, $T_g = 2\rmpi/\Omega_g$  be the orbital and apse precession time periods, respectively. Then we have   $(1-3\varepsilon)^{-1}\,T_b \,\geq\, T_w \,\geq\, T_b\,$ so that $T_w \sim\mbox{few}\times 10^8~\yr$ and $\,T_g \,\geq\, \varepsilon^{-1}\,T_b = 1.48\times 10^9~\yr$. The nodal regression period, $T_h = 2\rmpi/\omp\,$,
depends on $\rp$. In order to follow the In11 simulation it will
suffice to look at $150~\!\pc < \rp < 300~\!\pc\,$, in which range we have   
$\omp \simeq \omb$, as can be verified using equation~(\ref{omp-def}). 
Hence $\,T_h \simeq T_b \simeq 1.48\times 10^8~\yr\,$. 

Below we study resonances involving the three time periods, $T_w$, $T_g$ and $T_h\,$. Of these $T_w$ and $T_h$ are of comparable magnitudes, but $T_g$ is more than ten times longer than these. Hence we may expect the dominant resonances to involve close cancellation between $T_w$ and $T_h$, corresponding to resonant stars that nearly co-rotate with the GC, 
as discussed in \S~\!4.4.

\subsection{Resonance filtering radius $\rstar$}

\begin{figure}
\centering
\hspace{-0.5cm}
\includegraphics[width=0.52\textwidth]{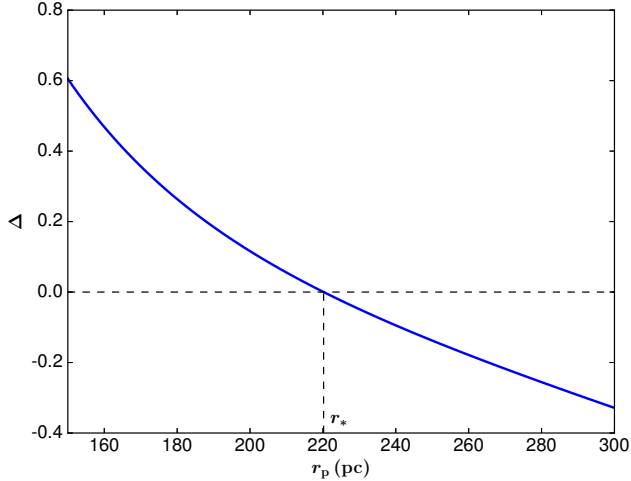}
%,trim={5cm 9.5cm 1cm 6cm}
\caption{\emph{The normalized fractional frequency difference} as a function of the GC's orbital radius.}
\vspace{0.2cm}
\label{Delta-rp} % delta_rp1.eps
\end{figure}

The core resonance conditions are obtained by substituting equations~(\ref{freq-core}) in (\ref{res-iso}): 
\beq
n\,\omb\!\left(\!1 \,-\, 3\,\frac{I\,}{I_b\,}\!\right) \;+\;
\ell\,\omb\!\left(\!\frac{L\,}{I_b\,}\!\right) \;-\; m\,\omp(\rp) \;=\; 0\,.
\label{res-core}
\eeq
For any given $(n,\ell,m > 0\,; \rp)$, this gives a straight line in the $(I, L)$-plane. However, we are interested in the response of core stars whose $I \leq \Imax\,$. Therefore core resonances are restricted to those $(n,\ell,m > 0\,; \rp)$ such that the resonant line passes through the triangle, $\,0\leq (L/I_b)\leq (I/I_b)\leq \varepsilon\,$, in the $(I, L)$-plane. Since $\omp \simeq \Omega_b$, we write $\omp = \Omega_b\left[\,1 + \varepsilon\Delta(\rp)\,\right]$. The normalized fractional frequency
difference, 
\beq
\Delta(\rp) \;=\; \frac{\omp(\rp) \,-\, \Omega_b}{\varepsilon\Omega_b}\,,
\label{Delta-def}
\eeq
determines which resonances are allowed at any given $\rp$. It can be computed using equation~(\ref{omp-def}) and is displayed in Figure~\ref{Delta-rp}. $\Delta(\rp)$ is a smooth, order-unity, decreasing function of $\rp$, varying from about $+0.6$ at $\rp = 150~\!\pc$ to $-0.33$ at $\rp = 300~\!\pc$, and passing through zero at $\rp = \rstar \simeq 220~\!\pc$.

The radius $\rstar$ has a simple interpretation in terms of the galaxy's core density profile. We recall that $\rstar$ is defined as the radius at which $\omp(\rp) = \Omega_b$, where $\omp(\rp)$ is the GC's circular orbital frequency and $\Omega_b$ is the orbital frequency of stars at the very center of the galaxy. Rearranging equations~(\ref{omp-def}) and  (\ref{omb-def}) we obtain:\footnote{The subscript `0' has been dropped for the central density and mass profile of the galaxy, to indicate that equation~(\ref{rstar-eqn}) is valid for any decreasing density profile, 
$\rho(r)$ with a finite central density, $\rho(0)$, and not just the Isochrone.}
\beq
\frac{4\pi}{3}\rho(0)\,r_*^{\,3} \;-\; M(\rstar) \;=\; \Mp\,.
\label{rstar-eqn}
\eeq
The left side of equation~(\ref{rstar-eqn}) is the difference in the mass enclosed within $\rstar$ (`mass deficit'), between a hypothetical constant density core and that given by the mass profile of the galaxy. Therefore:
\begin{itemize} 
\item[] $\rstar$ is the radius at which the 
galactic mass deficit is equal to the mass of the GC.  
\end{itemize}
The mass deficit vanishes for a constant density core, $\rho(r) = 
\rho(0) = \mbox{constant}$, so equation~(\ref{rstar-eqn}) cannot be satisfied for non-zero $\Mp\,$, emphasizing the importance of allowing for core density variation. It is precisely the deviation from a constant density core that enables resonances and associated torques.

Formulae for $\rstar$, for the Isochrone and the Burkert density profiles, are given in equations~(\ref{rstar-iso}) and (\ref{rstar-bur}) of
\S~\!7. In \S~\!5 we study in detail the role of $\rstar$ as a `filtering' radius for many low-order resonances, which is applied in \S~\!6 to the orbital decay of the GC. 

\subsection{Torques in dimensionless variables}

\begin{figure}
\centering
\hspace{-0.5cm}
\includegraphics[width=0.52\textwidth]{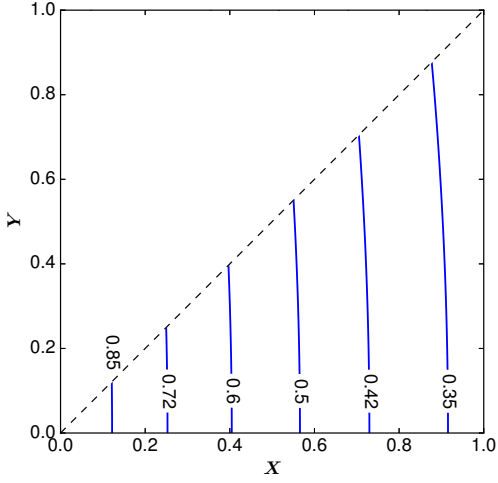} %figs/A_files/Axy2.eps
%,trim={5cm 9.5cm 1cm 6cm}
\caption{\emph{$A(X,Y)$ in the unit triangle}.}
\vspace{0.2cm}
\label{fig_A_T0}
\end{figure}

We now express the resonant torques of equation~(\ref{torq-res-iso}),  
using the dimensionless variables $(X, Y, Z)$, 
\beq
X \,=\, \frac{I}{\Imax}\,,\qquad
Y \,=\, \frac{L}{\Imax}\,,\qquad 
Z \,=\, \frac{L_z}{\Imax}\,,
\label{xyz-def}
\eeq
instead of $(I, L, L_z)$. The domain of these is restricted to the three dimensional wedge, $\,0 \leq Y \leq X \leq 1\,$, with $-Y \leq Z \leq Y\,$. 
Equation~(\ref{df-iso}) for the Isochrone DF can be written as: 
\beq
\frac{\rmd F_0}{\rmd E} \;=\; -\,19.05\,\frac{b}{\,GI_b^3\,}\;A(X, Y)\,,
\label{dfde}
\eeq
where $A(X, Y)$ is a dimensionless positive function with $A(0,0) = 1\,$. From Figure~(\ref{fig_A_T0}) it can be seen that $A(X,Y)$ is weakly dependent on $Y$. We also rescale and define the dimensionless Fourier coefficients,
\beq
\Phi_{n\ell m}(X,Y,Z) \;=\; \frac{(a^2 + \rpsq)^{1/2}}{G\Mp}\;\Phitilda_{n\ell m}(I,L,L_z)\,.
\label{new-fns}
\eeq
The torques depend on 
\beq
P_{n\ell m}(X, Y) \;=\; \int_{-Y}^Y\rmd Z\;\vert\Phi_{n\ell m}(X,Y,Z)\vert^2\,,
\label{power-def}
\eeq
which is a measure of the distribution, within the unit triangle, of the 
`power' in the $(n\ell m)$ Fourier-component. Using equations~~(\ref{res-core}), (\ref{dfde}) and (\ref{power-def}) in (\ref{torq-res-iso}) the resonant torque is:
\beq
\scrt_{n\ell m}(\rp) \,=\, -305\,\rmpi^4\varepsilon^2
\frac{\,GM^2_{\rm p}\,b\,\left[1 + \varepsilon\Delta(\rp)\right]}{a^2 + \rpsq}\;\scrf_{n\ell m}(\rp)\,,
\label{torq-res-core}
\eeq
where 
\begin{align}
\scrf_{n\ell m}(\rp) &\,=\, m^2\!\int_0^1\!\rmd X \int_0^X\!\rmd Y\, A(X,Y)\,P_{n\ell m}(X, Y)\,\times\nonumber\\[1ex]
&\quad\delta\!\left(\varepsilon^{-1}(n-m) -3nX + \ell Y - m\Delta\right) 
\label{res-fac-core}
\end{align}
is a dimensionless positive factor that measures the strength of a
resonance.

\subsection{Types of resonances}

The resonance condition of equation~(\ref{res-core}) can be rewritten 
in dimensionless form as:
\beq
\frac{\left(n \,-\, m\right)}{\varepsilon} \;+\; 
\left\{-3nX \,+\, \ell Y \,-\, m\Delta(\rp)\right\} \;=\; 0\,,
\label{res-xy}
\eeq
where we note that it is independent of $Z$. We recall that $m = 1,2,3,\mbox{etc}\,$, and only one of $(n,\ell)$ can be zero. Given $\rp$, for a triplet $(n,\ell,m > 0)$ to be a core resonance, the straight line of equation~(\ref{res-xy}) must pass through the unit triangle, $0 \leq Y \leq X \leq 1\,$. The magnitudes of the three terms inside the curly brackets of the resonance condition, equation~(\ref{res-xy}), are of order $\,3n$, $\ell$ and $m$, respectively. For the resonance condition to be satisfied, these three terms together must cancel $\varepsilon^{-1}(n-m) = 10(n-m)$. 

Of particular importance are resonances with small integers $(n,\ell,m>0)$. This is because the resonance strength, $\scrf_{n\ell m}$, of equation~(\ref{res-fac-core}) depends on $P_{n\ell m}$, which diminish rapidly for larger $(n,\ell,m>0)$. Hence it is natural to distinguish between the two main types of core resonances, accordingly as $n=m$ or $n\neq m\,$.

\smallskip
\noindent
{\bf 1.~Co-rotating (CR) resonances:} $\,n=m > 0\,$. Then  
equation~(\ref{res-xy}) reduces to 
\beq
-\,3mX \;+\; \ell Y \;-\; m\Delta(\rp) \;=\; 0\,.
\label{res-cr}
\eeq
Since this equation can be satisfied by many low-integers $\ell$ and $m>0\,$, we expect CR resonances to exert significant torques. A physical 
picture of the orbit of a CR resonant star can be obtained by setting $n=m$
in equation~(\ref{res-iso}), which is the primitive form of the resonance condition:
\beq
m\left[\Omega_w(I, L) - \omp\right] \,=\, -\ell\,\Omega_g(I, L)\,.
\nonumber
\eeq
Since $(\Omega_w, \Omega_g, \omp, m)$ are all positive quantities with 
$\Omega_w, \omp \gg \Omega_g$, we must have $\Omega_w \simeq \omp$, with 
the small difference between them resonating with $\Omega_g$, the small 
apse precession rate. So a resonant star nearly co-rotates with the GC, 
trailing or leading it slightly, depending on the sign of $\ell$.
Thus we have the two families of CR resonances:
\begin{itemize} 
\item[{\bf 1a.}] \emph{Trailing CR resonances}: $\ell > 0\,$, so $\Omega_w < \omp$ and the star trails the GC in its orbit. 
\smallskip

\item[{\bf 1b.}] \emph{Leading CR resonances}: $\ell \leq 0\,$, so $\Omega_w \geq \omp$ and the star leads the GC in its orbit.
\end{itemize}

\smallskip
\noindent
{\bf 2.~Non co-rotating resonances:} $\,n\neq m\,$. In this case $\Omega_w$ can differ from $\omp$ considerably. The resonance condition of equation~(\ref{res-xy}) retains its general form. The first term is now non-zero, with magnitude $\varepsilon^{-1}\vert n - m\vert =  \varepsilon^{-1}, \,2\varepsilon^{-1}, \,3\varepsilon^{-1},\mbox{etc} = 10,20,30,\mbox{etc}\,$. Each of the three terms in the curly brackets can be as large only if either $\vert n\vert \geq  4,7,10,\mbox{etc}\,$ or $\vert\ell\vert, m \geq 10,20,30,\mbox{etc}\,$. Therefore these are all higher-order resonances, 
whose torques will be much weaker than CR torques.  

\section{Co-rotating torques in the core}

Here we study Trailing and Leading CR resonances and compute the associated torques as functions of the GC's orbital radius for 
$150~\!\pc \leq \rp \leq 300~\!\pc$. We follow the progressive disappearances of CR resonances as $\rp$ decreases, and see the role of 
$\rstar\simeq 220~\!\pc$ as a characteristic `filtering radius'. 

The numerically intensive part of calculating the $\scrf_{m\ell m}(\rp)$
is in the evaluation of the Fourier coefficients, $\Phitilda_{n\ell m}(I,L,L_z)$, defined in equation~(\ref{phi-tilda}): for each $(I, L, L_z)$ 
we need to do a triple-integral over the angles $(w,g,h)$. Computation is 
expedited by noting that, by suitable transformation to new angle variables, one of the integrals can be evaluated analytically in terms of elliptic integrals, as given in the Appendix. The remaining two angle-integrals are then computed numerically. Once we have $\scrf_{m\ell m}(\rp)$ it can be substituted in equation~(\ref{torq-res-core}) to get the resonant torque, 
$\scrt_{m\ell m}(\rp)$. Then the net Trailing and Leading CR torque profile are:
\begin{subequations}
\begin{align}
\scrt_{\rm trail}(\rp) &\;=\; \sum_{m >0}\;\sum_{\,\ell > 0}\; \scrt_{m\ell m}(\rp)
\label{trail-tor}\,,\\[1ex]
\scrt_{\rm lead}(\rp) &\;=\; \sum_{m >0}\;\sum_{\,\ell \leq 0}\; \scrt_{m\ell m}(\rp)\,,
\label{lead-tor}
\end{align}
\end{subequations}
where the sums are over those $(m,\ell)$ for which, at given $\rp$, the resonant line lies within the unit triangle in $(X,Y)$--space. We discuss these in \S~\!5.1 and \S~\!5.2 for Trailing and Leading CR resonances, respectively. Each has two sub-cases, accordingly as the GC is outside or inside $\rstar\,$.

\begin{figure*}
\gridline{\fig{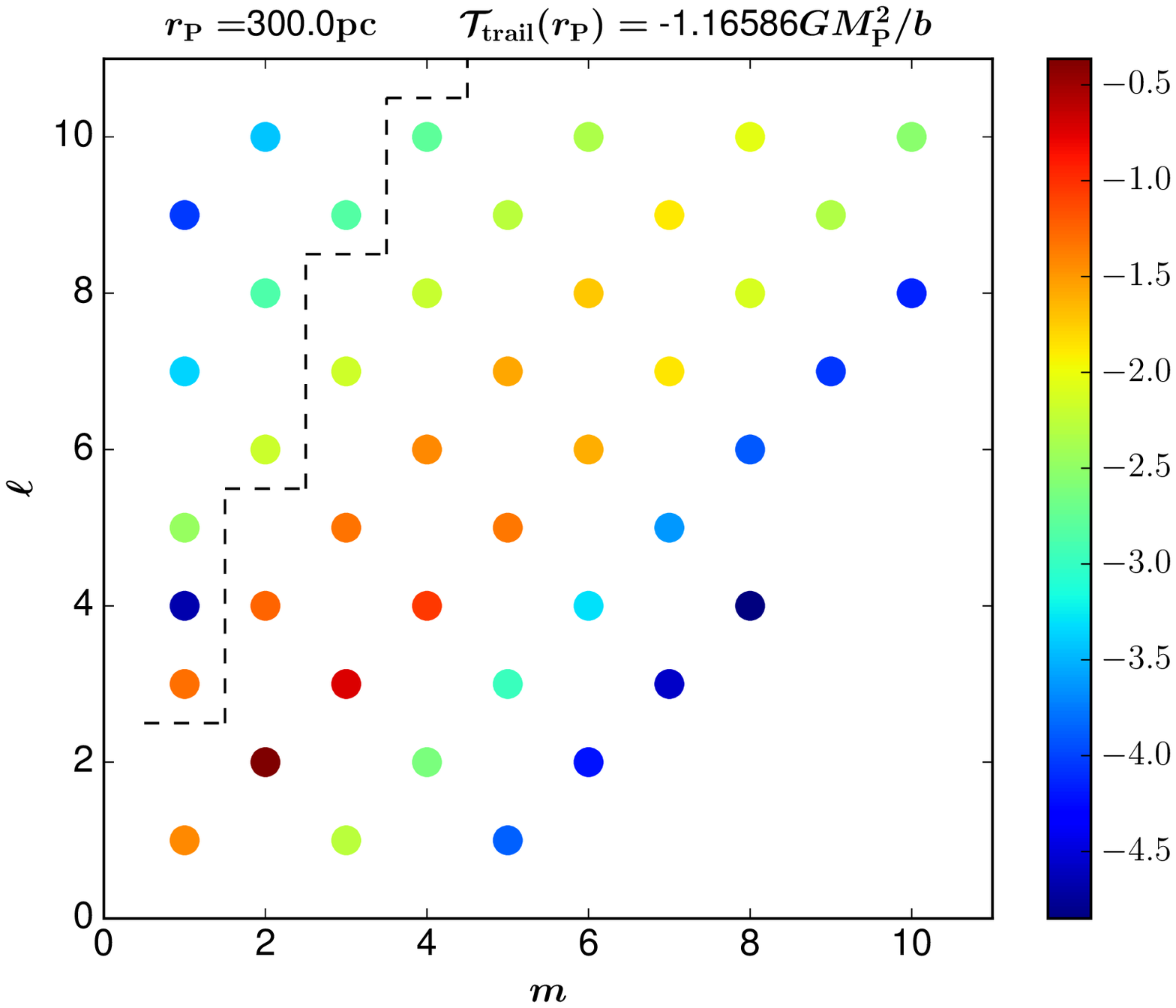}{0.3\textwidth}{} %figs/figs_trail_out/CRtrail_tor_10
         \hspace{-2.5cm} \fig{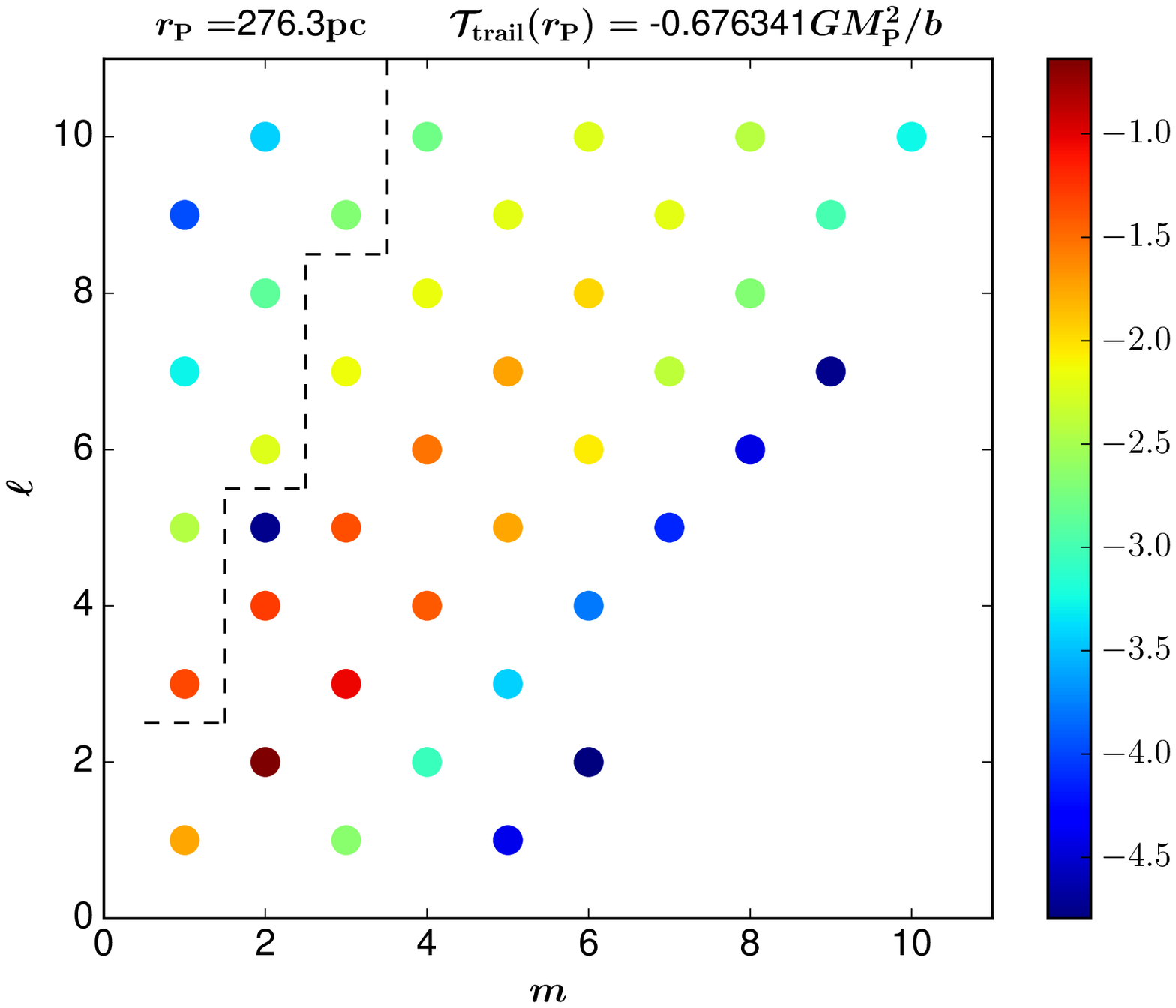}{0.3\textwidth}{}%figs/figs_trail_out/CRtrail_tor_7
	 \hspace{-2.5cm} \fig{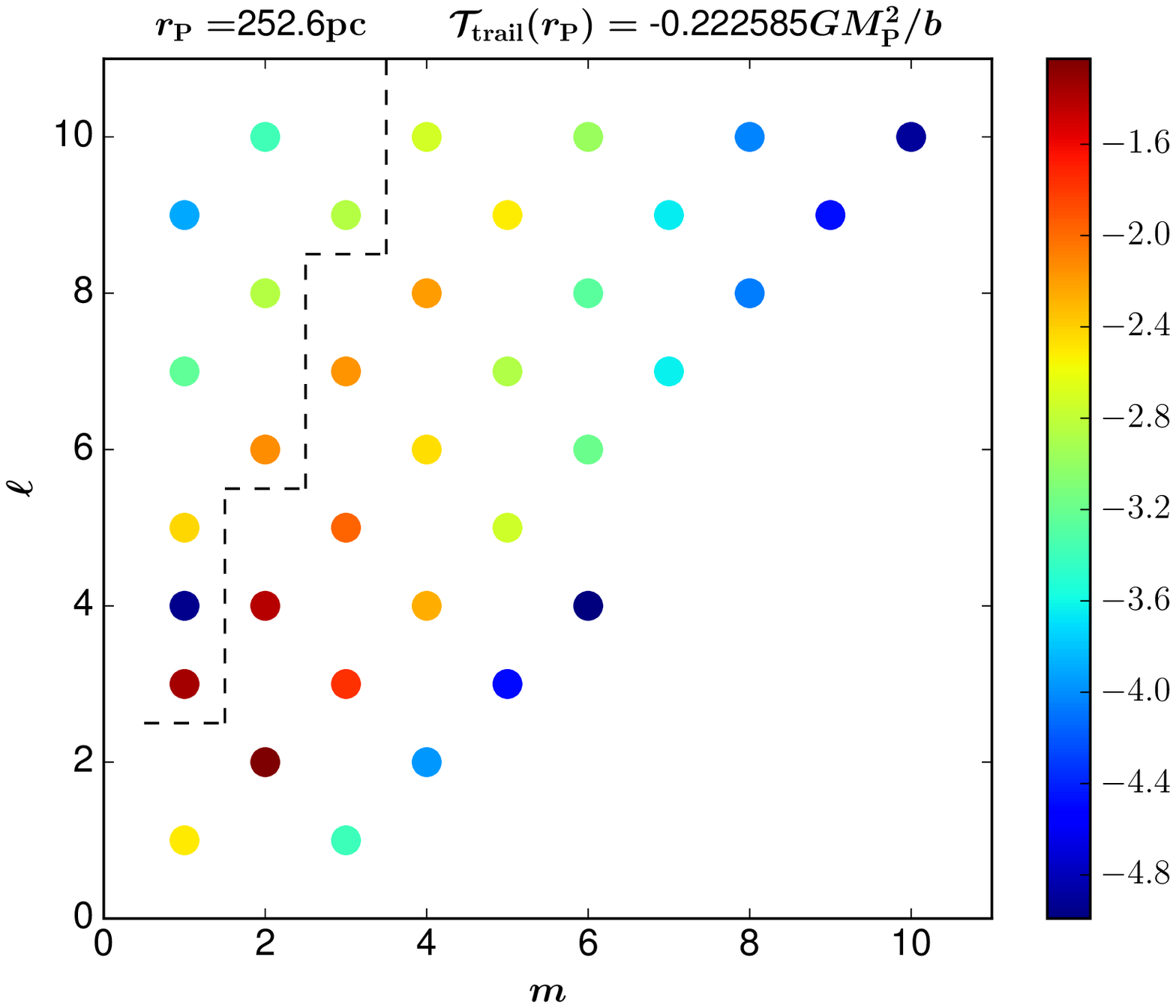}{0.3\textwidth}{}%figs/figs_trail_out/CRtrail_tor_4
	    }
	    \vspace{-1cm}
\gridline{\fig{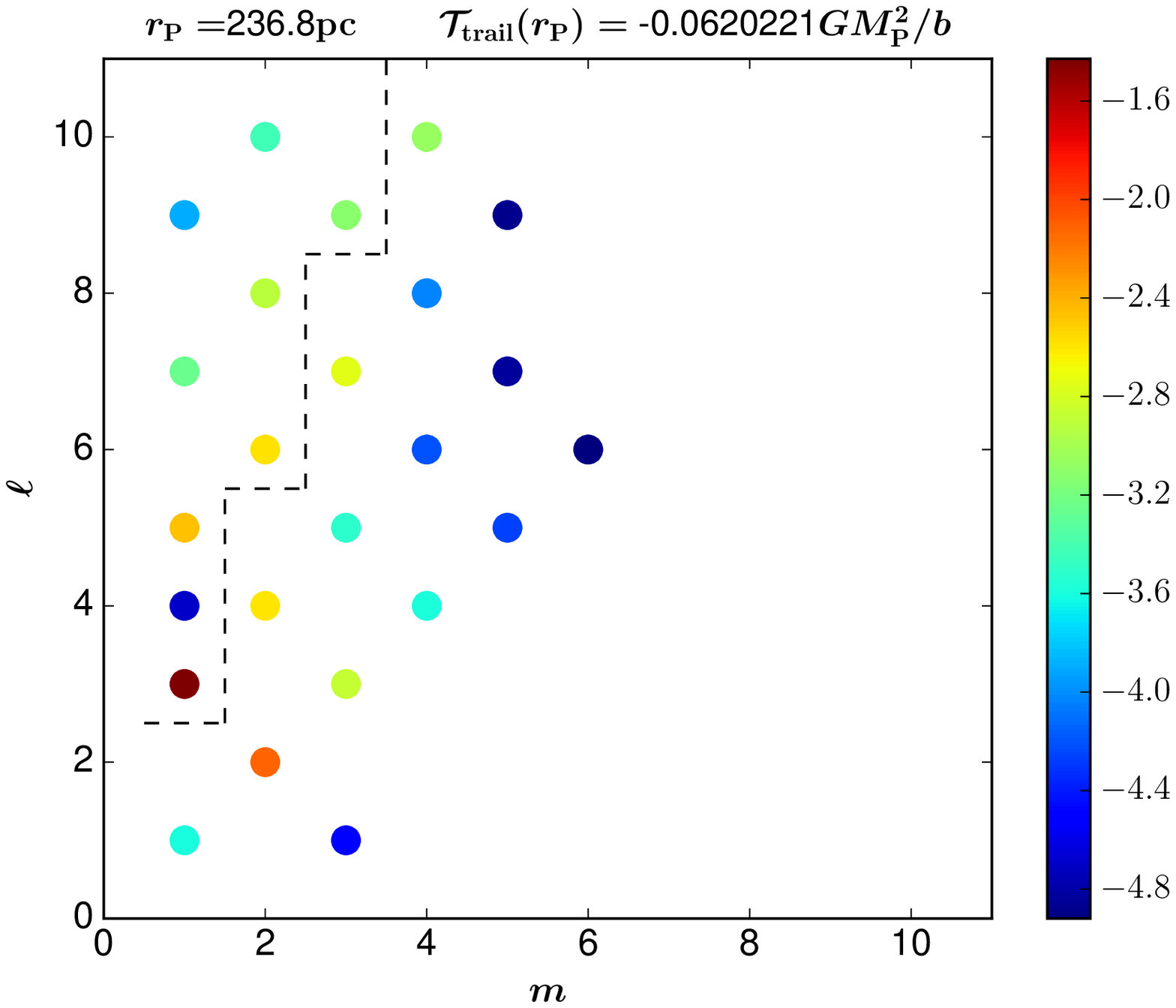}{0.3\textwidth}{}%figs/figs_trail_out/CRtrail_tor_2
         \hspace{-2.5cm} \fig{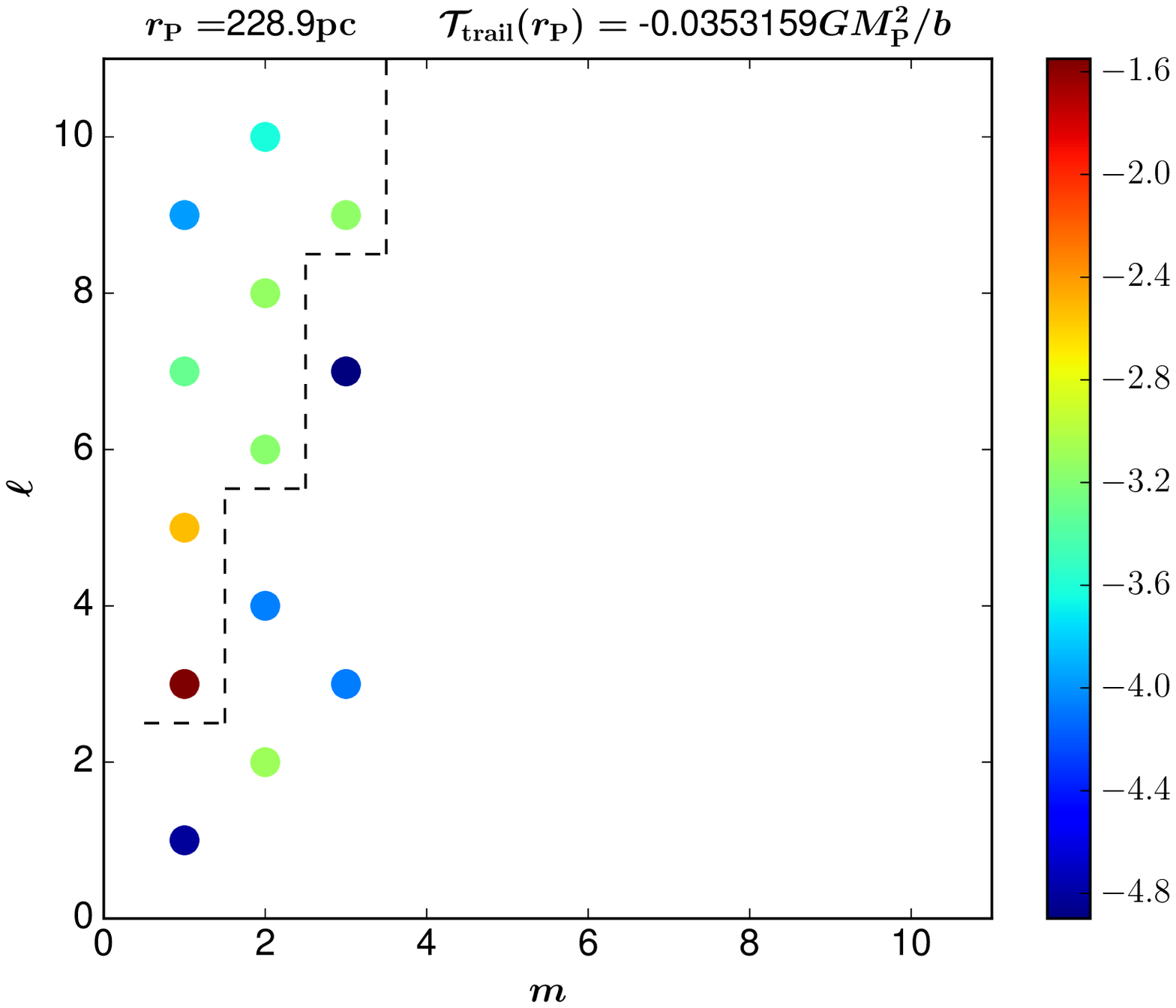}{0.3\textwidth}{} %figs/figs_trail_out/CRtrail_tor_1
	 \hspace{-2.5cm} \fig{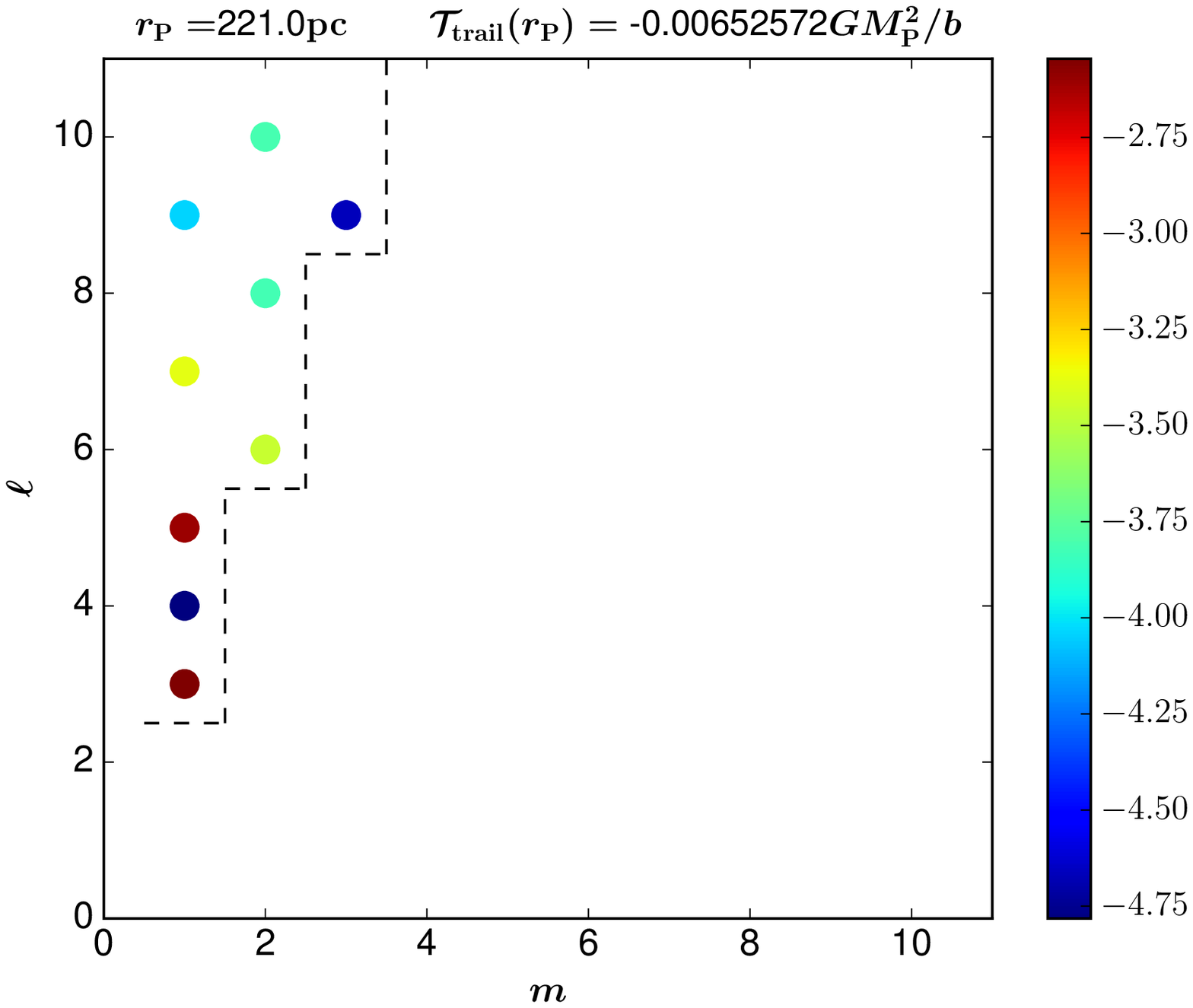}{0.3\textwidth}{} %figs/figs_trail_out/CRtrail_tor_0
	    }
	    \vspace{-0.7cm}
\caption{\emph{Trailing CR resonances and torques for GC outside $\rstar$}, 
for six different $\rp > \rstar$. All $m, \ell \geq 1$ are allowed, but only resonances whose $\vert\scrt_{m \ell m}\vert > 10^{-5}(G \Mp^2 b^{-1})$ are displayed. The color scale refers to $\log_{10}\!\left[\vert\scrt_{m \ell m}\vert/(G \Mp^2 b^{-1})\right]$. The dashed lines separate low $\ell < (3 - |\Delta|)m$ from high $\ell \geq (3 - |\Delta|) m $ resonances.}
\vspace{0.2cm}
\label{fig_ml_trail_out}
\end{figure*}

\subsection{Trailing Co-rotating Torques}

Since $\ell > 0$, equation~(\ref{res-cr}) implies that resonant lines have positive slopes in the $(X, Y)$ plane. 

\subsubsection{GC outside $\rstar\,$} 

When $\rstar \leq \rp < 300~\!\pc$, we have $\,-0.33 < \Delta \leq 0\,$. Equation~(\ref{res-cr}) gives,
\beq
X \;=\; X_{\rm r}(Y;m,\ell,\rp) \;\;\stackrel{{\rm def}}{=}\;\; 
\frac{|\Delta|}{3} \;+\; \frac{\,\ell\,}{3m}Y\,
\label{xres-cr-tr-out}
\eeq
as the resonant line, which intersects two of the three edges of the unit triangle. The line passes through the $Y=0$ edge at $X_1 = |\Delta|/3\,$. The second point lies on the edge $Y=X\,$ for $\,\ell < \left(3-|\Delta|\right)m\,$, and on the edge $X=1\,$ for $\,\ell \geq \left(3-|\Delta|\right)m\,$. The torques behave differently, as discussed below.

\medskip\noindent
\underline{Low $\ell$ resonances, $\,\ell < \left(3-|\Delta|\right)m\,$}: 
When $\rp = 300~\!\pc\,$, we have $\,|\Delta| = 0.33\,$, 
so $\,\ell < 2.67m\,$ are the only integer values of $\ell$ that are allowed. We list some of the low-integer $(m,\ell)$ values that are allowed:
\begin{align} 
&(1,1)\quad (1,2)\nonumber\\ 
&(2,1)\quad (2,2)\quad\ldots\quad (2,5)\nonumber\\
&(3,1)\quad (3,2)\quad\ldots\quad (3,8)\nonumber\\
&(4,1)\quad (4,2)\quad\ldots\quad (4,10)\nonumber\\
&(5,1)\quad (5,2)\quad\ldots\quad (5,13)\nonumber\\
&\;\;\ldots\quad\;\;\,\ldots\quad\;\;\ldots\quad\;\;\,\ldots
\label{res-list-tr-low}
\end{align}
As $\rp$ decreases $|\Delta|$ also decreases, so the range of allowed 
$\ell$ increases. But this has only a modest effect for small $m\,$.
The resonance strength factor of equation~(\ref{res-fac-core}) is:
\beq
\scrf_{m\ell m}(\rp) = 
\frac{\,m\,}{\,3\,}\!  
\int_0^{Y_a}\!\rmd Y\,A(X_{\rm r},Y)\,P_{m\ell m}(X_{\rm r}, Y)\,,
\label{torq-cr-tr-out1}
\eeq
where $Y_a = m|\Delta|/(3m - \ell)\,$. Since the upper limit of integration, $Y_a \propto |\Delta| \to 0$ as $\rp \to \rstar\,$, all $\scrf_{m\ell m}(\rstar) = 0\,$.

\medskip\noindent
\underline{High $\ell$ resonances, $\,\ell \geq \left(3-|\Delta|\right)m\,$}: When $\rp = 300~\!\pc\,$, we have $\,|\Delta| = 0.33\,$, so $\,\ell \geq 2.67m\,$ are the only integer values of $\ell$ that are allowed. The list of allowed $(m,l)$ values is complementary to list~(\ref{res-list-tr-low}):
\begin{align} 
&(1,3)\quad (1,4)\quad \ldots\nonumber\\ 
&(2,6)\quad (2,7)\quad\ldots\nonumber\\
&(3,9)\quad (3,10)\quad\ldots\nonumber\\
&(4,11)\quad (4,12)\quad\ldots\nonumber\\
&(5,14)\quad (5,15)\quad\ldots\nonumber\\
&\;\;\ldots\qquad\;\ldots\qquad\ldots
\label{res-list-tr-high1}
\end{align}
As $\rp$ decreases $|\Delta|$ also decreases, and the range of allowed 
$\ell$ decreases. Again, this has only a modest effect for small $m\,$.
The range in $\ell$ is narrowest when $\rp = \rstar$ for which $|\Delta| = 
0\,$; then $\ell \geq 3m$ and the list of allowed $(m,\ell)$ shrinks to: 
\begin{align} 
&(1,3)\quad (1,4)\quad \ldots\nonumber\\ 
&(2,6)\quad (2,7)\quad\ldots\nonumber\\
&(3,9)\quad (3,10)\quad\ldots\nonumber\\
&(4,12)\quad (4,13)\quad\ldots\nonumber\\
&(5,15)\quad (5,16)\quad\ldots\nonumber\\
&\;\;\ldots\qquad\;\ldots\qquad\ldots
\label{res-list-tr-high2}
\end{align}
The allowed $m=1,2,3$ resonances remain unaltered but $(4,11), (5,14), 
{\rm etc}$ have dropped out. The resonance strength factor of equation~(\ref{res-fac-core}) is: 
\beq
\scrf_{m\ell m}(\rp) = 
\frac{\,m\,}{\,3\,}\!  
\int_0^{Y_b}\!\rmd Y\,A(X_{\rm r},Y)\,P_{m\ell m}(X_{\rm r}, Y)\,,
\label{torq-cr-tr-out2}
\eeq
where $Y_b = \left(3-|\Delta|\right)m/\ell\,$. At $\rp = \rstar$, we have 
$\,\ell \geq 3m\,$, $\,X_{\rm r} = (\ell/3m)Y\,$ and $\,Y_b = 3m/\ell\,$. 
It is important to note that $\scrf_{m\ell m}(\rstar) \neq 0$, which is different from equation~(\ref{torq-cr-tr-out1}) of the low-$\ell$ case.

The $\scrf_{m\ell m}(\rp)$ of equations~(\ref{torq-cr-tr-out1}) and (\ref{torq-cr-tr-out2}) were computed numerically, as discussed at the beginning of \S~\!5, for all the $100$ resonances with $1 \leq m, \ell \leq 10$. Substituting these in equation~(\ref{torq-res-core}) we obtained the corresponding $\scrt_{m\ell m}(\rp)$. The six panels of Figure~(\ref{fig_ml_trail_out}) track Trailing CR resonances with $\vert\scrt_{m\ell m}(\rp)\vert > 10^{-5}(G \Mp^2/b)$, for $\rstar \leq \rp \leq 300~\!\pc$. Then $\scrt_{\rm trail}(\rp)$ was  calculated by summing over the $\scrt_{m\ell m}(\rp)$, as given in equation~(\ref{trail-tor}). A striking feature evident in the figures is the progressive loss of resonances and torque strengths, as $\rp$ decreases:
\begin{itemize}
\item[$\bullet$] At $\rp = 300~\!\pc$ there are $43$ resonances, with 
$10^{-5}(G \Mp^2/b) < \vert\scrt_{m \ell m}\vert \lesssim 10^{-0.36}(G \Mp^2/ b)$. Of these $34$ are low $\ell$ resonances and $9$ are high $\ell$ resonances. The strongest torque comes from the $(2,2)$ resonance. The net torque due to all the resonances is $\scrt_{\rm trail} \simeq -1.17(G \Mp^2/b)$.

\item[$\bullet$] At $\rp = 236.8~\!\pc$ there are $24$ resonances, with 
$10^{-5}(G \Mp^2/b) < \vert\scrt_{m \ell m}\vert \lesssim 10^{-1.4}(G \Mp^2/ b)$. Of these $15$ are low $\ell$ resonances and $9$ are the same high $\ell$ resonances. The strongest torque comes from the $(1,3)$ resonance. The net torque due to all the resonances is $\scrt_{\rm trail} \simeq -6\times 10^{-2}(G \Mp^2/b)$. 

\item[$\bullet$]
At $\rp = 221~\!\pc$ there are no low $\ell$ resonances of any strength to
speak of. $9$ high $\ell$ resonances survive, with $10^{-5}(G \Mp^2/b) < \vert\scrt_{m \ell m}\vert \lesssim 10^{-2.5}(G \Mp^2/ b)$. The strongest torque still comes from the $(1,3)$ resonance. The net torque due to all the resonances is $\scrt_{\rm trail} \simeq -7\times 10^{-3}(G \Mp^2/b)$. 
\end{itemize}
The torque profile, $\scrt_{\rm trail}(\rp)$, is given in Figure~(\ref{fig_torq-prof}b) and discussed in \S~\!6.

\begin{figure*}
\gridline{\fig{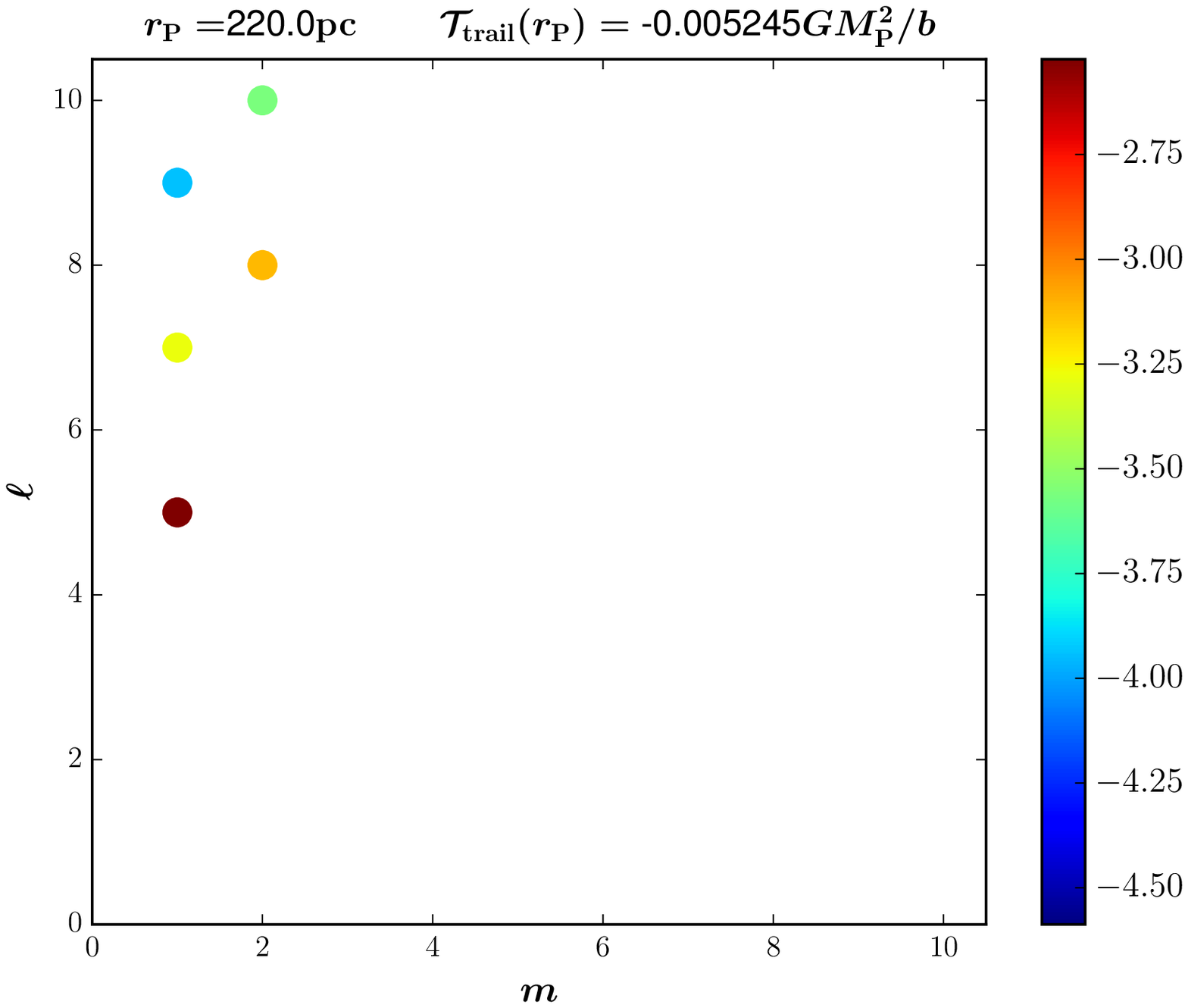}{0.3\textwidth}{}%figs/figs_trail_in/CRtrail_in_tor_10
         \hspace{-2.5cm} \fig{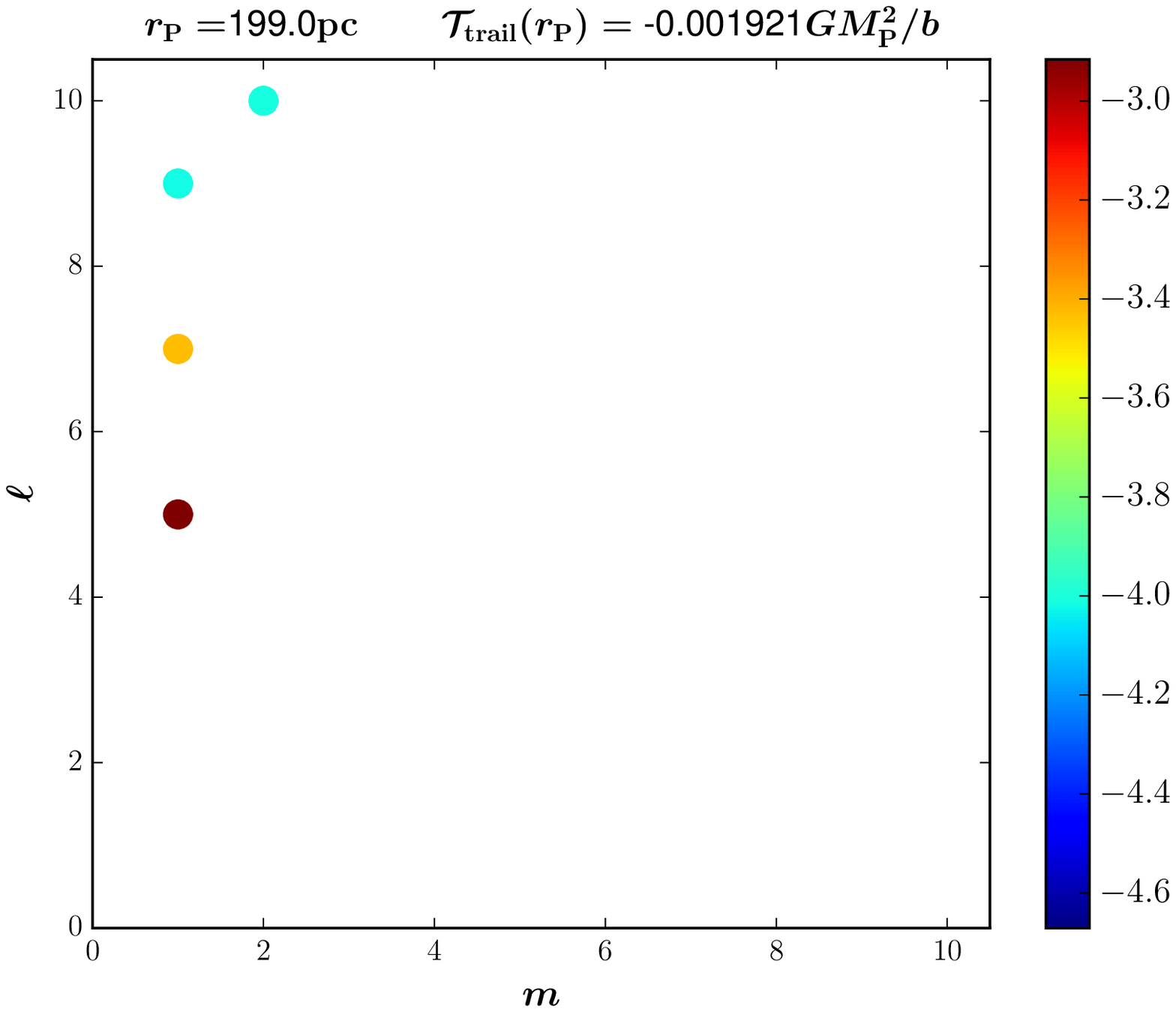}{0.3\textwidth}{}%figs/figs_trail_in/CRtrail_in_tor_7
	 \hspace{-2.5cm} \fig{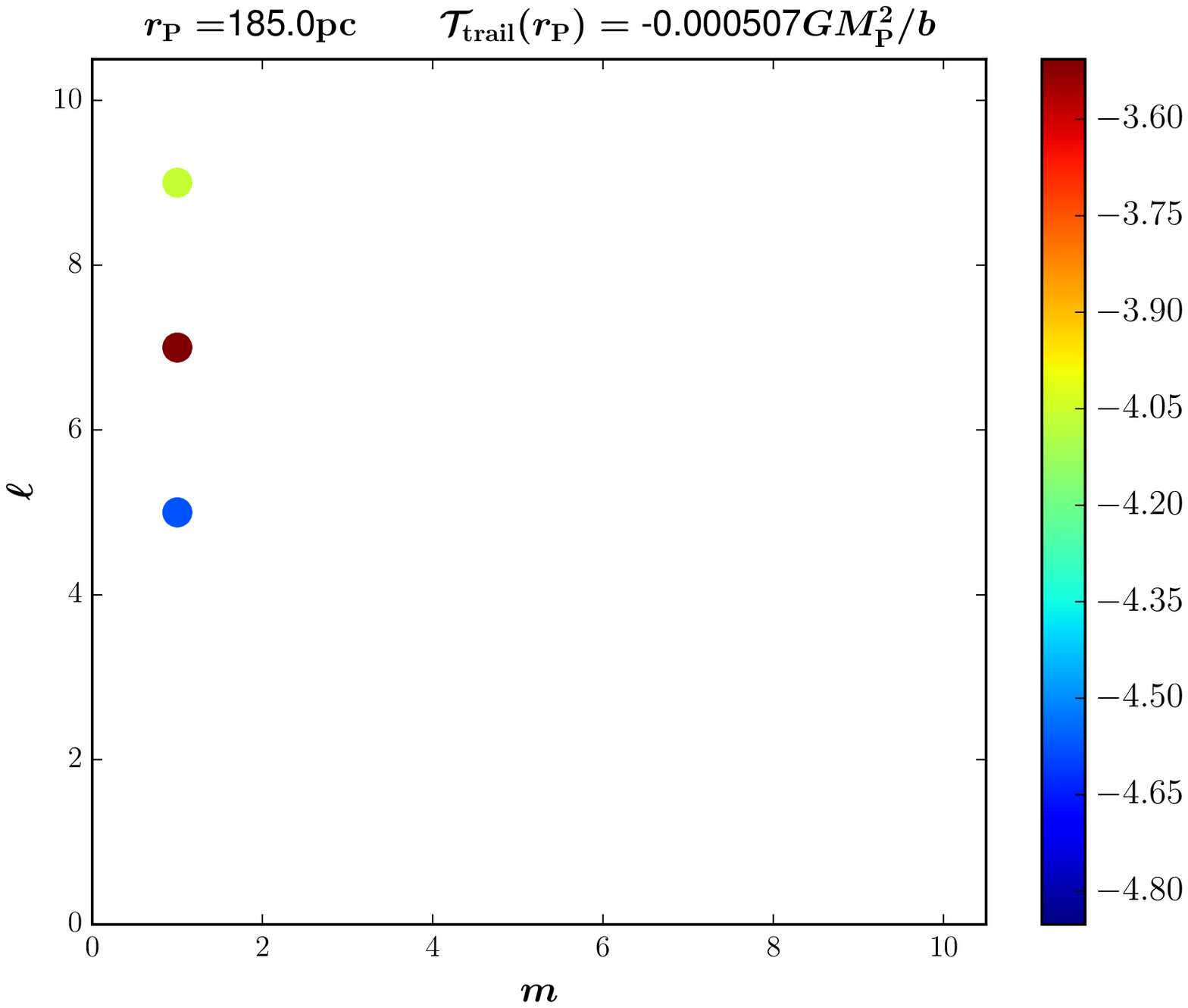}{0.3\textwidth}{}%figs/figs_trail_in/CRtrail_in_tor_5
	  }
\vspace{-1cm}	  

 \gridline{\fig{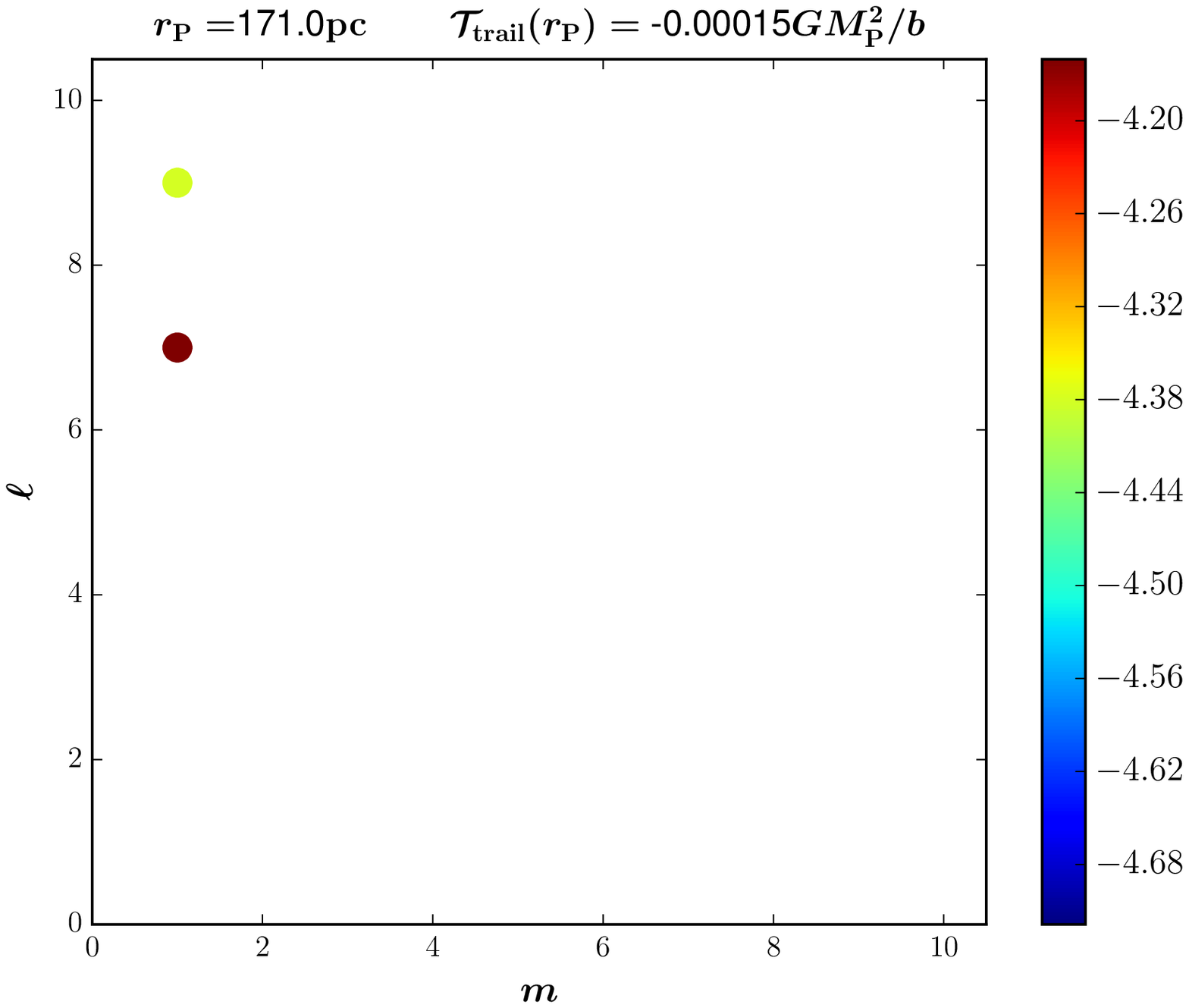}{0.3\textwidth}{}%figs/figs_trail_in/CRtrail_in_tor_3
         \hspace{-2.5cm} \fig{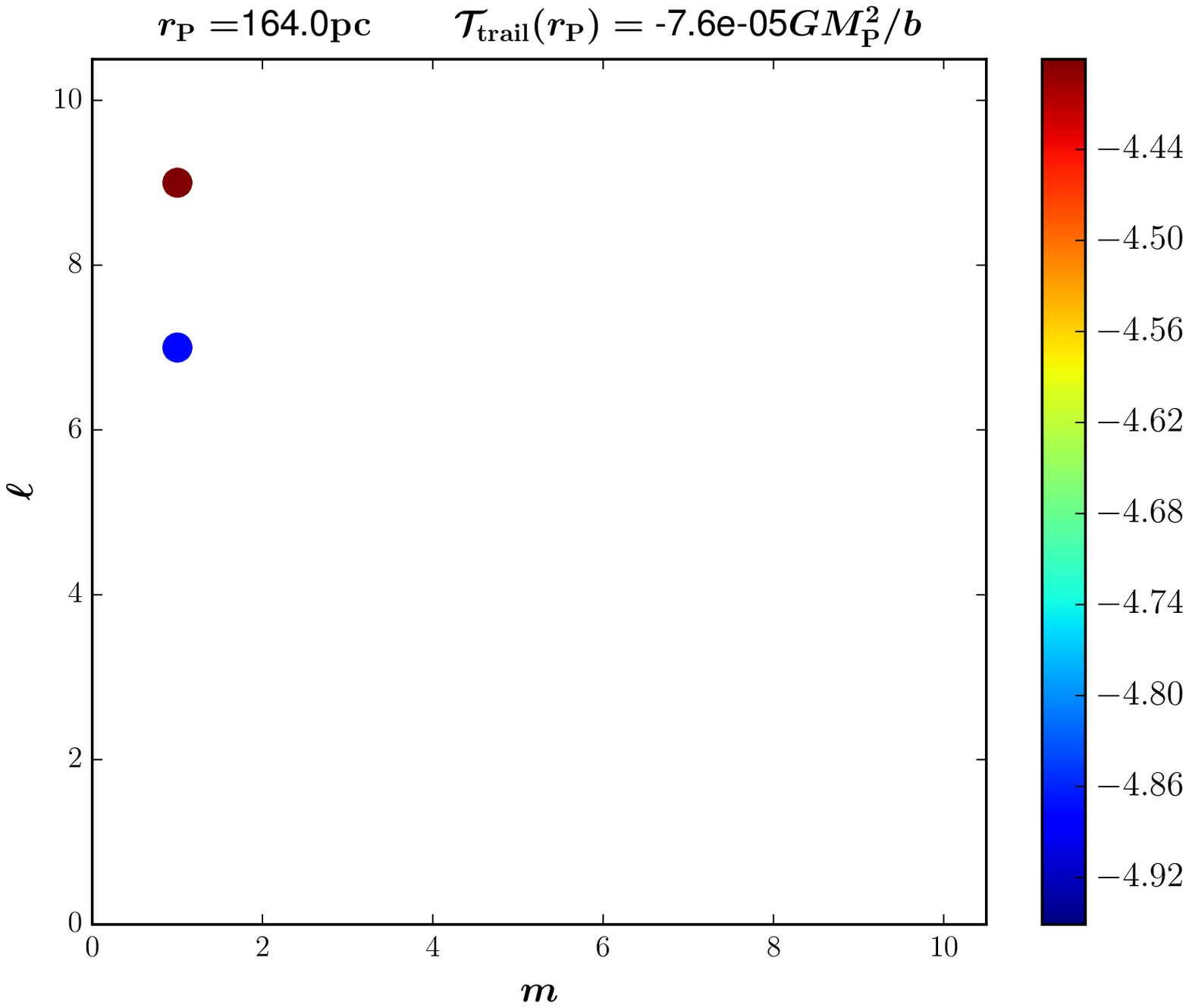}{0.3\textwidth}{}%figs/figs_trail_in/CRtrail_in_tor_2
	 \hspace{-2.5cm} \fig{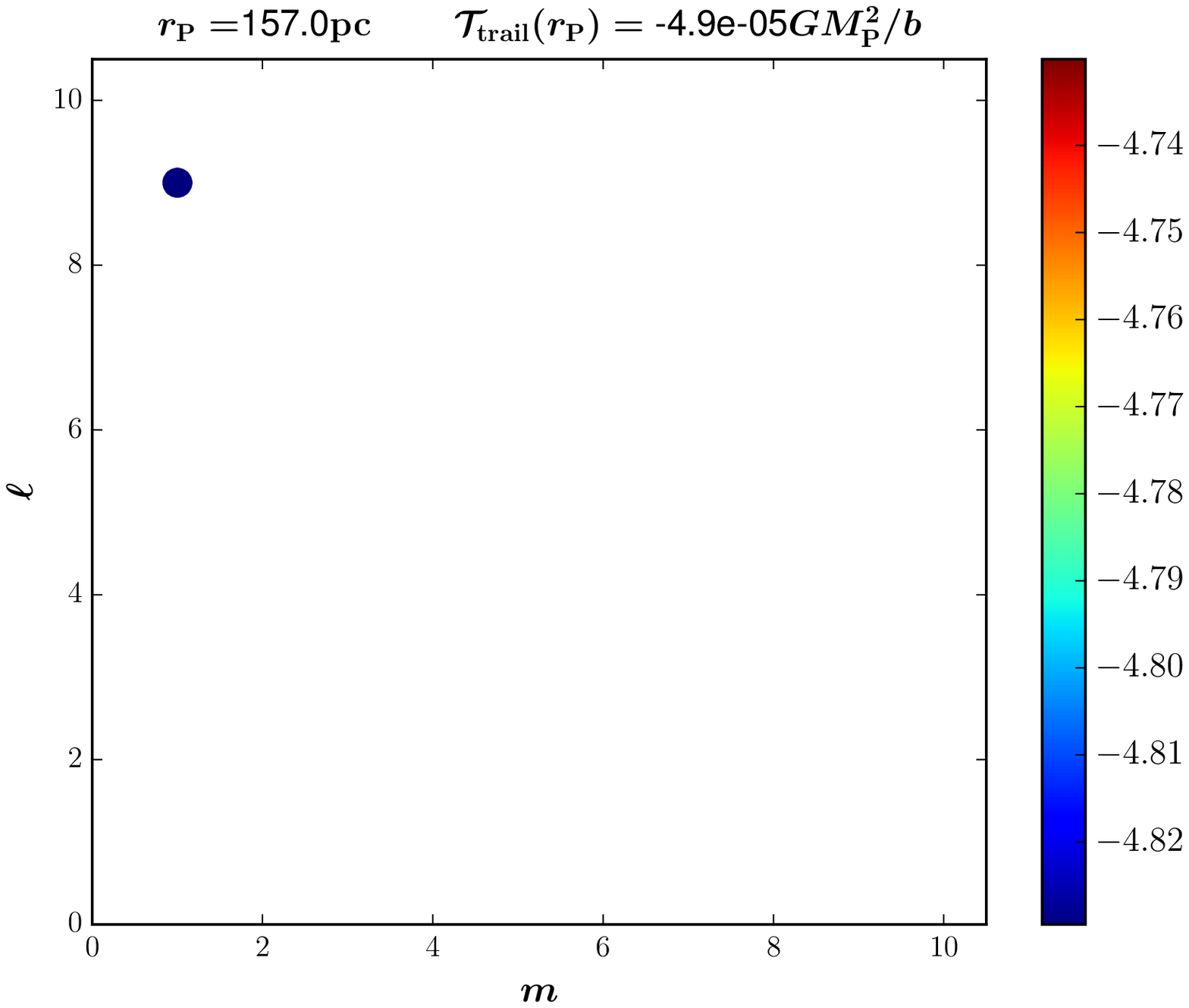}{0.3\textwidth}{}%figs/figs_trail_in/CRtrail_in_tor_1
	  }
\vspace{-0.7cm}	  
\caption{\emph{Trailing CR resonances and torques for GC inside $\rstar$},
for six different $\rp < \rstar\,$. All $m \geq 1$ and $\ell > (3 + \Delta)m$ are allowed but only resonances whose $\vert\scrt_{m \ell m}\vert > 10^{-5}(G \Mp^2 b^{-1})$ are displayed.}
\vspace{0.2cm}
\label{fig_ml_trail_in}
\end{figure*}

\bigskip
\subsubsection{GC inside $\rstar\,$}
 
When $150~\!\pc < \rp < \rstar$, we have 
$\,0 < \Delta \leq 0.6\,$. Equation~(\ref{res-cr}) gives,
\beq
Y \;=\; Y_{\rm r}(X;m,\ell,\rp) \;\;\stackrel{{\rm def}}{=}\;\; 
\frac{m}{\,\ell\,}\Delta \;+\; \frac{3m}{\,\ell\,}X\,
\label{xres-cr-tr-out}
\eeq
as the resonant line. This lies in the unit triangle only when 
$\ell > (3 + \Delta)m\,$. It intersects the $X=Y$ edge at $X_1 = m\Delta/(\ell -3m)\,$ and the $X=1$ edge at $Y_2 = (3 + \Delta)m/\ell$.

For $\rp\simeq\rstar$ we have $\Delta$ small but positive. The list of allowed $(m,\ell)$ values is:
\begin{align} 
&(1,4)\quad (1,5)\quad \ldots\nonumber\\ 
&(2,7)\quad (2,8)\quad\ldots\nonumber\\
&(3,10)\quad (3,11)\quad\ldots\nonumber\\
&(4,13)\quad (4,14)\quad\ldots\nonumber\\
&(5,16)\quad (5,17)\quad\ldots\nonumber\\
&\;\;\ldots\qquad\;\ldots\qquad\ldots
\label{res-list-tr-in1}
\end{align}
This must be compared with the list~(\ref{res-list-tr-high2}) of high $\ell$ resonances that survive at $\rp = \rstar$. We notice the absence of the $l=3m$ resonances, so $(1,3), (2,6),\mbox{etc}$ have dropped out just inside 
$\rstar$. As $\rp$ decreases $\Delta$ increases more steeply, leading to
increased loss of the lower of these high $\ell$ resonances. We calculated 
\beq
\scrf_{m\ell m}(\rp) = 
\frac{\,m^2\,}{\,\ell\,}\!  
\int_{X_1}^{1}\rmd X\,A(X\,,Y_{\rm r})\,P_{m\ell m}(X\,, Y_{\rm r})\,,
\label{torq-cr-tr-in}
\eeq
for all the $225$ resonances with $1 \leq m, \ell \leq 15$. Substituting these in equation~(\ref{torq-res-core}) we obtained the corresponding 
$\scrt_{m\ell m}(\rp)$. The six panels of Figure~(\ref{fig_ml_trail_in}) track the Trailing (high $\ell$) CR resonances with $\vert\scrt_{m\ell m}(\rp)\vert > 10^{-5}(G \Mp^2/b)$, for $150~\!\pc \leq \rp < \rstar$. Then $\scrt_{\rm trail}(\rp)$ was calculated by summing over the $\scrt_{m\ell m}(\rp)$, as given in equation~(\ref{trail-tor}). This set of figures should be seen as a continuation of the bottom right panel of Figure~(\ref{fig_ml_trail_out}). High $\ell$ resonances exist inside $\rstar$, and there is increased loss of both resonances and torque strengths, as $\rp$ decreases:
\begin{itemize}
\item[$\bullet$] At $\rp = 220~\!\pc$, which is just inside $\rstar$, there are just $5$ resonances with $10^{-5}(G \Mp^2/b) < \vert\scrt_{m \ell m}\vert \lesssim 10^{-2.5}(G \Mp^2/ b)$. The strongest torque comes from the 
$(1,5)$ resonance. The net torque due to all the resonances is $\scrt_{\rm trail} \simeq -5\times 10^{-3}(G \Mp^2/b)$.

\item[$\bullet$] At $\rp = 185~\!\pc$, there are only $3$ resonances with 
$10^{-5}(G \Mp^2/b) < \vert\scrt_{m \ell m}\vert \lesssim 10^{-3.5}(G \Mp^2/ b)$. The strongest torque now comes from the $(1,7)$ resonance. The net torque due to all the resonances is $\scrt_{\rm trail} \simeq -5\times 10^{-4}(G \Mp^2/b)$. 

\item[$\bullet$]
At $\rp = 157~\!\pc$, there is just one resonance left, $(1, 9)$, with 
$\vert\scrt_{m \ell m}\vert > 10^{-5}(G \Mp^2/b)$. The net torque due to 
this, together with contributions from weaker resonances, is $\scrt_{\rm trail} \simeq -5\times 10^{-5}(G \Mp^2/b)$. 
\end{itemize}

The torque profile, $\scrt_{\rm trail}(\rp)$, is given in Figure~(\ref{fig_torq-prof}a) and discussed in \S~\!6.

\subsection{Leading Co-rotating Torques}

\begin{figure*}
\gridline{\fig{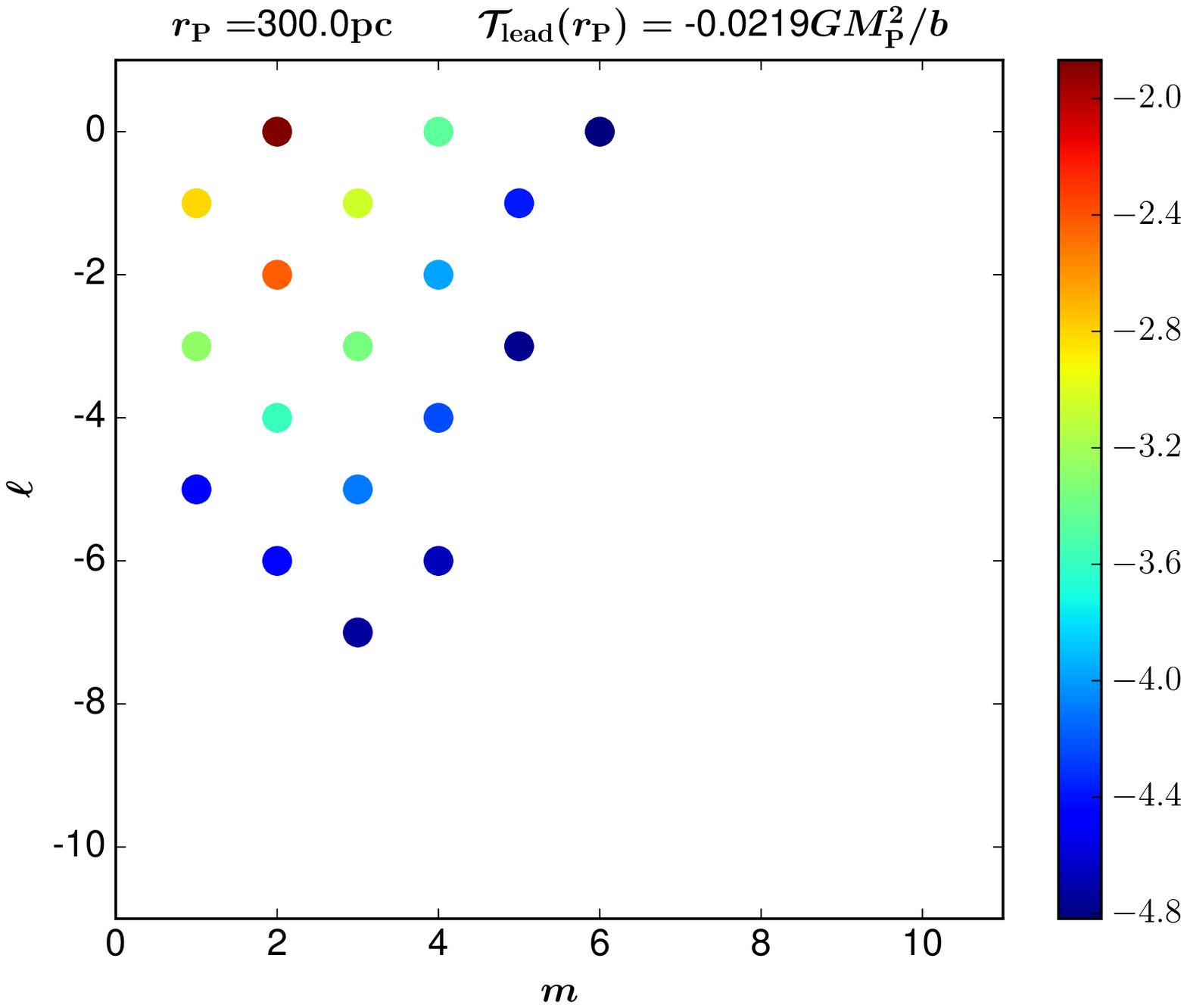}{0.3\textwidth}{}%figs/figs_lead/CRlead_tor_10
	  \hspace{-2.5cm}\fig{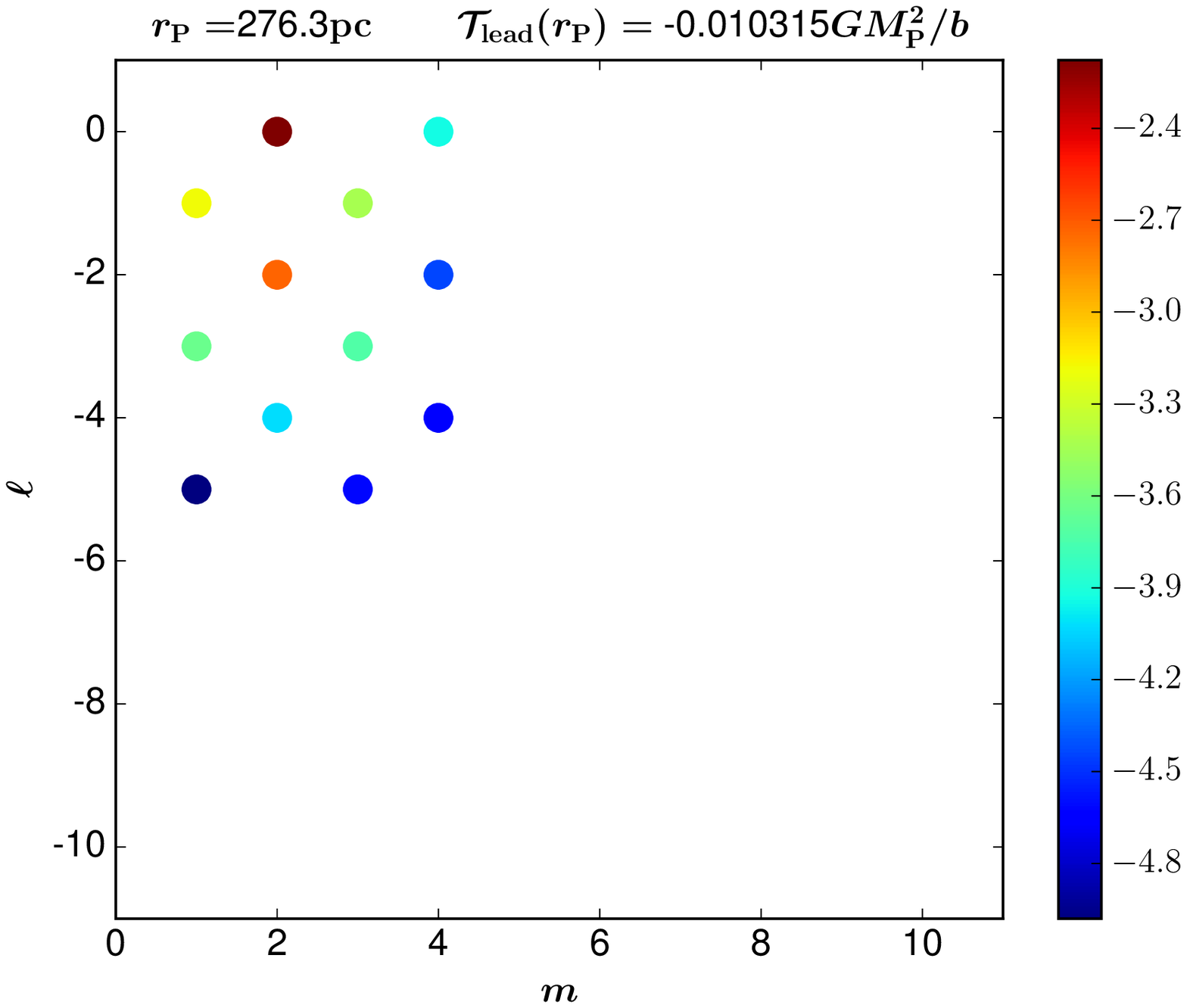}{0.3\textwidth}{}%figs/figs_lead/CRlead_tor_7
	  \hspace{-2.5cm}\fig{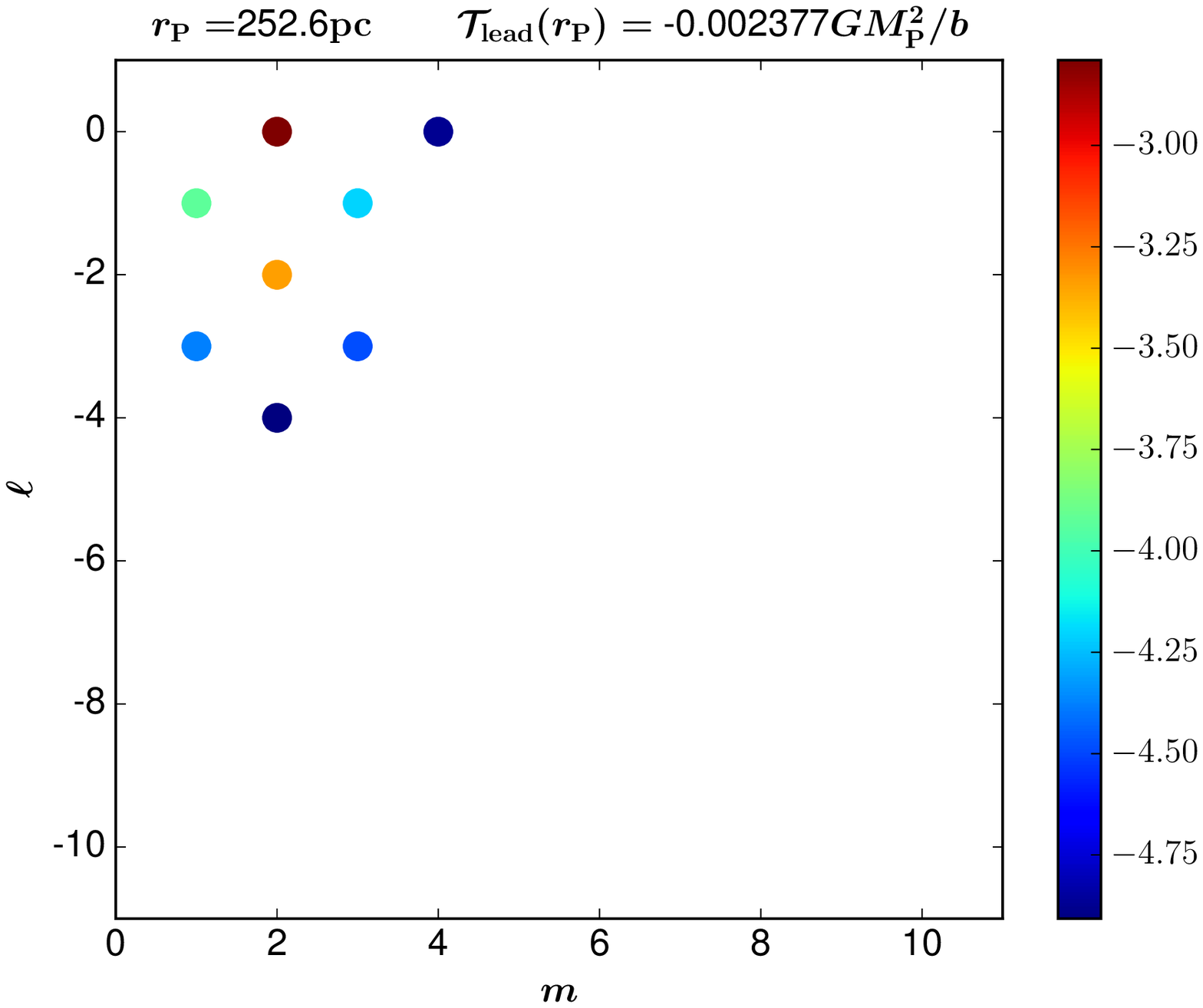}{0.3\textwidth}{}%figs/figs_lead/CRlead_tor_4
	  }
	  \vspace{-1cm}
\gridline{\fig{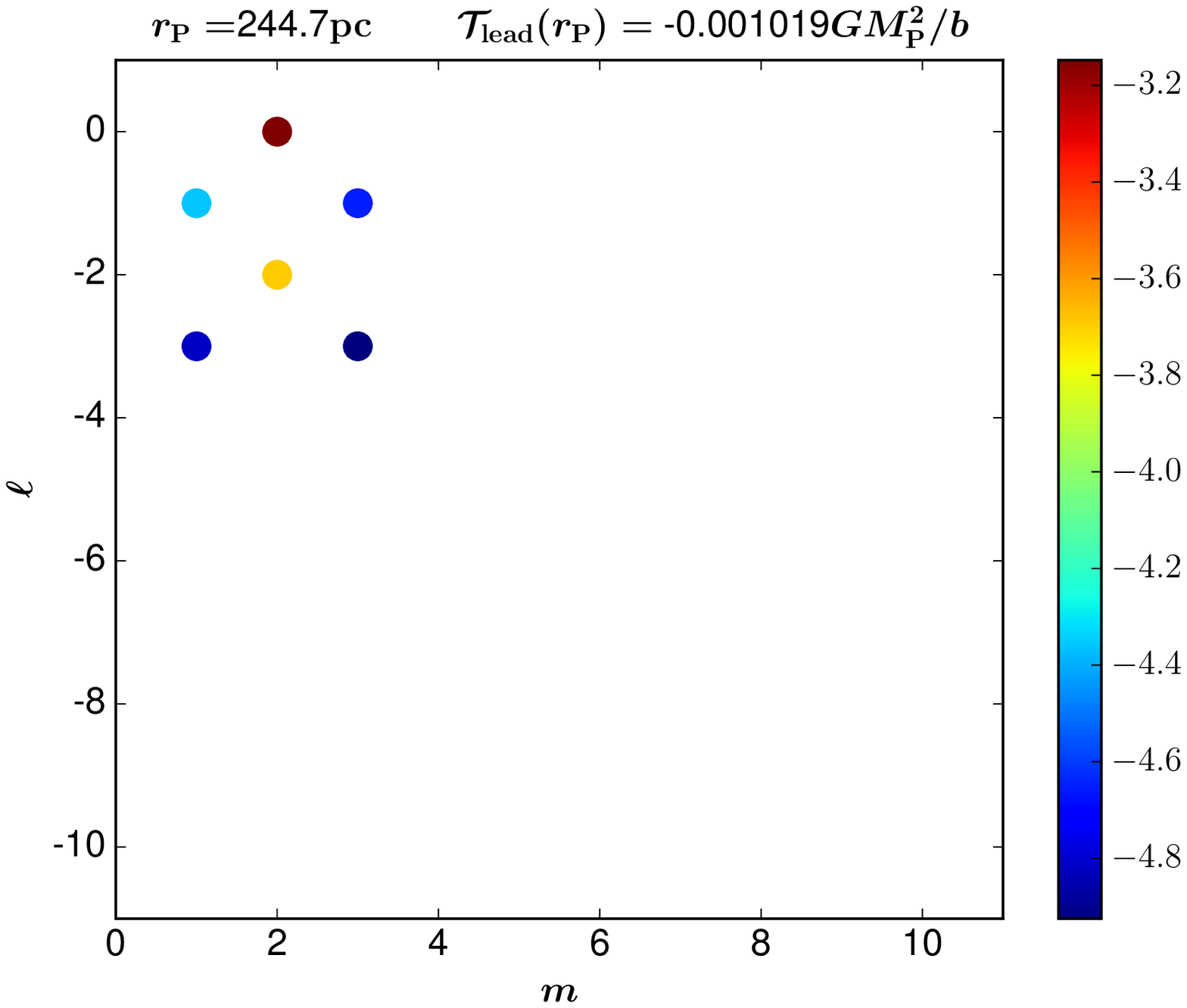}{0.3\textwidth}{}%figs/figs_lead/CRlead_tor_3
	 \hspace{-2.5cm} \fig{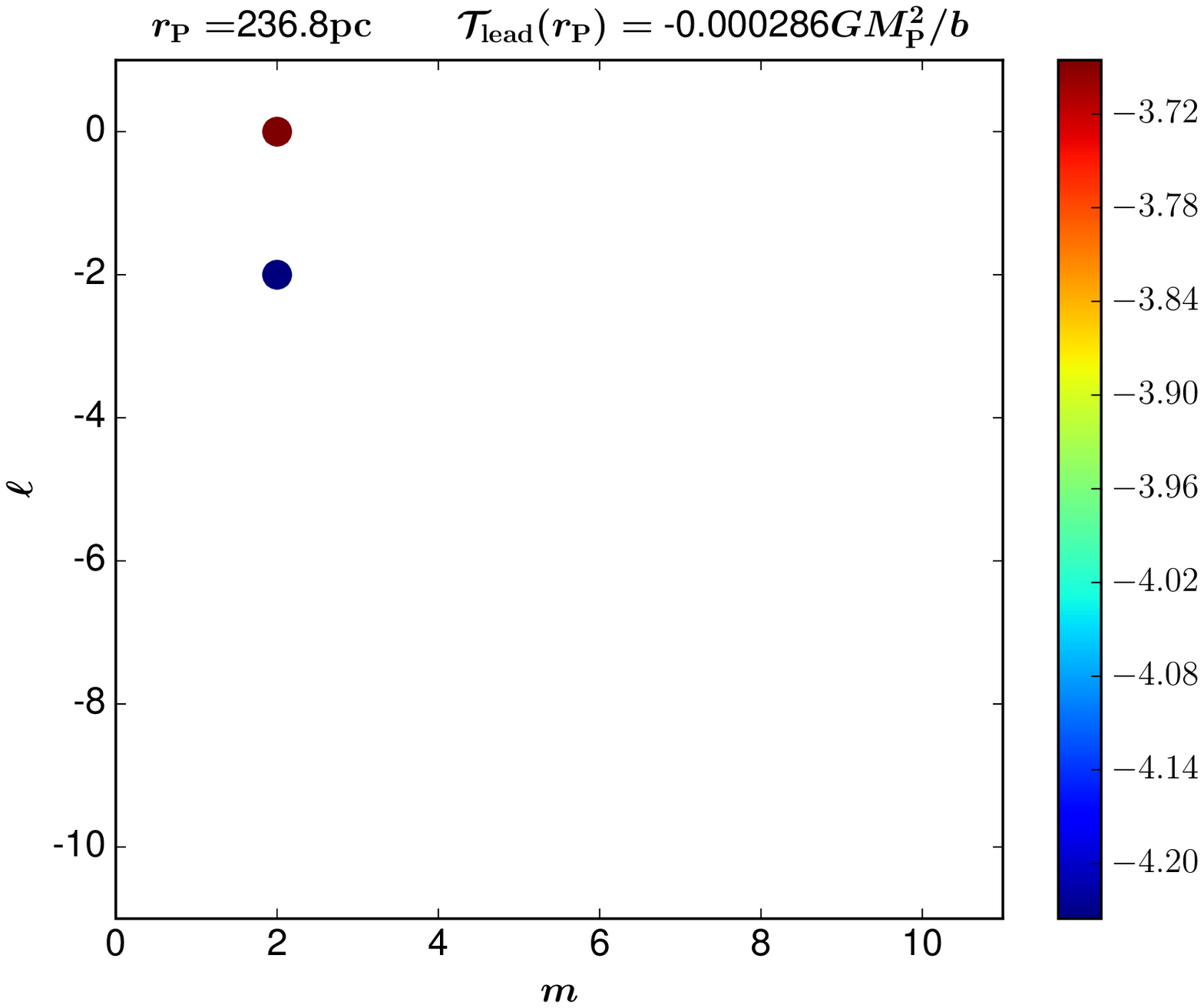}{0.3\textwidth}{}%figs/figs_lead/CRlead_tor_2
	 \hspace{-2.5cm} \fig{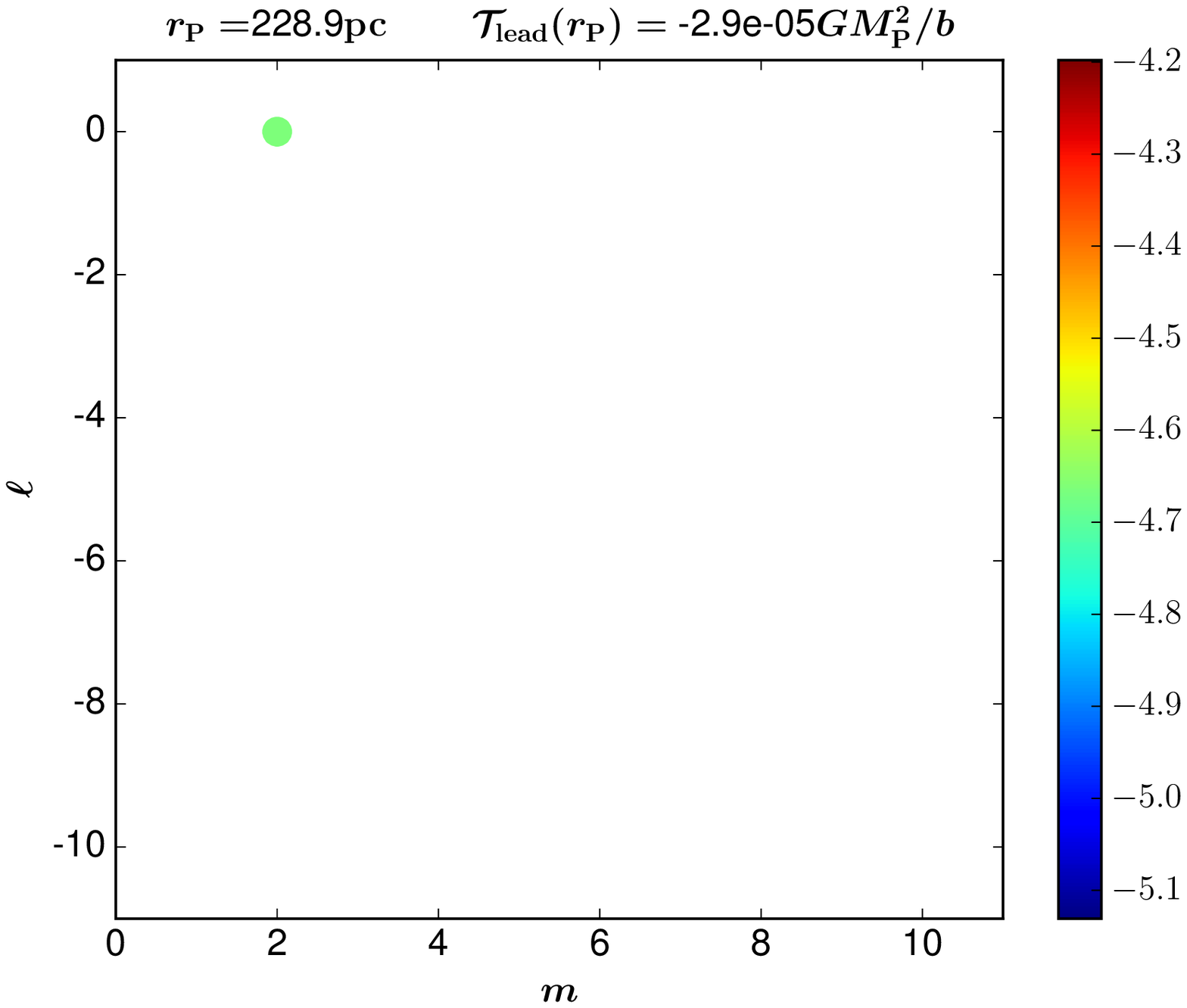}{0.3\textwidth}{}%figs/figs_lead/CRlead_tor_1
	  }
	  \vspace{-0.7cm}
\caption{\emph{Leading CR resonances and torques for GC outside $\rstar$}, 
for six different $\rp > \rstar$. All $m \geq 1$ and $\ell \leq 0$ are allowed, but only resonances whose $\vert\scrt_{m \ell m}\vert > 10^{-5}(G \Mp^2 b^{-1})$ are displayed.}	  
\vspace{0.2cm}
\label{fig_ml_lead} 
\end{figure*}

When $\ell \leq 0$ the resonance condition of equation~(\ref{res-cr}) is:
\beq
3mX \;+\; |\ell| Y \;+\; m\Delta(\rp) \;=\; 0\,.
\label{res-cr-lead}
\eeq
Resonant lines have negative slopes in the $(X, Y)$ plane. 

\subsubsection{GC outside $\rstar\,$} 
When $\rstar \leq \rp < 300~\!\pc$, we have 
$\,-0.33 < \Delta \leq 0\,$, and equation~(\ref{res-cr-lead}) gives
\beq
X \;=\; X_{\rm r}(Y;m,\ell,\rp) \;\;\stackrel{{\rm def}}{=}\;\; 
\frac{|\Delta|}{3} \;-\; \frac{\,|\ell|\,}{3m}Y\,
\label{xres-cr-lead}
\eeq
as the resonant line, which connects the points $(X_1\,, Y_1 = X_1)$ and $(X_2\,, 0)$, where 
\beq
X_1 \;=\; \frac{|\Delta|}{3}\left(1 \,+\, \frac{|\ell|}{3m}\right)^{-1}\,,
\qquad
X_2 \;=\; \frac{|\Delta|}{3}\,.
\label{x1x2}
\eeq
We note that $X_2$ is independent of $(m, \ell)$, whereas the ratio
$X_2/X_1 = \left(1 + |\ell|/3m\right) > 1\,$ is independent of $\rp\,$.
The resonance strength factor is  
\beq
\scrf_{m\ell m}(\rp) = 
\frac{\,m\,}{\,3\,}\!  
\int_0^{X_1}\rmd Y\,A(X_{\rm r}\,,Y)\,P_{m\ell m}(X_{\rm r}\,, Y)\,.
\label{torq-cr-lead-out}
\eeq
As $\rp \to \rstar$ the upper limit of integration, $X_1 \propto |\Delta| \to 0\,$, and all the $\scrf_{m\ell m}$ vanish. 

\begin{figure*}
\gridline{
\fig{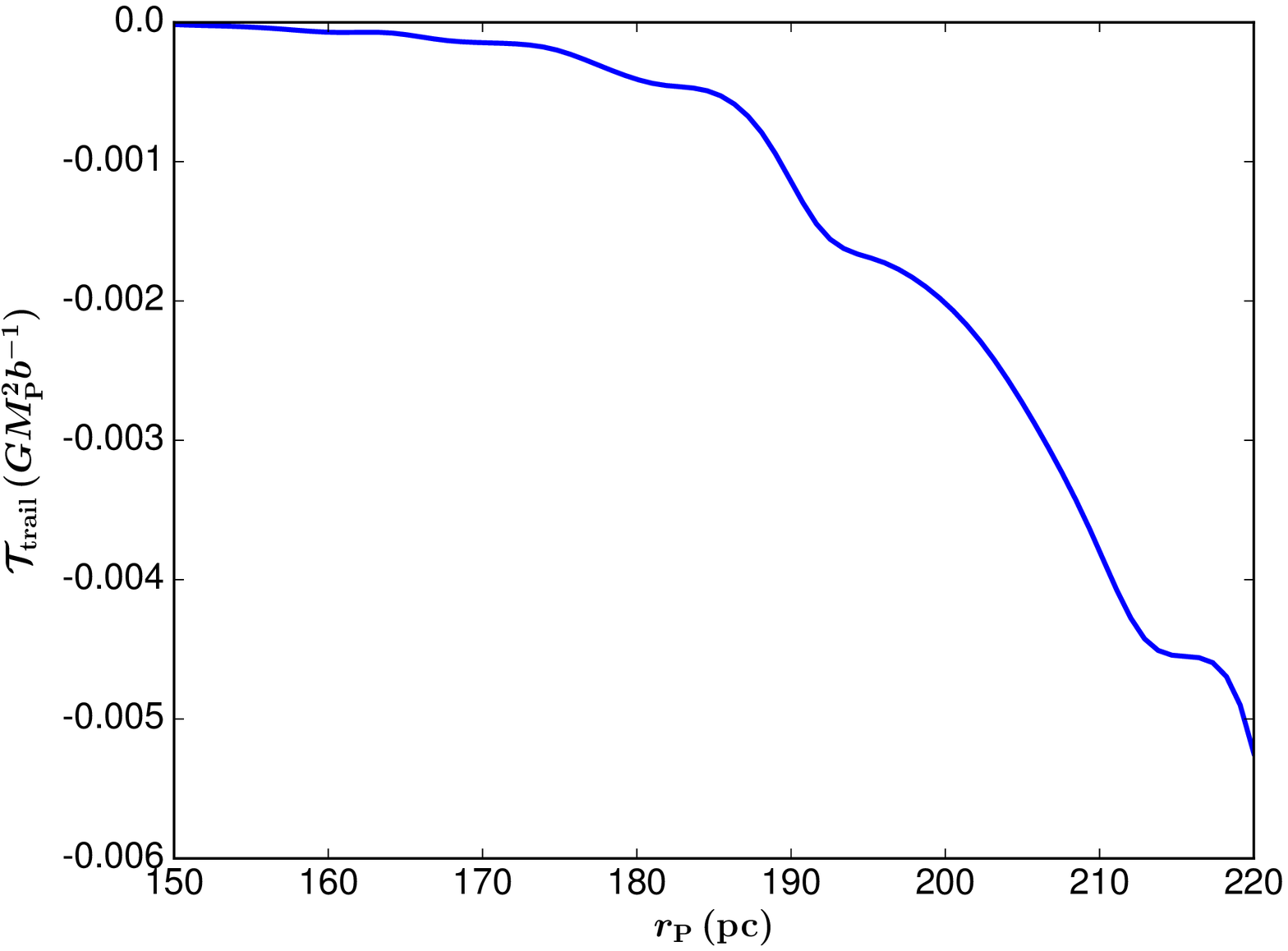}{0.34\textwidth}%figs/figs_trail_in/CRtrail_in_torq
{(a) Trailing, GC inside $\rstar$.}\hspace{-0.3cm}
\fig{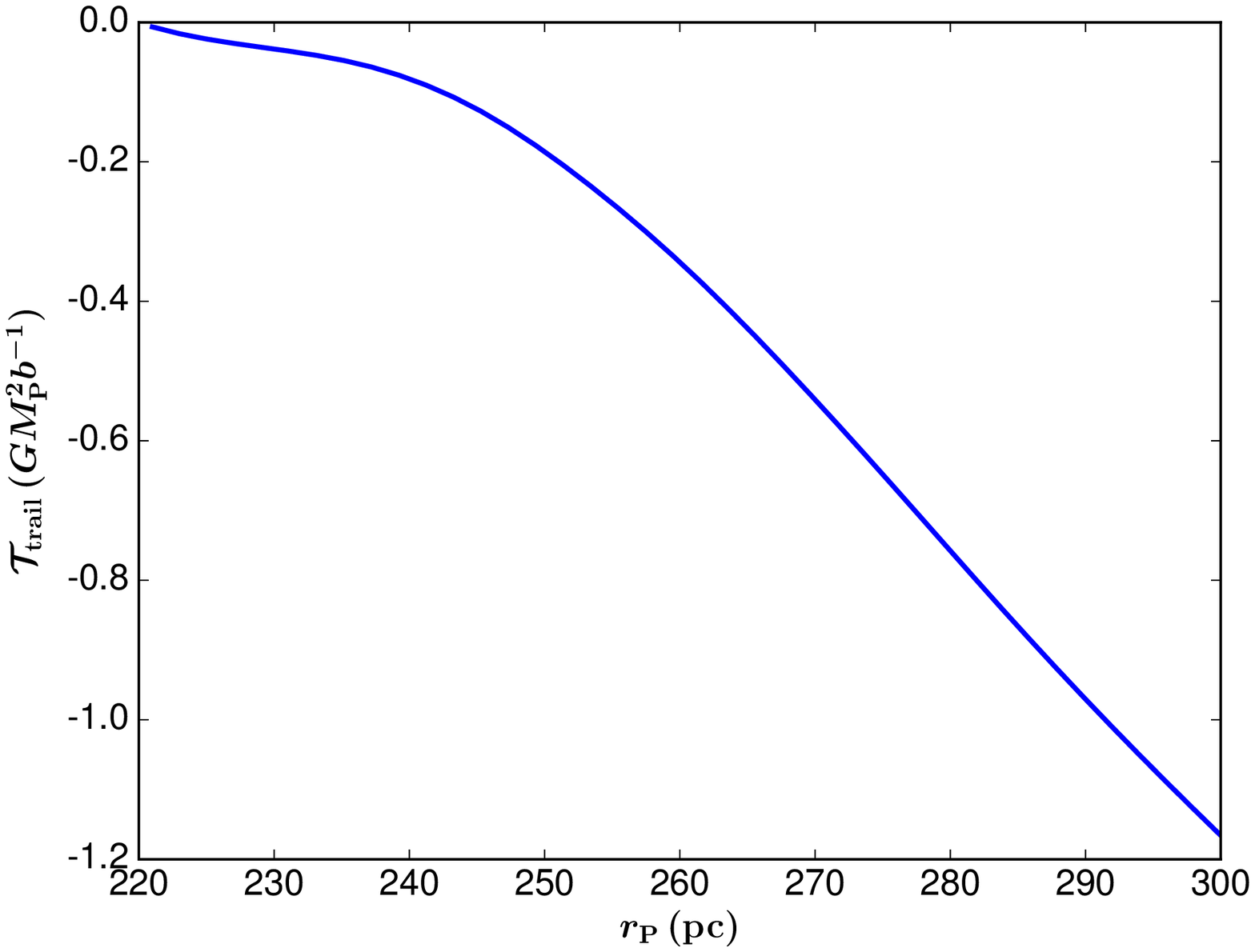}{0.34\textwidth}%figs/figs_trail_out/CRtrail_torq
{(b) Trailing, GC outside $\rstar$.}\hspace{-0.3cm}
\fig{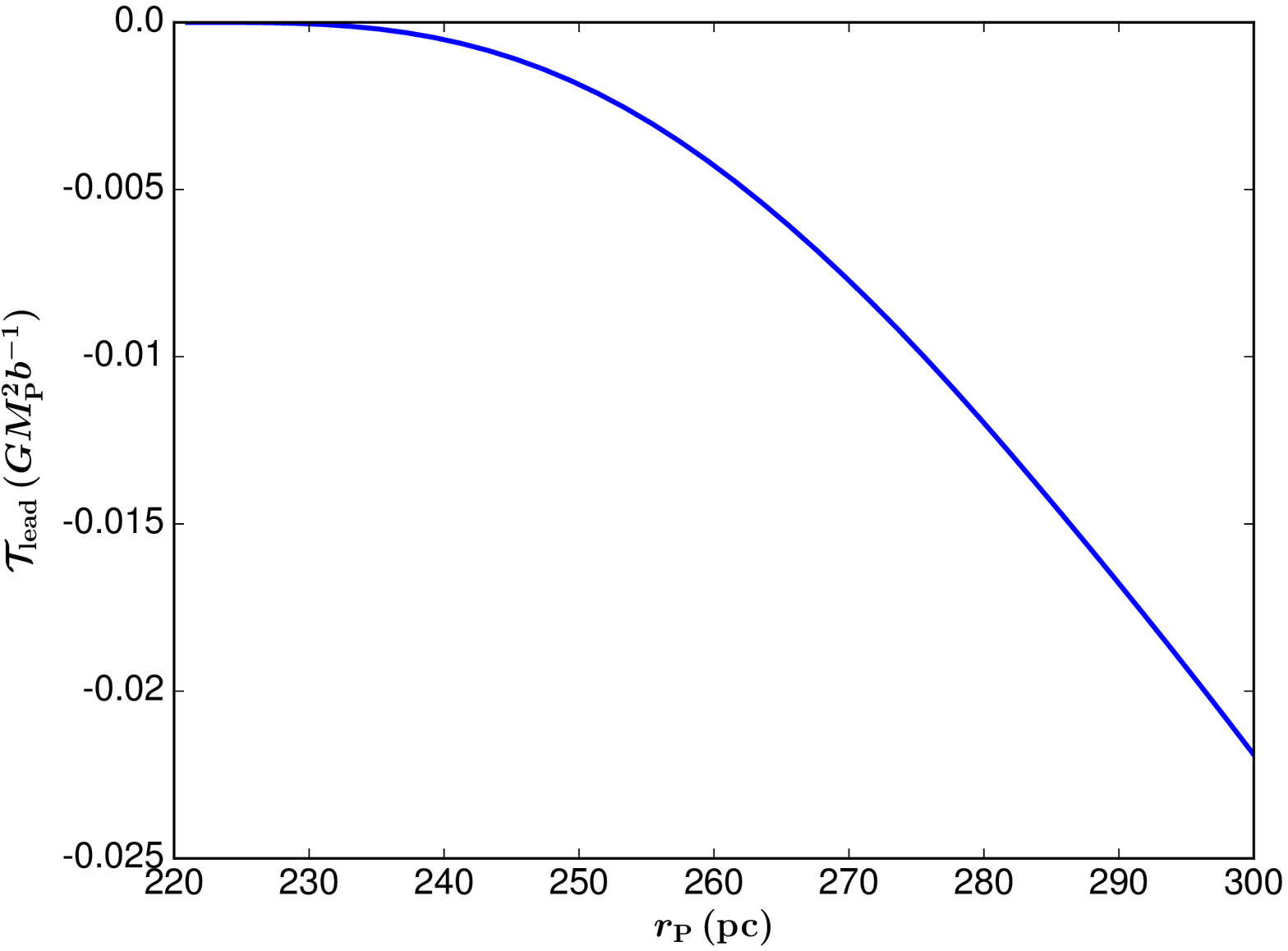}{0.34\textwidth}%figs/figs_lead/CRlead_torq
{(c) Leading, GC outside $\rstar$.}         
          }
\caption{\emph{Profiles of net Trailing and Leading CR Torques}.}
\vspace{0.2cm}
\label{fig_torq-prof}
\end{figure*}

We calculated $\scrf_{m\ell m}(\rp)$ for all the $110$ resonances with $1 \leq m \leq 10$ and  $-10 \leq \ell \leq 0\,$. Substituting these in equation~(\ref{torq-res-core}) we obtained the corresponding $\scrt_{m\ell m}(\rp)$. The six panels of Figure~(\ref{fig_ml_lead}) track the Leading CR resonances with $\vert\scrt_{m\ell m}(\rp)\vert > 10^{-5}(G \Mp^2/b)$, for $\rstar \leq \rp < 300~\!\pc $. Then $\scrt_{\rm lead}(\rp)$ was calculated by summing over the $\scrt_{m\ell m}(\rp)$, as given in equation~(\ref{lead-tor}). As in the case of the Trailing torques discussed in \S~\!5.1.2., there is progressive loss of resonances and torque strengths, as $\rp$ decreases. But the Leading resonances are fewer in number and weaker. $(2, 0)$ and $(2, -2)$ are the dominant resonances throughout the range of $\rp$.   
\begin{itemize}
\item[$\bullet$] At $\rp = 300~\!\pc$ there are $18$ resonances, with 
$10^{-5}(G \Mp^2/b) < \vert\scrt_{m \ell m}\vert \lesssim 10^{-2}(G \Mp^2/ b)$. The net torque due to all the resonances is $\scrt_{\rm lead} \simeq -2\times 10^{-2}(G \Mp^2/b)$.

\item[$\bullet$] At $\rp = 245~\!\pc$ there are only $6$ resonances, with 
$10^{-5}(G \Mp^2/b) < \vert\scrt_{m \ell m}\vert \lesssim 10^{-3.2}(G \Mp^2/ b)$. The net torque due to all the resonances is $\scrt_{\rm lead} \simeq -
10^{-3}(G \Mp^2/b)$. 

\item[$\bullet$]
At $\rp = 229~\!\pc$ there is just one resonance left, $(2, 0)$, with 
$\vert\scrt_{m \ell m}\vert > 10^{-5}(G \Mp^2/b)$. The net torque due to 
this, together with contributions from weaker resonances, is $\scrt_{\rm lead} \simeq -3\times 10^{-5}(G \Mp^2/b)$.
\end{itemize}
The torque profile, $\scrt_{\rm lead}(\rp)$, is given in Figure~(\ref{fig_torq-prof}c) and discussed in \S~\!6.

\bigskip
\subsubsection{GC inside $\rstar\,$} 
When $\rp < \rstar\,$, we have $\Delta > 0\,$,
and there are no solutions of equation~(\ref{res-cr-lead}) that lie in 
the unit triangle. Hence Leading CR resonances do not exist for $\rp < \rstar$, and the associated strength factors must vanish:
\beq
\scrf_{m\ell m}(\rp) \;=\; 0\,,\quad\mbox{when $\,\ell \leq 0\,$
and $\,\rp \leq \rstar\,$.} 
\label{torq-cr-lead-in}
\eeq
Hence all the $\scrt_{m\ell m}(\rp) = 0$, and the net Leading CR torque,
$\scrt_{\rm lead}(\rp) = 0$ when the GC is inside $\rstar\,$.\footnote{A characteristic feature revealed in Figures~(\ref{fig_ml_trail_out})---(\ref{fig_ml_lead}) is that $\scrt_{m\ell m}(\rp)$, with 
$(m, \ell)$ both even or both odd, have larger magnitudes than those corresponding to the even-odd or odd-even cases.} 

\bigskip
\section{Suppressed dynamical friction}

\subsection{Torque profiles and suppression factors}

The Trailing and Leading net torque profiles, $\scrt_{\rm trail}(\rp)$
and $\scrt_{\rm lead}(\rp)$, were calculated for $150~\!\pc \leq \rp \leq 300~\!\pc$, by summing over $\scrt_{m\ell m}\,$, as discussed in \S~\!5.
These are plotted in Figure~(\ref{fig_torq-prof}), whose salient features can be understood with reference to Figures~(\ref{fig_ml_trail_out})---(\ref{fig_ml_lead}). In this section all torque values are referred to in units of $G\Mp^2b^{-1}$.
\begin{itemize}
\item[$\bullet$] Reading panels (a, b) of Figure~(\ref{fig_torq-prof}) from right to left, we see that $\vert\scrt_{\rm trail}\vert$ decreases from about $1.17$ at $\rp = 300~\!\pc$, to about $0.22$ at $\rp = 252~\!\pc$. The curve is smooth because there are several low and high $\ell$ resonances in operation throughout, counting $43$ at $\rp = 300~\!\pc$ and $35$ at $\rp = 252~\!\pc$ with $\vert\scrt_{m \ell m}\vert > 10^{-5}$. The strongest of these is the $(2,2)$, but there are a handful of others of near-comparable strengths. $\vert\scrt_{\rm trail}\vert$ continues to decrease with $\rp$, with $\vert\scrt_{\rm trail}\vert \simeq 6\times 10^{-2}$ at $\rp = 237~\!\pc$.  The number of significant resonances has thinned out; there are only $24$ with $\vert\scrt_{m \ell m}\vert > 10^{-5}$, of which the $(1,3)$ is the strongest. We are on the verge of a transition, where the low $\ell$ resonances are rapidly losing strength and cease to exist for $\rp \leq \rstar \simeq 220~\!\pc$. $\vert\scrt_{\rm trail}\vert$, which now comes from only high $\ell$ resonances, is small. How small this is cannot be discerned from the left end of panel (b), but is seen to be about $5\times 10^{-3}$ from the right end of panel (a). $\vert\scrt_{\rm trail}\vert$ declines more rapidly inside $\rstar$. But in contrast to the nearly featureless behavior outside $\rstar$, $\vert\scrt_{\rm trail}\vert$ shows steep falls interspersed with plateaus. The reason for this is the transition to a state in which the main contribution to  $\vert\scrt_{\rm trail}\vert$ comes from just one or two resonances, whose dominance is transitory. The dominant resonances inside $\rstar$ are: the $(1,5)$ just inside $\rstar$, the $(1, 7)$ at $\rp = 185~\!\pc$, and the $(1, 9)$ at $\rp = 164~\!\pc$. The progressive shift to higher $\ell$ as $\rp$ decreases is mainly responsible for the declining torque strength, with $\vert\scrt_{\rm trail}\vert \simeq 1.6\times 10^{-5}$ at $\rp = 150~\!\pc$.

\item[$\bullet$] Panel (c) of Figure~(\ref{fig_torq-prof}) shows that $\vert\scrt_{\rm lead}\vert$ decreases smoothly as $\rp$ decreases.
When compared with panel (b), we see that $\vert\scrt_{\rm lead}(\rp)\vert \ll \vert\scrt_{\rm trail}(\rp)\vert$, because Leading resonances are 
weaker and fewer in number. Similar to low $\ell$ Trailing resonances, the 
Leading resonances exist only when the GC is outside $\rstar$, with 
$(2, 0)$ and $(2, -2)$ being the dominant ones. At $\rp = 300~\!\pc$ there are $18$ resonances with $\vert\scrt_{m \ell m}\vert > 10^{-5}$, 
contributing to $\vert\scrt_{\rm lead}\vert \simeq 2\times 10^{-2}$. At $\rp = 245~\!\pc$, there are only $6$ resonances with $\vert\scrt_{m \ell m}\vert > 10^{-5}$ giving $\vert\scrt_{\rm lead}\vert \simeq 10^{-3}$. Thereafter 
the torque is highly suppressed, with $\vert\scrt_{\rm lead}\vert \simeq 3\times 10^{-5}$ at $\rp = 229~\!\pc$, and $\vert\scrt_{\rm lead}\vert = 0$ at $\rp = \rstar$.
\end{itemize}

The torque profiles in Figure~(\ref{fig_torq-prof}) should be compared with 
those of the Chandrasekhar torque, $\scrt_{\text{C}}(\rp)$, of Figure~(\ref{decay-ch}a). At $\rp = 300~\!\pc$, the four different 
versions of the Chandrasekhar formula give values of $\scrt_{\text{C}}$ ranging between $-1.14$ and $-1.7\,$. At $\rp = 300~\!\pc$ the net LBK torque is $\scrt \,=\, \scrt_{\rm trail} + \scrt_{\rm lead} \simeq -1.17 - 0.02 \,=\, -1.19\,$, and hence $\scrt \approx\scrt_{\text{C}}$. This is reassuring, and    
serves as a consistency check: when several resonances of comparable strengths are active, which is the case at $\rp = 300~\!\pc$, 
we expect $\scrt \approx \scrt_{\text{C}}$, as indeed we found. In order to demonstrate how really suppressed (inside $300~\!\pc$) 
both $\scrt_{\rm trail}(\rp)$ and $\scrt_{\rm lead}(\rp)$ are \emph{vis}-$\grave{a}$-$vis$ $\,\scrt_{\text{C}}(\rp)$, we compare them with the solid blue curve of Figure~(\ref{decay-ch}a), whose torque has the least magnitude among the four curves. In this case $\vert\scrt_{\text{C}}\vert$ decreases gradually from about $1.14$ at $\rp = 300~\!\pc$, to $0.6$ at $\rp = 
\rstar$, and $0.29$ at $\rp = 150~\!\pc$.

\begin{figure*}
\gridline{\fig{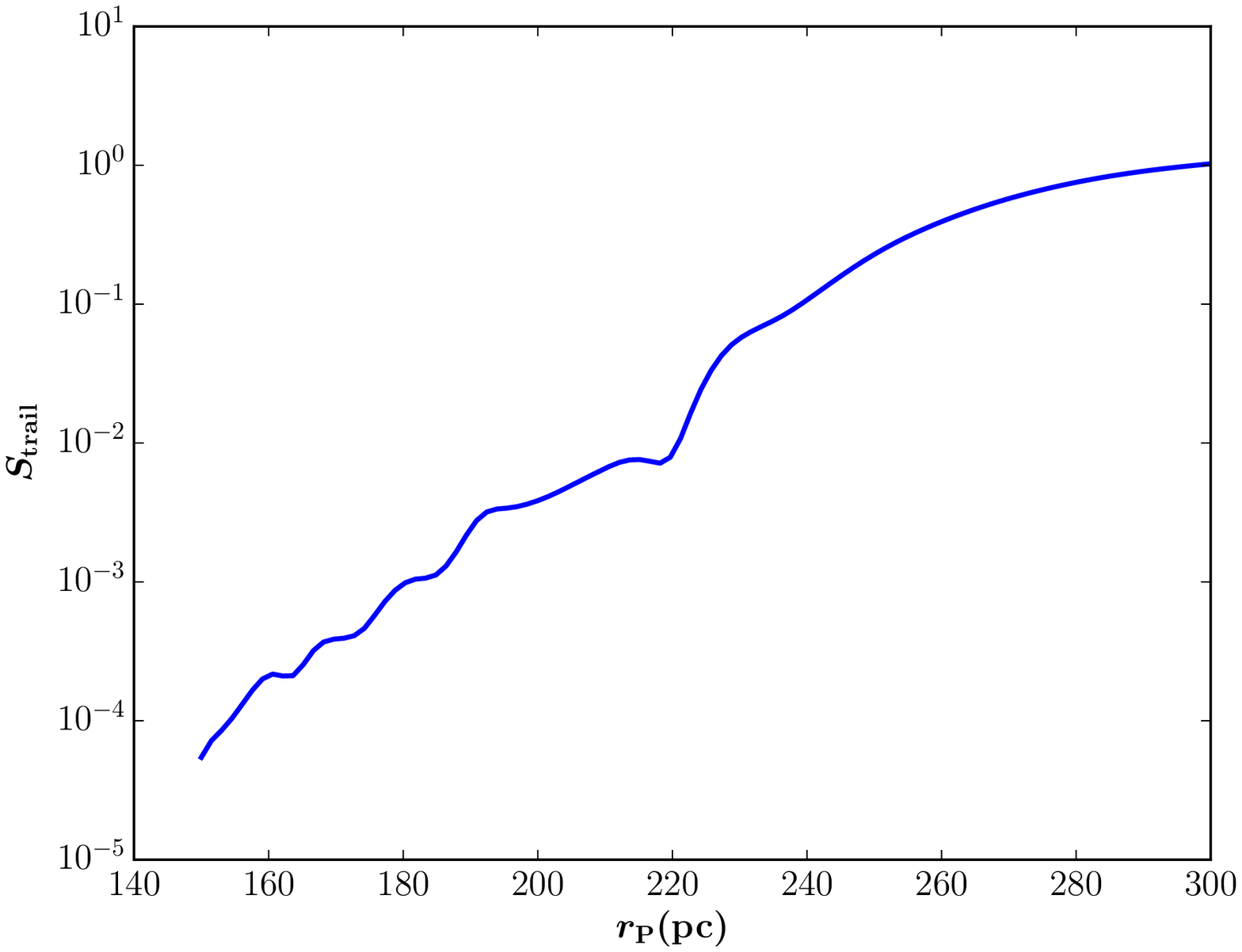}{0.5\textwidth}{(a) Trailing. }%figs/fin_S_trail
         \fig{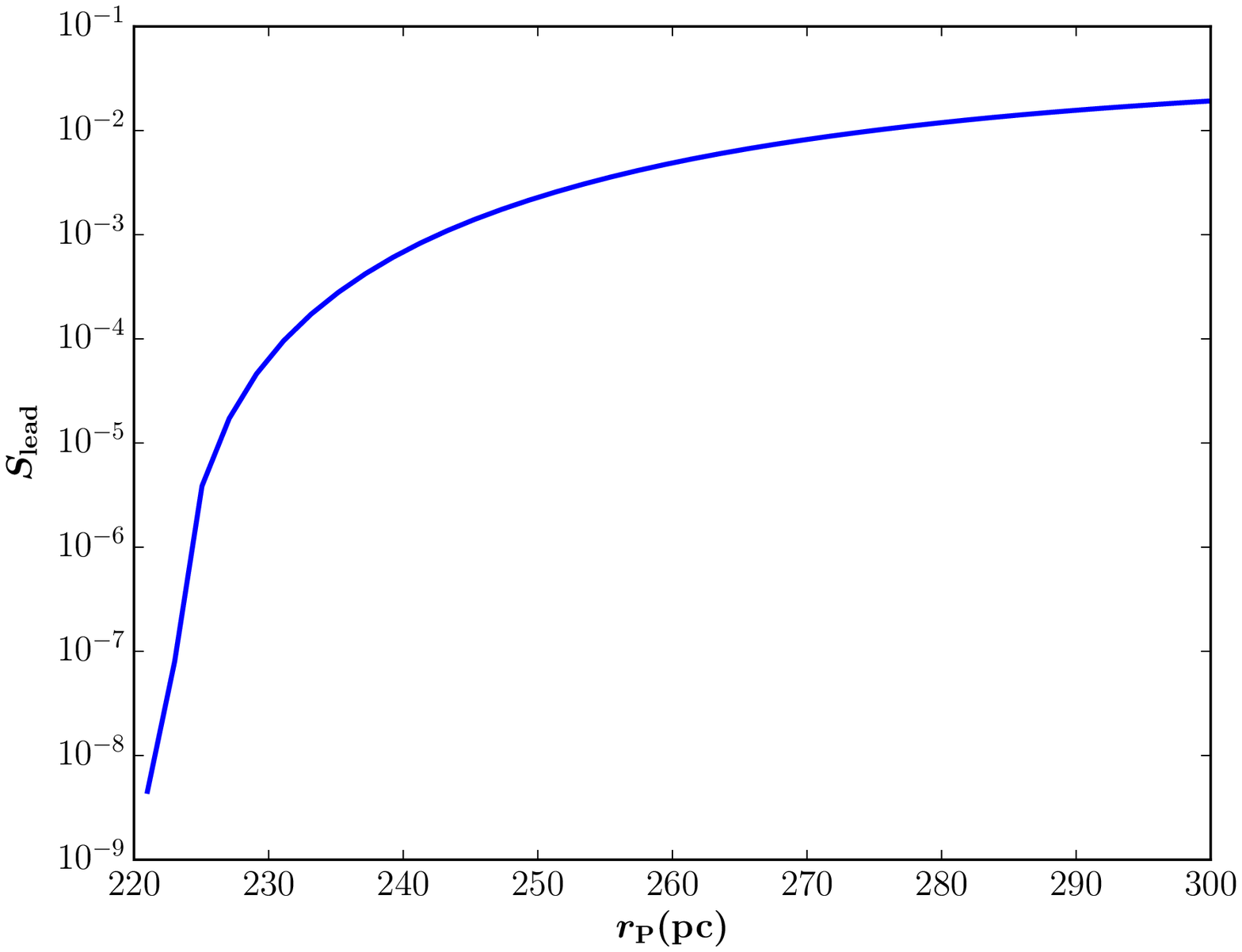}{0.5\textwidth}{(b) Leading. }%figs/fin_S_lead
          }
\caption{\emph{Suppression factors for Trailing and Leading torques}.} 
\label{fig_S}
\vspace{0.2cm}
\end{figure*}

The Trailing and Leading torque suppression factors,
\beq
S_{\rm trail}(\rp) = \frac{\scrt_{\rm trail}(\rp)}{\scrt_{\text{C}}(\rp)}\,,\quad\;
S_{\rm lead}(\rp) = \frac{\scrt_{\rm lead}(\rp)}{\scrt_{\text{C}}(\rp)}\,,
\label{supp-fac}
\eeq
are plotted in Figure~\ref{fig_S}, where it should be noted that the 
ordinates are displayed in a logarithmic scale. Both $S_{\rm trail}$ and $S_{\rm lead}$ decrease as $\rp$ decreases, with $S_{\rm trail} \gg S_{\rm lead}$. As Figure~(\ref{fig_S}a) shows, $S_{\rm trail} \simeq 1$ at $\rp = 300~\!\pc$ and decreases rapidly as $\rp$ decreases. There is a steep drop as $\rp$ approaches $\rstar$, when the low $\ell$ Trailing resonances begin losing strengths, and cease to exist at $\rp = \rstar$ where $S_{\rm trail} < 10^{-2}$ is mostly due to the $(1, 3)$ and $(1, 5)$ resonances. Inside $\rstar$, $S_{\rm trail}$ continues to decline rapidly, showing the steep falls and plateaus of Figure~(\ref{fig_torq-prof}a). As noted earlier these features are due to the transitory dominance of a succession of high $\ell$ resonances, the $(1,5)\,, (1,7)\,, (1,9)~\mbox{etc}$, as $\rp$ decreases, until $S_{\rm trail} < 10^{-4}$ at $\rp = 150~\!\pc$. Figure~(\ref{fig_S}b) shows $S_{\rm lead}$ falling steadily from $2\times 10^{-2}$ at $\rp = 300~\!\pc$, to about $6\times 10^{-5}$ at $\rp = 230~\!\pc$, followed by a steeper decline until it vanishes at $\rp = \rstar$.

\subsection{Stalling of the GC's orbit}

Let us suppose that the GC was set on a circular orbit of radius 
$\rp = 750~\!\pc$ at some initial time. We expect that the net LBK torque,
$\scrt \approx\scrt_{\text{C}}$ for $\rp \gtrsim 300~\!\pc$,
so $\rp(t)$ decays initially according to the Chandrasekhar formula, as discussed in \S~\!3.2. Figure~(\ref{decay-ch}b) shows $\rp(t)$ for four different versions of $\scrt_{\text{C}}$. The time for $\rp$ to decay from $750~\!\pc$ to $300~\!\pc$ varies from $5.6~\!\Gyr$ to $7.3~\!\Gyr$. Thereafter $\scrt(\rp)$ departs from $\scrt_{\text{C}}(\rp)$, as discussed in detail in \S~\!6.1. So we must calculate further orbital decay by using the LBK torque. The equation governing $\rp(t)$ is:
\beq
\Mp\,\frac{\rmd}{\rmd t}\!\left(\omp\rpsq\right) \,=\, \scrt(\rp) \,=\,
\scrt_{\rm trail}(\rp) + \scrt_{\rm lead}(\rp)\,.
\label{rp-eqn-lbk}
\eeq

\begin{figure}[h]
\centering
  \includegraphics[width=0.5 \textwidth]{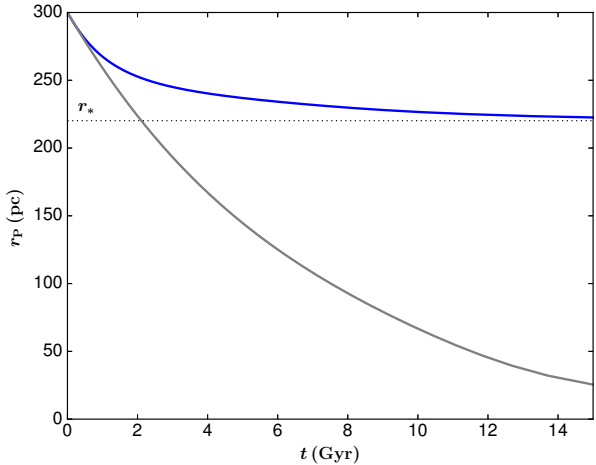} %CRorb_decay
  \caption{\emph{Orbital decay $\rp(t)\,$}. The blue curve is for the LBK torque, and the gray curve is for the Chandrasekhar torque discussed in the text.}
  \vspace{0.2cm}
\label{orb-decay-fin}  
\end{figure}

Equation~(\ref{rp-eqn-lbk}) was integrated numerically, with initial condition $\rp = 300~\!\pc$ at $t=0$, using the torque profiles of Figure~(\ref{fig_torq-prof}). The resulting $\rp(t) = R_{\mbox{\scriptsize{\,LBK}}}(t)$ is shown as the blue curve in Figure~(\ref{orb-decay-fin}). It is evident that orbital decay
is highly suppressed. We compare this with the gray curve, $\rp(t) = R_{\text{C}}(t)$, for orbital decay with the most highly suppressed of the Chandrasekhar torques, used in equation~(\ref{supp-fac}).  
For small $t < 0.5~\!\Gyr$ the two curves overlap. This is because 
$\scrt \simeq \scrt_{\text{C}}$ near $\rp = 300~\!\pc$. 
Thereafter they depart:
\begin{itemize}
\item[$\bullet$] After $3~\!\Gyr$, $\,R_{\mbox{\scriptsize{\,LBK}}}\simeq 245~\!\pc$, while $R_{\text{C}}$ drops below $\rstar$ and reaches close to
$190~\!\pc$.

\item[$\bullet$] After $6~\!\Gyr$, $\,R_{\mbox{\scriptsize{\,LBK}}}\simeq 235~\!\pc$ and $R_{\text{C}}\simeq 125~\!\pc$. Decay slows down.
 
\item[$\bullet$] After $14~\!\Gyr$, $\,R_{\mbox{\scriptsize{\,LBK}}}$ hovers just above $\rstar$, while $R_{\text{C}}$ has plunged to $30~\!\pc$. 
\end{itemize}
The blue curve, $R_{\mbox{\scriptsize{\,LBK}}}(t)$, appears like an asymptote to 
$\rp = \rstar$, but this is not really the case. Eventually $\rp(t)$ will drop below $\rstar$, because $\scrt_{\rm trail}(\rp)$ is non-zero for $\rp < \rstar$. But the time scales are much longer than is astronomically interesting. 

\section{Conclusions}

Dynamical friction on a globular cluster (GC) of mass $\Mp = 2 \times 10^5~\!\msun$ --- set on an initially circular orbit inside an Isochrone model of 
a dwarf galaxy --- is highly suppressed when the GC's orbit enters an inner-core region. For a galaxy with core radius $b = 1000~\!\pc$ and core mass 
$M_{\rm c} = 4 \times 10^8~\!\msun$ (see \S~3.1.1 for choice of parameter values) this corresponds to the GC's orbital radius $\rp \leq 300~\!\pc$. We found that, when $\rp = 300~\!\pc$, the retarding LBK torque on the GC, 
$\scrt \approx\scrt_{\text{C}}$, the Chandrasekhar torque. Inside $300~\!\pc$, the LBK torque is highly suppressed because of progressive transition 
to states in which there are fewer and weaker resonances in operation, 
so $\vert\scrt\vert \ll \vert\scrt_{\text{C}}\vert$. The orbital decay slows down drastically and, over astronomically interesting time scales, the GC appears to stall at a radius $\rp \gtrsim \rstar \simeq 220~\!\pc$.

In the In11 simulation, the GC's orbital decay slows down
when $\rp < 300~\!\pc$, and it appears to  stall around a mean value of about $225~\!\pc$ until the end of the simulation about $6~\!\Gyr$ hence. In11 also studied the energy gained/lost by the stellar orbits and identified the sort of orbits that interact the strongest with the GC. Figure~10 of In11 displays some sample orbits, where it can be seen that these are nearly co-rotating with the GC and lagging it slightly.
We found similar behavior in our calculations of torques and orbital decay in \S~\!5 and \S~\!6. From Figure~(\ref{orb-decay-fin}) we see that the GC's orbital decay indeed slows down inside $300~\!\pc$, with $\rp \simeq 245~\!\pc$ after $3~\!\Gyr$, and not reaching $\rstar$ even after $14~\!\Gyr$.
Moreover, the strongest torques are exerted by Trailing Co-rotating resonances.  

The close agreement between the stalling radius in In11, and the range 
$245 - 220~\!\pc$ we obtained is somewhat fortuitous, because the Isochrone and Burkert density profiles behave differently near the center. The Isochrone has an analytic core density profile, $\rho_0(r) \simeq \rho_0(0)\left[1 - 5r^2/3b^2\right]$ for $r \ll b$, falling quadratically with $r$. The Burkert has a non-analytic core density profile, $\rho_{\rm B}(r) \simeq \rho_{\rm c}\left[1 - r/r_{\rm c}\right]$ for $r \ll r_c$, which falls linearly with $r$; the three dimensional density gradient is singular at the origin. The corresponding mass profiles are $M_0(r) \simeq (4\pi/3)\rho_0(0)r^3\left[1 - r^2/b^2\right]$ and $M_{\rm B}(r) \simeq (4\pi/3)\rho_{\rm c}r^3\left[1 - 3r/4r_{\rm c}\right]$. 
Solving equation~(\ref{rstar-eqn}),
\begin{subequations}
\begin{align}
\rstar &\;\simeq\; \left(\frac{\Mp}{M_{\rm c}}\right)^{1\!/5}b
\;\simeq\; 220~\!\pc\,,\qquad\mbox{Isochrone}
\label{rstar-iso}\\[1em]
\rstar &\;\simeq\; \left(\frac{4\Mp}{3M_{\rm c}}\right)^{1\!/4}
r_{\rm c} \;\simeq\; 160~\!\pc\,.\qquad\mbox{Burkert}
\label{rstar-bur}
\end{align}
\end{subequations}
where $M_{\rm c}$ is the core mass of equation~(\ref{core-mass}). 

The Burkert profile has a smaller $\rstar$ because its mass deficit, 
defined in equation~(\ref{rstar-eqn}), increases more strongly with $r$. 
For the Isochrone, the `stalling' radius $\lesssim 240~\!\pc$ is not 
too far away its $\rstar \simeq 220~\!\pc$. But why does the GC in the In11 simulation appear to stall at $\sim 225~\!\pc$, which is about half-way between $300~\!\pc$ and the Burkert's $\rstar \simeq 160~\!\pc$~\!? The first step toward addressing this question would be to repeat the calculations of this paper for the Burkert profile, and see how resonances drop off and torques weaken as $\rp$ drops below $300~\!\pc$.
Some differences can surely be expected, because $\rho_{\rm B}(r)$ is 
non-analytic at the center and falls steeper than $\rho_0(r)\,$. But this would probably not compass the entire story. 

Figure~13 of In11 plots energy transfer --- which is proportional to the angular momentum transfer --- between the stars and the GC. Most of the stars absorb angular momentum from the GC, whereas a very small fraction of the stars (few thousands out of ten million) lose angular momentum to the GC. The former behave like the stars we studied in this paper; they absorb angular momentum from the GC and contribute to the net retarding LBK torque in proportion to the resonant torques strengths. But the latter population --- referred to as `horn particles' in In11 --- cannot be so described because, in the TW84 theory, angular momentum is always absorbed by the stars for any galaxy with DF $F(E)$ with $(\rmd F/\rmd E) < 0\,$. In Figure~13 of In11 the two populations of stars are seen to exert almost equal and opposite torques on the GC, so that the total torque is effectively zero. How do we understand this in terms of our exploration of resonances in the inner-core of the Isochrone?

Our calculations are a direct application of the LBK torque formula of TW84,  
which is derived from the linear theory of the collisionless Boltzmann equation (CBE). In order to describe the `horn particles' it is necessary to go beyond linear theory and take into account the non-linear theory of adiabatic capture into resonance, which is discussed in TW84. Non-linear corrections to the CBE create `islands' in the neighborhood of resonant surfaces in phase space. An isolated island consists of (`captured') orbits librating about a parent resonant orbit, and bounded by separatrices on which the libration period is infinite.\footnote{Beyond the separatrices are (`free') circulating orbits that are reasonably well described by the linear CBE underlying the calculation of the LBK torque.} The `horn particles' 
in In11 must be a population of stars captured in one or more resonant islands in phase space. Only a small fraction of the stars are `horn particles' because resonant islands occupy small phase volumes, of order the square-root of the perturbation. As the GC's orbit decays, the locations of resonant surfaces in $(I, L)$-space will drift slowly with time, as will the sizes and shapes of the resonant islands.  

For well-separated resonant islands the non-linear perturbation to the galaxy's DF can be calculated using the theory of \citet{st96}. Such a calculation must demonstrate that the captured stars lose angular momentum to the GC, thereby pushing it away so that it stalls somewhat farther away out from $\rstar$, nearer to $300~\!\pc$, as seen in In11. We recall from TW84 that, when many resonances of comparable strengths are active, the net effect is incoherent so the LBK torque is reasonably approximated by the Chandrasekhar torque. In contrast when only a few Trailing Co-rotating resonances dominate --- as in the inner core region --- the net effect could be cooperative, as suggested in In11. We have seen that the LBK torque itself is highly suppressed, so the oppositely directed torque due to the small number of resonantly captured stars may well suffice to cancel it. Then the galaxy and the GC will no longer exert torque on each other, and we would have an approximate self-consistent solution of the collisionless Boltzmann equation describing the galaxy and the GC in a state of frictionless rotation. 
 
\appendix
\section{Fourier coefficients of the tidal potential of the GC}

The Fourier coefficients $\Phitilda_{n \ell m}$ of the tidal potential 
$\phip$ of GC are given as in equation~(\ref{phi-tilda}),
\beq
\Phitilda_{n\ell m}(I,L,L_z) \;=\,
\oint \frac{\rmd w}{2\rmpi}\,\frac{\rmd g}{2\rmpi}\,
\frac{\rmd h}{2 \rmpi}\;
\phip(w,g,h;I,L,L_z) 
  \exp\left\{-\,\rmi\left(n w + \ell g + m h\right)\right\}\,,
\label{phi-tilda2}
\eeq
in terms of a three dimensional Fourier integral over the angles
$(w, g, h)$. For Co-rotating resonances $n=m$, so $w$ and $h$ occur
only in the combination $w + h$. Transforming to new integration variables, $\left( w' , g , k  \right)$, where $w'= w + h$ and $k= - h$ we get
\beq
\Phitilda_{m\ell m}(I,L,L_z) \;=\; \oint \frac{\rmd w'}{2 \rmpi} \, \frac{\rmd g}{2 \rmpi} \, \left<\phip\right>_k  \exp\left\{-\,\rmi\left(m w' + \ell g \right)\right\}\,,  
\label{phi-tilda2}
\eeq
where 
\beq
\left<\phip\right>_k \;=\; \oint \frac{\rmd k}{2\rmpi}\,\phip
\label{phi-kav}
\eeq
is the $k$-averaged tidal potential. Below we show that this can be evaluated analytically for core orbits. Then $\Phitilda_{m\ell m}$ is given as a two dimensional Fourier transform over $w'$ and $g$ of a known function. The Fourier integrals were evaluated numerically using {\tt Mathematica} with a relative tolerance of $1~\!\%$. Integrals of small magnitudes converge slowly when the relative error is specified, so 
we used absolute tolerances of $10^{-6}$ and $10^{-8}$ for $\rp > \rstar$ and $\rp < \rstar$, respectively. 

\smallskip
\noindent
{\bf Calculation of $\left<\phip\right>_k\,$:}
Let ($x$,\, $y$,\, $z$) be cartesian coordinates in the rotating frame
in which the GC is quasi-stationary. Without loss of generality we assume that the GC lies on the $x$-axis. Then the tidal potential of equation~(\ref{pert-tid}) is: 
\beq
\phip \;=\; -\,GM_p\left[\,\frac{1}{\sqrt{a^2 + \rp^2 +r^2 - 2 \, \rp \, x }}  \,\;-\;\, \frac{\rp \, x}
{\,\left(a^2 + \rpsq\right)^{3/2}\,}\,\right]\,. 
\label{pert-tid2}
\eeq
For core orbits equation~(\ref{rpsi-core}) gives, 
\begin{subequations}
\begin{align}
r^2 &\;\simeq\; \frac{I}{\omb}\left[\,1 - e\cos(2w)\,\right]\,,\\[1em]
x  &\;\simeq\; \sqrt{ \frac{\, I \,}{\, \omb \,}}\,\left[ \sqrt{1-e}\, C_w \,( C_g \, C_h - S_g \, S_h \, C_i ) - \sqrt{1+e} \, S_w \, (S_g \, C_h + C_g \, S_h \, C_i)  \right]\,,
\end{align}
\end{subequations}
where $C \equiv \text{cosine}\,$ and $\,S \equiv \text{sine}$. Using these in equation~(\ref{pert-tid2}), $\,\phip$ can be expressed in terms of action-angle variables. The next task is to express quantities in terms of $(w', g, k)$ and average $\phip$ over $k$. It is more convenient, and mathematically equivalent, to average over the angle $\beta = 2 k + w'$, instead of over $k$. Rewriting 
\begin{subequations}  
\begin{align}
r^2 &\;\simeq\; \frac{I}{\omb} \left( 1 - e \, C_{w'} C_{\beta} + e \, S_{w'} S_{\beta}  \right)\,, \\[1ex]
x &\;\simeq\; \sqrt{ \frac{I}{\omb} } \; \bigg[ \frac{ \sqrt{1-e}}{2} \big( C_g C_{w'} - S_{g} S_{w'} C_i  \big) - \frac{ \sqrt{1+e}}{2} \big( S_g S_{w'} - C_g C_{w'} C_i \big) \;+\; \notag \\[1em]
 & \qquad \qquad\qquad \frac{1}{2} C_g C_{\beta} \big( \sqrt{1-e} - \sqrt{1+e} \, C_i  \big) + \frac{1}{2} S_g S_{\beta} \big( \sqrt{1-e}\, C_i - \sqrt{1+e}\, \big) \bigg]\,,
 \label{coord_beta}
\end{align}
\end{subequations}
we have
\beq
a^2 \,+\, \rp^2 \,+\, r^2 \,-\, 2\rp x \;\simeq\; A \,+\, B\,C_{\beta} \,+\,
D\,S_{\beta}\,,
\eeq
where $A$, $B$ and $D$ are $\beta$-independent functions, given by
\begin{subequations}
\begin{align}
A &\;=\; a^2 + \rp^2 + \frac{I}{\omb} + \rp \sqrt{\frac{I}{\omb}}\, \bigg[ \sqrt{1+e} \, \big\{ S_g \, S_{w'} - C_g \,C_{w'} \, C_i \big\} -  \sqrt{1-e} \, \big\{ C_g \, C_{w'} - S_g \, S_{w'} \, C_i  \big\} \bigg]\,, \\[1ex]  
B &\;=\; -e \, C_{w'} \, \frac{I}{\omb} - \rp \sqrt{\frac{I}{\omb}} \, C_g \, \big\{ \sqrt{1-e} - \sqrt{1+e} \, C_i  \big\}\,, \\[1ex]
D &\;=\; e \, S_{w'} \, \frac{I}{\omb} - \rp \sqrt{ \frac{I}{\omb} } \, S_g \, \big\{ \sqrt{1-e}\, C_i - \sqrt{1+e} \, \big\}\,.
\end{align}
\end{subequations}
Then the integral
\beq
\mathscr{I}_1 \;=\; \int_{0}^{2 \pi} \frac{\rmd k}{2 \pi} \frac{1}{\sqrt{a^2 + \rp^2 + r^2 - 2\, \rp \, x} \,} 
\;\simeq\; \int_{0}^{2 \pi} \frac{\rmd \beta}{2 \pi} \frac{1}{ \sqrt{ A + B \, C_{\beta} + D \, S_{\beta} } }\,.
\eeq
Using equation~2.580(1) of \citet{gr07}, we have
\beq
\mathscr{I}_1 \;\simeq\; \frac{2 }{\pi \sqrt{A + \sqrt{ B^2 + D^2 } }} \; \mathcal{K}\bigg( \frac{ 2 \sqrt{B^2 + D^2}  }{ A + \sqrt{ B^2 + D^2 }} \bigg)
\eeq
where 
\beq
\mathcal{K}(s) \;=\; \int_{0}^{\pi/2} \rmd \theta \; \frac{1}{\sqrt{ 1 - s \, \sin^2{\theta}}} 
\eeq
is the complete elliptic integral of the first kind. We also need
\beq
\oint \frac{\rmd k}{2 \pi}\,x \;=\; \oint \frac{\rmd \beta}{2 \pi}\,x \;\simeq\; \frac{1}{2} \sqrt{ \frac{I}{\omb} } \bigg[ \sqrt{1-e}\, \left( C_g \, C_{w'} - S_g \, S_{w'} \, C_i  \right) - \sqrt{1+e} \, \left( S_g \, S_{w'} - C_g \, C_{w'} \, C_i \right)  \bigg]\,.
\eeq
Therefore the $k$-averaged tidal potential is:
\beq
\begin{split}
\left<\Phi_{\rm P}\right>_k &\;\simeq\; G\Mp \Bigg[- \frac{2}{\pi \sqrt{ A+\sqrt{B^2 + D^2}}} \, \mathcal{K}\bigg(  \frac{ 2 \sqrt{B^2 + D^2}  }{ A + \sqrt{ B^2 + D^2 }}\bigg)  \;+\; \\[1em]
&\frac{\rp}{2} \sqrt{\frac{I}{\omb \,(a^2 + \rp^2)^3}} \, \bigg\{ \sqrt{1-e} \,\big( C_g \, C_{w'} - S_g \, S_{w'} \, C_i \big) - \sqrt{1+e}\, \big( S_g \, S_{w'} - C_g \, C_{w'} \, C_i \big) \bigg\}
\Bigg] \, .
\end{split}
\eeq

\pagebreak

\end{document}